\documentclass[11pt,twoside,a4paper,reqno,intlimits]{amsart}
\usepackage[cp1251]{inputenc}
\usepackage[russian,english]{babel}
\usepackage{amssymb,graphicx,color}
\usepackage{mathrsfs,dsfont}
\usepackage{mathpazo}

\usepackage[pdfauthor={Brezhnev}, colorlinks=true, linktocpage=false,
            pdfwindowui=false, pdfmenubar=false, 
            citecolor=blue, urlcolor=blue, 
            pdfkeywords={quantum foundations, non-axiomaticity, 
            detector clicks, ensembles, superposition principle, 
            arithmetic, numbers, vector space, abstracting, 
            interpretations, self-referentiality},
            hyperfootnotes=true, pdfstartview=FitH,
            bookmarksopen=false, bookmarks=false]{hyperref}

\usepackage{CutPage}
\textcenter{155mm}{217mm}
\usepackage{YURA,mScripts,TOC}
\usepackage{Brackets,Symbols,Ulem}
\usepackage[red]{Foots}

\theoremstyle{remark}%
\newtheorem{comm}{
\indent\blue\textup{R\,\,\,e\,\,\,m\,\,\,a\,\,\,r\,\,\,k}~~\ignorespaces}
\newenvironment{comment}%
{\par\smaller[1]\begin{comm}}{\end{comm}}

\newcommand{\TRANS}[2][2]{\mathrel{\Over[1]{#2}
	{\REPEAT{\mathchar"439\mkern-1mu}{#1}\mkern-7mu\leadsto}}}
\newcommand{\GOTO}[2][2]{\mathrel{
	\Over[1]{#2}{\REPEAT{\mathchar"439}{#1} \mathchar"044B}}}

\def\Tplus{\mathbin{\Over [1.3]{\sss\smallfrown}{+}}}
\def\Uplus{\mathbin{\Under[0.65]{\ds\smash\smallsmile}
	{\raise0.15ex\hbox{$+$}}}}
\def\+{\mathbin{\hat+}}
\def\tplus{\ensuremath{\mathbin{
	\vcenter{\hbox{\larger\usefont{OT1}{pxtt}{m}{n}+}}}}}
\def\Approx{\mathrel{\setbox0=\hbox{$\sim$}
	\rlap{\copy0} \rlap{\raise0.4ex\copy0} \lower0.4ex\box0}}

\newcommand{\DOWN}[1][2]{\rotatebox[origin=cc]{-90}{$\goto[#1]$}}
\newcommand{\UP}[1][2]{\rotatebox[origin=cc]{90}{$\goto[#1]$}}

\def\ans#1{\{\state{#1}\}}
\def\state#1{{\,\underline{\!#1\!}\,}{}}
\def\ket#1{|\bo#1\rangle}
\def\bra#1{\langle\bo#1|}
\def\bet#1{|#1\)}

\def\obj#1{\raise0.4ex\hbox{$\sss\mkern-1mu\left\lgroup\right.
	\mkern-4mu$}#1\raise0.4ex\hbox{$\sss\mkern-2.5mu\left.
	\right\rgroup\mkern-4mu$}}

\def\fr{\mathtt{f\!}}
\def\zf{{\smaller[1]ZF}}
\def\boT{{\frak T}}
\def\Aleph{\oper[-4]{\bo\aleph}}
\def\SL{\textup{\texttt{Stat\-Length}}}

\def\statePsi{\state\Psi}
\def\statePhi{\state\Phi}
\def\ansPsi{\ans\Psi}
\def\ketPsi{\ket\Psi}
\def\ketPhi{\ket\Phi}

\def\SM{{\hbox to0ex{$\langle$}
	\hbox{$\mkern2mu\langle\mkern1mu\cal S$,}\hskip2pt%
	\hbox{$\mbf M$,}\hskip2pt\hbox{\ldots\hskip-1pt}%
	\hbox to0pt{$\rangle$}\hbox{$\mkern2mu\rangle$}}}

\def\hA{\oper[17]{\scr A}}
\def\hB{\oper[6] {\scr B}}

\makeatletter
\def\section{\@startsection{section}{1}%
  \z@{.7\linespacing\@plus\linespacing}{.5\linespacing}%
  {\normalfont\scshape\centering\color[rgb]{0,0,1}}}

\author[\textsf{Yu.~Brezhnev}]{Yurii V.~Brezhnev$\blue^*$}

\title[\textsl{L\lowercase{inearity of quantum superposition}}]
{Linear superposition\\[0.5ex]
as a core theorem of quantum empiricism}

\address{Department of Quantum Field Theory,
Tomsk State University, Russia}
\email{\Courier{brezhnev@phys.tsu.ru}}

\def\datename{\hspace{-1.5em}\relax}
\date{\setlength{\arraycolsep}{0em}%
$\begin{array}[b]{rll}
&\text{\emph{Key words}: }&
\makebox[0em][l]{quantum foundations, non-axiomaticity,
detector clicks, ensembles, superposition principle,}\\
&&\makebox[0em][l]{arithmetic, numbers, vector space, abstracting,
interpretations, self-referentiality}\\[1ex]
{\blue^*}&\makebox[0em][l]{\textsf{Department of Quantum Field Theory,
Tomsk State University, Russia}}\\
&\makebox[0em][l]{\emph{E-mail}: \Courier{brezhnev@phys.tsu.ru}}
\end{array}$
\hfill
$\begin{array}[t]{r}\text{\smaller[2]13 April 2022}\end{array}$}

\begin{document}

\hfill
{\href{https://arxiv.org/abs/1807.06894}%
{\smaller[2]\blue\Courier{https:/\!\!\!/arxiv.org/abs/1807.06894}}}

\vspace{-0.5ex}\hfill
{\smaller[2](\Russian русская версия по email-запросу)}\\

\begin{abstract}
Clarifying the nature of the quantum state $\ketPsi$ is at the root
of the problems with insight into counter"=intuitive quantum
postulates. We provide a direct---and math-axiom free---""empirical
derivation of this object as an element of a vector space.
Establishing the linearity of this structure""---""quantum
superposition""---is based on a set"=theoretic creation of ensemble
formations and invokes the following three principia:
(\hyperlink{I}{\red\textbf{\textsf{I}}}) quantum statics,
(\hyperlink{II}{\red\textbf{\textsf{II}}}) doctrine of the number in the
physical theory, and (\hyperlink{III}{\red\textbf{\textsf{III}}})
mathematization of matching the two observations with each other
(quantum covariance).

All of the constructs rest upon a formalization of the minimal
experimental entity---the registered micro"=event, detector click.
This is sufficient for producing the $\bbC$"=numbers, axioms of
linear vector space (superposition principle), statistical mixtures
of states, eigenstates and their spectra, and non"=commutativity of
observables. No use is required of the spatio"=temporal concepts. As
a result, the foundations of theory are liberated to a significant
extent from the issues associated with physical interpretations,
philosophical exegeses, and mathematical reconstruction of the
entire quantum edifice.
\end{abstract}
\maketitle
\thispagestyle{empty}

\vfill

\hfill \parbox{9cm}{\smaller{\slshape%
\ldots\ Instead of presenting a clear argument about states and
superpositions and their relation to detector clicks,
\ldots\hfill $(!)$ \\[1.5ex]
\ldots\ an impenetrable mush of philosophical reflections about \ldots,
draped around a huge amount of quotations
that seem only indirectly pertinent to the subject\\[1.5ex]
\ldots\ discourse is of a highly theoretical nature, without contact
with the experimental practice of physics\\[1.5ex]
\ldots\ The treatment of interference is not adequately
discussed \hfill $(!)$\\[1.5ex]
\ldots\ I cannot imagine that the manuscript is readable for a physics
audience}

\hfill (excerpts from review reports)}
\vspace{3ex}

\newpage
\tableofcontents

\vbox{
\section{Introduction and Summary}\label{intro}

\flushright\tiny
\textsl{\ldots\ somewhat curious that, even after\\
nearly a full century, physicists still do not\\
quite agree on what the theory tells us \ldots}\\
\textsc{--- G.~'t~Hooft} \cite[p.~5]{Hooft}%
\medskip

\textsl{It is almost a crying shame that we are\\
nowhere close to that with quantum mechanics,\\
given that it is over 70 years old now}\\
\textsc{--- C.~Fuchs} \cite[p.~32]{fuchs4}}
\smallskip\nopagebreak

The contradiction between the fundamental nature of quantum theory
(\qt) and the phenomenological feature of its mathematics
\cite{mermin} is likely to never cease instigating the attempts to
overcome it. As H.~Putnam had said, ``Human curiosity will not rest
until \ldots\ questions [of the nature of the \qt"=formalism] are
answered''.

The subject"=matter and leitmotiv of what follows is that the linear
superposition and theory's axioms have an origin""---they are
derivable, and it is entirely empirical. The theory is thereby
demystified, and the interpretative challenges that accompany the
exegeses of \qt\ are a \emph{nonexistent} problem coming from ``a
confusion of categories'' \cite[p.~89]{muynck}, \ie, from the
``semantic confusion'' \cite[p.~10]{silverman}. A direct outgrowth of
this ideology is not only a derivation of the superposition
principle (page~\pageref{princ}) but also the axiom"=free production
of the `chief' quantum formula""---the Born rule $p=|\frak a|^2$
\cite{br2}.

\subsection{On foundations of quantum theory}

The debates concerning the foundations of quantum mechanics (\qm)
hitherto ``show no sign of abating'' \cite[p.~222]{zeilinger3},
\cite{saunders, laloe}, and despite widespread scepticism
\cite{landau, faddeev, englert, gottfried, nahman, lipkin} it is
generally acknowledged that the problem is a real one
\cite{schloss3, accardi, laloe, mittel, mittel2}---it is not
something made up or ``just a dispute over words''
\cite[p.~5]{aaronson+}---and sometimes ``has been regarded as a very
serious one'' \cite[p.~418]{lipkin}, \cite{nielsen, zeilinger1}. Say,
R.~Penrose has expressed (2004) an even more radical ``conviction
that present"=day quantum mechanics has no credible ontology, so that
it \emph{must} be seriously modified''.

In recent decades the discussions have even worsened \cite{fuchs4,
fuchs3}, and current research has intensified, due to the
tremendously increased and formerly inconceivable technological
possibilities of operating with individual micro"=objects and the
urge to implement the idea of quantum computing \cite{deutsch,
aaronson+}.

The reason for this state of affairs remains the same as it was
before. Unlike the classical theories""---\eg, thermodynamics or
relativity theory, ``Ma di assiomatizzazioni della teoria quantistica
ce ne sono moltissime'' \cite[p.~30]{accardi} and the \qm"=axiomatics
itself seems wholly divorced from human language \cite{neumann,
dirac, accardi, greenstein, bell, espagnat, home, laloe, ney,
saunders, silverman, slavnov3, lipkin}. Quantum postulates are not
merely formal. They are phrased in terms of linear operators on a
complex Hilbert space $\bbH$ \cite{auletta, ballentine2, david,
gottfried, jauch1, muynck, neumann, klyshko, landau} and, with that,
literally not a single word here can be brought into conjunction
with reality by means that have \emph{at least some kind} of
relationship with the classical description. What is more, it is
very well known that the abstract character of these terms is
required by the essence of the point (covariance) and, at the same
time, that the attempt to link them with physical images is imposed
by a decree and results in the famous paradoxes associated with such
concepts as causality, (non)""locality, and realism \cite{zeilinger,
zeilinger2, allah, alter, ansmann, bell, fuchs1, greenberg, laloe,
london, greenstein}. All of that causes a problem with
interpretations of~\qm.

It is well known that the theory has steadfastly resisted any unique
ontological reading and, in particular, reconciliation between
interpretations. This is reflected not only in the voluminousness of
the literature. The differences in viewpoint are often based on
points of principle \cite{ballentine3, schloss2, schloss5, halat,
saunders, lipkin, nahman, fuchs2, mermin, spekkens2}, and even
highly qualified publications face criticism \cite{everett1,
lundeen, pusey, mensk}. Among other things, we encounter appeals
\cite{kampen, englert, fuchs1, accardi, mermin, aaronson, ludwig5,
schloss3} (there is even a manifesto(s) \cite{touzalin},
\cite[p.~990]{fuchs2}), striking titles such as ``scandal of quantum
mechanics'' \cite{kampen2, henry}, ``\textsc{quantum outcome: allah
willed it?}'' \cite[p.~188; Wheeler]{wheeler}, ``the Oxford Questions
\ldots\ to two clouds'' \cite[p.~6]{briggs}, ``The Canon for Most of the
Quantum Churches'' \cite[p.~988]{fuchs2}, ``Quantum mechanics for the
Soviet naval officers'' \cite[p.~161]{hren1}, ``the patron saint of
heretics in the One True Church of Copenhagen'' (about D.~Bohm), ``A
Feminist Approach to Teaching Quantum Physics''
\cite[p.~182]{fuchs4}, ``Church of the Larger Hilbert Space''
(J.~Smolin) \cite{fuchs4, englert}, and also April Fools'
\cite{englert2} and the medical jokes about ``the `state of health of
the quantum patient'\,'' \cite[p.~vii]{hren0}, political parallels
with ``Marxism \ldots\ [and] the Cold War'' \cite{freire}, and many more
\cite{ballentine0, bub, laloe, mermin, stacey, stapp2, greenstein}.

An interesting fact. Cambridge University Press has published a
500-page-long book \cite{fuchs4} containing an arresting electronic
correspondence""---D.~Mermin called it ``samizdat''
(self"=published"")---""between C.~Fuchs and modern researchers and
philosophers in the field of quantum foundations. This
correspondence has continued \cite[over 2300~pages]{fuchs3} and now
covers the 1995--2011 years. It characterizes the state of affairs
in the field, and does not merely add to one's impression of the
unending discussions about quantum matters (see introductory
sections in \cite{fuchs2}\,(!\@) and in \cite{hren1}), it also
represents, due to the lack of formality, a plentiful source of
ideas and of valuable thoughts. Schlosshauer's very informative
`quantum interviews' \cite{schloss3} pursue the same goals.

It is worth mentioning that the quantum challenges had led, quite a
while ago, to attempts to revise even the formalizing the logic of
our thinking \cite{beltrametti, stairs}---a very nice mathematical
theory dating back to von~Neuman in the 1930's
\cite[sect.~III.5]{neumann}; termed quantum logic \cite{foulis0}.
There are handbooks on that subject \cite{engesser} and this topic
is still under intensive investigation now. See also the last
paragraph in sect.~\ref{cat}.

The lack of transparent motivations for mathematics""---a pressing
requirement of physics""---means that \qm"=formalism is hard to
distinguish from a ``cook book of procedures and rituals'' (J.~Nash),
a ``user"=manual'' \cite[p.~1690]{slavnov3}, \cite{slavnov2},
\cite[p.~xiii]{greenstein}, or from ``a library of \ldots\ tricks and
intuitions'' \cite{nielsen}. Therefore the ``dissatisfaction regarding
comprehension'' and the ``need for interpretation that is alien to an
exact science'' \cite[pp.~7--8]{kadom} lead to the fact that ``we
admit, be it willingly or not, that quantum mechanics is not a
physical theory but a mathematical model'' \cite[p.~1701]{slavnov3}
or that ``nature imitates a mathematical scheme'' \cite[p.~347;
Heisenberg]{jammer}. De facto, \qt\ ``is in a sense like a
traditional herbal medicine used by ``witch doctors''. We don't
\textsc{really} understand what is happening'' (J.~Nash) and ``we have
essentially \emph{no}\footnote{Throughout the text, the
\emph{italic} and \textsl{slanted} type in ``quotations'' is original,
unless otherwise indicated.} grasp on why the theory takes the
precise structure that it does'' \cite[p.~32]{fuchs4}, which raises
the suspicion that ``something is wrong with the theory'' (H.~Putnam)
and that ``this quantum skyscraper is built on very shaky ground''
\cite[p.~8]{hren1}.

At the same time, well"=founded opinions have long been known to the
effect that ``quantum theory needs no `interpretation'\,''
\cite{fuchs1}, \cite{englert, kampen2, mermin} or that ``only
consequences of the basic tenets of quantum mechanics can be
verified by experiment, and not its basic laws''
\cite[p.~16]{faddeev}. In other words, the discrepancies between
opinions are significant, and often radical: from epithets such as
``schizoid, \ldots\ situation is desperate'' \cite[p.~420]{lipkin},
\cite[p.~424]{pilan} to supporting the rationale for quantum
computations \cite{deutsch} and whole books written on the subject
\cite{saunders}. Concerning the ``schizoid'', the case in point is the
many-world conception by Everett--DeWitt. See also pages~158, 161,
176--179 in \cite{dewitt} regarding the ``state of schizophrenia'' and
`explanations' as to why ``schizophrenia cannot be blamed on quantum
mechanics'' \cite[p.~182]{dewitt}.

In any case, the controversy between ``the warring factions, \ldots,
many [quantum] faiths, \ldots\ and instrumentalist camps''
\cite[pp.~60--61]{schloss3}, \cite[sect.~5.5]{sudbery},
\cite{auletta, home}---``[t]hey all declare to see the light, the
ultimate light'' \cite[p.~987]{fuchs2}---cannot be considered as an
acceptable state of affairs (see also sect.~\ref{phil}), for the
simple reason that the quantum philosophy issues turn into an
`industry' of interpretations""---an unhealthy state of
affairs""---while at the same time the very same philosophers call
for its denial: ``interpretation of \qm\ emerged as a growth
industry'' \cite[p.~92]{mackin}.

\subsection{Formula of superposition}\label{super}

Contrariwise, the ``dominant role of mathematics in constructing
quantum mechanics'' has lead to that mathematical ``assumptions are
usually considered to be physical'' \cite[p.~1691]{slavnov3}. That is
to say, ``there has been a substitution of concepts''
\cite[p.~295]{slavnov2} and ``one of the consequences of quantum
revolution was the replacement of explanations of physical phenomena
by their mathematical description'' \cite[p.~296]{slavnov2}. These
characteristics convey, in the best possible way, the
dissatisfaction with the fact that quantum physics ``was actually
reduced to a physical interpretation of the Hilbert space theory''
\cite[p.~1690]{slavnov3}. The $\bbH$"=space in itself is a fairly
cumbersome mathematical structure, and even determines a crucial
principle""---""superposition of states \cite{dirac}. It is thus not
surprising that this principle becomes ``one of the vague points \ldots\
the [Dirac] argument is difficult to consider as rational \ldots\ the
physical principle simply fits underneath it'' (excerpt from the
preface to the Russian edition of~\cite{shwinger}).

Mathematics of the $\bbH$"=space contains three constituents: a
vector space, the inner"=product add-on over it, and topology. The
two latter ones invoke the first one which is completely independent
(algebra) and begins with the formula
\begin{equation}\label{1}
\ket\psi=\frak a\cdot\ket\varphi+\frak b\cdot\ket\chi\;.
\end{equation}
This is the pivotal expression of quantum theory. Comprehending its
genesis is tantamount to comprehending the nature of the
\emph{linearity} of \qm.

In \eqref{1}, there occur the complex numbers $\frak a,
\frak b\in\bbC$, symbols of operations $\cdot$ and~$+$, and also
vectors $\ket\psi$, $\ket\varphi$, $\ket\chi\in\bbH$. It is clear
that until an empirical basis for all these devices is found, the
interpretation of abstraction \eqref{1} and questions of the kind
``Quantum States: What the Hell Are They?\@'' (55~times in
\cite{fuchs3}) will remain a problem. To all appearances, the
problem is considered so difficult""---``quantum states \ldots\ cannot
be `found out'\,'' \cite[p.~428]{saunders}---that the non"=axiomatic
meaning of these symbols was not even discussed in the literature.
In the meantime, not only is the situation far from hopeless, but it
also admits a solution. The present work is devoted to gradual
progress towards an understanding of formula \eqref{1}. Stated
differently, equality \eqref{1} becomes an empirical `theorem'
(p.~\pageref{Theorem}).

\begin{itemize}
\item The main part of the challenge is not only to ascertain what is
 being added/"!multiplied in \eqref{1}, but also to realize primarily
 \emph{what `to add/"!multiply' is}, and ``Where Mathematics Comes
 from'' \cite{lakoff} at all.
\end{itemize}
``What does it mean, physically, to ``add'' things?\@'' \cite[p.~178;
D.~Darling]{fuchs4}. More than that, aside from the symbols
$\{\frak a$, $\frak b$, $\ket\psi$, $\ket\varphi$, $\ket\chi$,
$\cdot\,$, $+\}$ the expression \eqref{1} contains the sign of
equality $=$ (see also \cite[pp.~29, 30\,(!)]{baez+}, \cite{mazur})
and, surprising as it may seem, it conceals one of the key
points---the 3"~rd principium of quantum theory
(\hyperlink{III}{\red\textbf{\textsf{III}}}, p.~\pageref{III}).

The guiding observation is based on the fact that the only thing
that we have access to are the microscopic events, and therefore ``we
have little to begin with other than what an experimental physicist
would call experiments with a single microsystem''
\cite[p.~5]{ludwig1}.
\begin{quote}
``[W]e must recognize that the focusing on individual elements
whatever these may be is absolutely indispensable for all our
thinking. \ldots\ What may be regarded as an individual event?''
\\\phantom.\hfill R.~Haag \cite[p.~302]{haag2}
\end{quote}

Consequently, we must begin from individual events and from
collecting them into ensemble formations. It is precisely in this
context that we will use the word empiricism""---""quantum
empiricism of micro-acts of perception""---and it is in this respect
that \qt\ has a statistical nature. As Einstein had put it, ``It may
be a correct theory of statistical laws, but an inadequate
conception of individual elementary processes'' \cite[p.~156;
Einstein]{home}; see also \cite[pp.~38--40]{einstein},
\cite[Chs.~7--8]{home}, \cite[p.~40]{einstein}, \cite{neumann,
stacey}. Such a viewpoint has been long championed by L.~Ballentine
\cite{ballentine} and H.~Groenewold \cite[p.~468]{jammer2} and
justified in detail by G.~Ludwig \cite{ludwig1, ludwig2, ludwig3,
ludwig4}. A.~Leggett proposes accordingly the ``extreme statistical
interpretation'' \cite{leggett1}, \cite[p.~79]{schloss3} in the sense
that ``to seek any further ``meaning'' in the formalism is pointless
and can only generate pseudoquestions''. With that, he overtly
applies such characteristics as ``complete gibberish''
\cite[p.~70]{leggett1} and ``verbal window dressing''
\cite[p.~79]{schloss3}.

The difficulty is, of course, in creating the object $\ket\psi$
itself. A step-by-step characterization of this procedure
(sects.~\ref{Ans}--\ref{statespace}) and key words to what follows
have been reflected in the (sub)""section titles listed in the
Contents.

\subsection{Physics {\smaller[1]$\rightleftarrows$} mathematics.
Doctrine of numbers}\label{physmath}

Thus the situation appears to be one whereby the physics itself
faces inconsistencies in its foundations and the mathematical
superstructures are difficult to reconcile with its motivations
(physical principles) \cite{benioff}. But, on the other hand,
attempts to axiomatize an interface between them \cite{hart} only
conceal a deeper insight \cite{zeilinger1}. M.~Born had called
attention to the fact that ``probable refinements of mathematical
methods will not suffice to produce a satisfactory theory, but that
somewhere in our doctrine is hidden\eLab{born} a concept'' and
T.~Maudlin was more definite: ``physicists have been misled by the
mathematical language they use to represent the physical world''.

In other words, we observe an overemphasis on the role of the
ready"=made math"=structures""---""algebras, spaces, \thelike---and an
under"=evaluation of `seemingly na\"{\i}ve' empirical aspects voiced
in the ordinary language \cite{heisenberg2}. The situation is no
different from that which H.~Weil had characterized in the
introductory section to \cite[p.~10]{weyl} as follows.
\begin{quote}
``All beginnings are obscure. Inasmuch as the mathematician operates
with his conceptions along strict and formal lines, he, above all,
must be reminded from time to time that the origins of things lie in
greater depths than those to which his methods enable him to
descend''.
\end{quote}
The ``origins'' are expressible of course only in the natural
language; sect.~\ref{S0} is devoted to this.

What we propose below is an implementation of the idea that the
postulational view must be abandoned and replaced by a negation of
the prior existence of both the physical ``preconceived notions''
\cite[p.~328]{ludwig2} and the mathematical structures. Physics and
mathematics should be created `from scratch'. Paul Benioff calls
this idea ``a~coherent theory of physics and mathematics''
\cite[p.~639]{benioff}, \cite[p.~33; P.~Benioff]{fuchs4}. Then, due
to the initial absence of mathematics, the introducing of
mathematical structures is almost ruled out, proofs must get
replaced by an empirical inference, and semantics of physics""---the
language of physical reasoning""---is initially under a linguistic
ban. It cannot exist a~priori. That is to say, even the
natural"=language conjunction of mathematical terms with physical
adjectives (and verbs \cite[p.~3102; ``to happen, to be, to
exist'']{hartle2}) becomes far from being free, as with the classical
description's language (sects.~\ref{S}, \ref{SS}, \ref{phys},
\ref{2slit}). R.~Haag, on 1"~st page of the work \cite{haag},
emphasizes:
\begin{itemize}
\item ``we\eLab{haag} should not consider [``vocabulary of Quantum
 Theory''] as sacrosanct. \ldots\ every word in the vocabulary is
 subject to criticism''.
\end{itemize}

Returning to the ensemble formations, it is only they that have to
come to the fore, and argumentation should be subordinated only to
the low"=level microscopic empiricism. Predominance of the empirical
over the theoretical will then immediately touch on the closest
creature of the latter""---the notion of a number---since numbers do
not come `from the sky', and the theory will have to be a
quantitative one.

Despite the overflow of abstracta in \qt, the \emph{doctrine of
number}---\lrceil{num\-ber$\:\times\:$unit}, to be precise
(sects.~\ref{Numbers} and \ref{minus})---has, it seems, not yet
entered foundational discussions \cite{gryb}. Consequently, the
numbers turn into a kind of `problem of numbers'
(principium~\hyperlink{II}{\red\textbf{\textsf{II}}}), and we are thus led
to the necessity of revising the take on the foundations themselves:
\begin{equation*}
\lceil\text{quantum fundamentals}\rceil\quad\goto[3]\quad
\lceil\text{the problem/"!doctrine of the number}\rceil\;.
\end{equation*}
This paradigm shift is a unique trait of the quantal (not the
classical) view of things and a substantial part of the following is
devoted specifically to that.

In the outline of the present work, the workflow will constitute
re"=creating the structure of a linear vector space. More precisely,
the producing an \emph{a~priori unknown} mathematics, which
\emph{will be found to be} an algebra of such a space with a complex
conjugation. As a matter of fact, we provide an answer to Haag's
question ``How do we translate the description of an experimental
arrangement into mathematical symbols?'' in the context of his own
``idea of basing the interpretation of quantum theory on the concept
of ``events'' which may be considered as facts independent of the
consciousness of an observer'' \cite[p.~295]{haag2}.

The main point to be immediately emphasized is that the
mathematization of the discrete micro-acts of observations is quite
a nontrivial procedure \eqref{stream}, and the further strategy,
along with the structure of this article, can be schematized as
follows.
\begin{equation*}
\framebox[9cm]{$\begin{array}{c}
\text{natural language, prolegomena to the quantum}\\
\text{(no math and ontology here)}
\end{array}$} \tag{sect.~\ref{S0}}
\end{equation*}
\begin{equation*} \downarrow\qquad\qquad\downarrow \end{equation*}
\begin{equation*}
\hspace{-2cm}\framebox[9cm]{$\begin{array}{c}
\text{accumulation of micro-events,}\\
\text{low-level quantum empiricism, ensemble mixtures}
\end{array}$} \hspace{-2cm} \tag{sects.~\ref{Ans}--\ref{whyC}}
\end{equation*}
\begin{equation*} \downarrow\qquad\qquad\downarrow \end{equation*}
\begin{equation*}
\framebox[9cm]{$\begin{array}{c}
\text{mathematization of ensembles' empiricism}\\
\text{(how the math comes about)} \tag{sect.~\ref{EmpMath}}
\end{array}$}
\end{equation*}
\begin{equation*} \downarrow\qquad\qquad\downarrow \end{equation*}
\begin{equation*}
\framebox[9cm]{$\begin{array}{c}
\text{quantum superposition and \qm-linearity}\\
\text{(no physical concepts and numbers here)}
\end{array}$} \tag{sect.~\ref{SP}}
\end{equation*}
\begin{equation*} \downarrow\qquad\qquad\downarrow \end{equation*}
\begin{equation*}
\framebox[9cm]{$\begin{array}{c}
\text{algebraic structure `the numbers $\bbR$ and $\bbC$',}\\
\text{binary and unary operations}
\end{array}$} \tag{sect.~\ref{Numbers}}
\end{equation*}
\begin{equation*} \downarrow\qquad\qquad\downarrow \end{equation*}
\begin{equation*}
\framebox[9cm]{$\begin{array}{c}
\text{\lrceil{quantum states} $=$
\lrceil{linear vector space}}\\
\text{formula
$\ket\psi=\frak a\cdot\ket\varphi+\frak b\cdot\ket\chi$}
\end{array}$} \tag{sect.~\ref{statespace}}
\end{equation*}
\begin{equation*} \downarrow\qquad\qquad\downarrow \end{equation*}
\begin{equation*}
\hspace{-2cm}\framebox[9cm]{$\begin{array}{c}
\text{naturalness and inevitability of abstracta,}\\
\text{observable quantities and their values,}\\
\text{quantum statistics and Born's rule \cite{br2}}\\
\text{(\emph{no} interpretations here)}
\end{array}$} \hspace{-2cm} \tag{sects.~\ref{minus}--\ref{interpret}}
\end{equation*}

\noindent This box"=diagram cannot be reduced or restructured. For
example,
\begin{itemize}
\item \emph{superposition foregoes numbers}, and measurement and
 physical properties follow \emph{strictly after} the
 $\ket{{}\Courier{ket}}$"=vectors have been created.
\end{itemize}

By and large, the aforesaid ideology is supported by the common
belief---often certainty even \cite{englert}---that \qm\ is
\emph{not} perturbative, its linearity is \emph{not} associated with
linear approximation of something else, and, in general, it is
\emph{not} extensible (ultimate \cite{colbeck} and non"=deformable)
and must be free of interpretations \cite{englert, fuchs1}. All of
these concerns, in one way or another, are directly related to the
derivation of formula~\eqref{1}.

\vbox{
\section{Points of departure}\label{S0}

\flushright\tiny
\textsl{In the Beginning was the Word}\\
\textsc{--- A.~Zeilinger} \cite[$01{:}05'47''$]{zeilinger2}%
\medskip

\textsl{Most of the time the apparatus is empty and\\
sometimes you have a photon coming through}\\
\textsc{--- A.~Zeilinger} \cite[$12'39''$]{zeilinger2}}
\smallskip\nopagebreak

Since empiricism is in essence supra"=mathematical \cite{bueno}, \ie,
it is concerned with \emph{meta}\-mathematics \cite{kleene, raseva},
its mathematization""---""theory construction""---should begin not
with postulates and definitions, but rather with the semantic
formation of an object language (of `the Quanta'), of vocabulary,
and ``may only be described by ``words'' and not by a theory''
\cite[p.~106]{ludwig1}, \cite{ludwig3, ludwig5}. As A.~Peterson and
K.~Popper had observed, ``Math can never be used in phys until have
words'' \cite[p.~209]{barrett}, \ie, ``we cannot construct theories
without using words'' \cite[p.~12]{popper}. Therefore, relying on the
established understanding of the underlying causes for the quantum
eye on physics \cite{neumann, auletta, gottfried, greenstein}, up
until the end of this section we will adopt the natural"=language
meaning of the words observation, system, state, numbers~(!), plus
and to divide, physical influence, transition, large/small,
micro/macro, \etc. Their contents will later be defined more
precisely or entirely changed. For instance, the sense of the word
`state' will be drastically transformed, to which we are drawing
attention in advance. Accordingly, a degree of informality""---it
has been clarified in Remark~\ref{meta}---is inevitable here, but
``the lack of precision \ldots\ is a necessity'' \cite[p.~48]{heisenberg}
at the moment.

\subsection{Variations as micro-level transitions}\label{S}

We will (and `must' \cite[Ch.~3]{kleene}) first view the concept of
a system at an intuitive level \cite[sect.~1.1]{edwards}---there is
what is referred to as an isolated system~$\cal S$.
\begin{itemize}
\item[\hypertarget{S}{\red\textsf{S}}] Let us tentatively (a~priori) relate
 the concept of a \emph{state} to the associated context describable
 by the words `the system $\cal S$ can \emph{vary}, \emph{be
 different} or \emph{in different states}'. That is to say, system
 $\cal S$ is always in a certain state $\statePsi$ belonging to the
 set $\boT=\{\statePsi,\statePhi,\ldots\}$, each element of which is
 admissible for $\cal S$, and all of them are different from each
 other: $\statePsi\ne\statePhi$.
\end{itemize}
In other words, the concept of a quantum system may not have a
precise/"!axiomatical definition at the moment. Or, if it comes to
that, the system is what's being constantly varied when observed,
and ``varied'' is the key word here.

The statement ``states are different'' does not require a
consideration when $\statePsi$ and $\statePhi$, referred to as
state, are the abstract elements of an abstract set
$\{\statePsi,\statePhi,\ldots\}$. However, in order to tie its elements
to reality, we have to introduce the criteria of
coincidence/"!distinguishability of one from the other. Criteria may
not come otherwise than from observational procedures, without which
it is impossible to either detect states, or claim that they differ,
coincide, or that they are, if any.

On the other hand, the nature of micro"=phenomena shows that
observations are always associated with irreducible intervention in
the system, manifesting in what is known as \emph{transition}
$\statePsi \rightsquigarrow \statePsi'$ (or destruction). As an
example, observations at accelerators are literally the
destructions, and bulk at that. Due to a lack of criteria, there is
no sense in attributing to this concept the adjectives small/large,
(in)""significant/"!partial, or collocations such as `comparison of
destructions at instants $t_1$, $t_2$'. Let us proceed from the idea
that initially there is nothing but the transition. Transitions may
actually occur without destructions
$\statePsi\rightsquigarrow\statePsi$, however.

Two different $\statePsi$, $\statePhi$ may be destroyed into new
$\statePsi'$, $\statePhi'$, as well as into the combinations of the
old/new. Thus, strictly speaking, the sense of words `different,
new, \ldots' eludes us in this case, which is why even the
identification of $\statePsi$"=elements and the $\boT$ itself, as a
set, becomes questionable. Therefore, besides the formal writings
$\statePsi=\statePhi$ and $\statePsi\ne\statePhi$ for
$\statePsi,\statePhi\in\boT$, the \emph{physical}
distinguishability/"!equivalence (recognizability
${\not\approx}{/}{\approx}$) needs to be established. As to the
identification (and to the identity) in this regard, see
von~Neumann's reasoning ``\Courier{One might object against} \textsf{II}
\ldots'' on page~302 of his book \cite{neumann}. The sole thing that
distinguishability may rely on is the transition acts. In turn,
variation is a key element in transitions, which is why we will
begin constructing with distinguishability.

Let us take the still virtually unlimited way $\scr A$ of
intervening $\TRANS[1]{\sss\scr A}$ in $\cal S$ and attempt to
introduce distinguishability $\statePsi\not\approx\statePhi$ as
$\scr A$"=distinguishability. Due to the fact that micro"=transition
$\statePsi\TRANS[1]{\sss\scr A}\statePsi'$ is not pre"=determined,
initial states $\statePsi$ undergo arbitrarily free changes. Next
time, the results will be different and absolutely arbitrary (the
term ``different'' is understood to mean $\ne$), and each act is
indiscernible from a case in which it contains ones similar to
itself within itself. It would be natural to associate such a case
to the absurd, which is unrelated to the meaning of the words
`physical observation', and to discard the given~$\scr A$.

Non"=meaninglessness arises only if we impose the negation of random
combinations of $\ne$ and $=$ in transitions, at least for a part of
$\boT$, \ie, introduce the preservation acts
$\statePsi\TRANS[1]{\sss\scr A}\statePsi$. The `preservation' should
be read here as indestructibility of state, \ie, as a
($=$)"=coincidence under the secondary act $\statePsi
\rightsquigarrow \statePsi \rightsquigarrow\statePsi$. Otherwise,
the vanishing difference between `preservation' and `variation'
leads to a linguistic chaos \cite[p.~232]{foulis}. This means that
the destruction $\statePsi \rightsquigarrow \statePsi'$ may not be
considered as a 1-fold one. State $\statePsi'$ on the right should
be examined for changeability and transform into the left part of
the subsequent transition: $\statePsi \rightsquigarrow \statePsi'
\rightsquigarrow \statePsi’’$. Thereby the structure $\statePsi'
\rightsquigarrow \statePsi’’$ with the \emph{binate} entity
`before/"!after' or `on the left/right' becomes the key one, and we
consider it an initial object in subsequent constructs. The
preserved states are, by definition, those that pass the
reproducibility test.

Thus logic requires to begin with the transition compositions
\begin{equation*}
\statePsi\TRANS{\scr A}\statePsi'
\TRANS{\scr A}\statePsi’’\TRANS{\scr A}\cdots\;,
\end{equation*}
wherein the cases like
\begin{equation}\label{nons}
\cdots\quad \statePsi' \TRANS{\scr A} \statePsi' \TRANS{\scr A}
\statePsi’’\quad\cdots
\end{equation}
are ruled out (a ban on changing of what has been unchanged), and
the never"=ending sequence
\begin{Align}\label{chaos}
&\statePsi\TRANS{\scr A}\statePsi' \TRANS{\scr A}\cdots
\TRANS{\scr A} \statePsi’’\TRANS{\scr A}\cdots
\intertext{(non"=recognisability of states) must be terminated}
\label{seq} &\statePsi\TRANS{\scr A}\statePsi'
\TRANS{\scr A}\cdots\TRANS{\scr A}
\statePsi’’\TRANS{\scr A}\state\alpha\TRANS{\scr A}\state\alpha\;,
\end{Align}
yielding a `finiteness' (\,$=$~realisticness) and the concept of
conserved/"!distinctive $\state\alpha$"=states. The terminology
$\state\alpha$"=event \cite{englert} could be used instead.

Freedom of elements in sequence~\eqref{seq}, including the choice of
$\state\alpha$"=states, is not limited by anything besides the ban
on~\eqref{nons}. Therefore this arbitrariness, which is physically
never recognizable, curtails the generic chain \eqref{seq} into the
shortened one
\begin{equation}\label{box}
\def\fboxsep{1.5\dimen36}
\statePsi\TRANS{\scr A}
\boxed{\cdots\cdots}\TRANS{\scr A}\state\alpha
\TRANS{\scr A}\state\alpha\;,
\end{equation}
which is identical to the scheme
\begin{equation}\label{Box}
\def\fboxsep{0.8\dimen36}
\boxed{\cdots\statePsi\cdots}\TRANS{\scr A} \state\alpha
\end{equation}
with certain $\state\alpha\in\boT$.

Discussions on ``what happens \ldots\ [and] ``how''\,''
\cite[p.~217]{neumann} at the very microscopic level are extremely
widespread in the literature \cite{greenberg, alter, kampen, london,
lundeen, mittel, kadom} (see \cite{allah, auletta, schloss4,
mittel2} for the exhaustive references), although it is not
difficult to predict the fact that the attempts to understand the
inner structure of box~\eqref{Box} will only lead back to an
identical box; so, the ``turtles all the way down'' (ascribed to
W.~James), followed by the great Wheeler's slogan ``No tower of
turtles'' (1989).

Indeed, the uncontrollability of micro"=changes is universally known,
yet describing them as a process in time $t\mapsto t+\eps$ will
start employing the language terminology""---""functions, arithmetic
operations, the physical words, \etc---that has not yet been created
even for the fixed instants $t_1$, $t_2$. However, what may be
associated with fixed time are only non"=temporal entities, for which
we have nothing but transitions~\eqref{box}. The attempt to manage
them, \ie, to control intervention in $\cal S$, results in looping
or `measuring the measurement', let alone the ambiguity of this term
itself.
\begin{quote}
``[I]t is not meaningful to speak of a measurement ``at time
$\tilde t$.'' \ldots\ the real physical meaning of the time parameter
\ldots\ has nothing to do with the notion ``time of measurement.''\,''.
``[T]he description of the measurement process in quantum mechanics
in terms of ``pre"=theories'' is not possible''
\\\phantom.\hfill G.~Ludwig \cite[p.~288]{ludwig1},
\cite[p.~340]{ludwig2}
\end{quote}
See also \cite[p.~365]{ludwig2}, \cite[p.~150]{ludwig4},
\cite[p.~100]{ludwig5}, \cite[pp.~644--646]{peres0}, and
\cite{reynolds}. Just as before, the physical assessments such
as`abrupt', `(ir)""reversible', '(non)""simultaneous', ``immediately
following \ldots'' \cite[pp.~231, 410]{neumann}, or the
`weak/"!nondemolishing' (measurements \cite{lundeen}), \etc\ are
unacceptable here. No temporal process may be present in the
foundations of the theory \cite[sects.~VII.4,~6]{ludwig1},
\cite[Chs.~III, XVII]{ludwig2}, \cite{ludwig3}, since it is
immediately not clear `And what exactly are we having at instants
$t_1$ or~$t_2$?'. In the reverse direction""---\lrceil{time $\goto$
measurement}---the situation is also rather indefinite since the
``\,``Time'' is not an entity to which the operations of measurement,
direct or indirect, apply'' \cite[p.~5]{reynolds}.

\begin{comment}
All said above means that attempts to deduce \qm\ dynamically
\cite[10~$\bcdot$~Re\-con\-struc\-tions]{schloss3} are beforehand
doomed to vicious circles `round the boxes' and time $t$ like
attempts to `vindicate' dynamically Lorenz's contraction instead of
kinematic postulates of the relativity theory \cite{bub}. Consistent
theory must rest either on `irreducible' elements \eqref{Box} or
upon `boxes' of a different kind. In the latter case, the theory
becomes a particular \emph{model with interpretation}; \eg, the
Lindblad equations \cite{weinberg, weinberg2}, decoherence
\cite{joos, zurek, schloss4, schloss}, stochastic dynamics, and
other statistic"=dynamical models \cite{allah, kadom}. Anyway, an
ability to model and understanding are not the same thing, and this
point was repeatedly emphasized in the literature
\cite[sect.~I.2]{accardi}, \cite{slavnov3, schloss3} with regard to
\qt.
\end{comment}

That said, if theory is built \emph{as a fundamental one, rather
than as a model} \cite[p.~144]{schloss3}, with a primary entity
\emph{changeability} $\TRANS[1]{\sss\scr A}$, the box \eqref{Box}
may only be involved in it as the initial starting point and as an
\emph{indescribable}\eLab{bbox} object with the absolute rather than
with a relative sense. Elements of reality, in whatever
understanding""---say, Bell's ``\!\emph{be}ables'' \cite{bell}---may
not exist before/"!after/"!inside/"!outside of the box. It can be only the
\emph{structureless abstractio}. Accordingly, the notions of
preparation, of measurement, of `interaction with', and of a
physical process are meaningless without construction \eqref{Box}.

These statements are clearly in agreement with the fact that any
reasoning must not contradict the formal logical rules
\cite{kleene}, hence, there must exist \cite{raseva, shenfield,
benioff} the empirically undefinable logical atoms. A.~Peres writes
\cite[p.~173]{peres}: ``While quantum theory can in principle
describe \textsl{anything}, a quantum description cannot include
\textsl{everything}. In every physical situation \textsl{something} must
remain unanalyzed''. Or as Pauli had put it, ``Like the ultimate fact
without any cause, the individual outcome of a measurement is \ldots\
not comprehended by laws''. Specifically, the set $\boT$ and
transitions"=arrows~$\TRANS[1]{\sss\scr A}$ are also the atoms. ``\ldots\
preexisting concept \ldots\ We cannot formulate the theory without this
concept'', concludes B.~Englert \cite[p.~2]{englert}. From the
aforesaid, we may formulate the following tenet.
\begin{itemize}
\item[\hypertarget{I}{{\red\textbf{\textsf{I}}}}] Quantum statics should
 forego quantum dynamics.\\
 \phantom.\hfill(\embf{The $\bo1$-st principium of quantum theory})
\end{itemize}

The rationales do not end here, and will be later amplified once we
begin to exploit the terminology that is usually taken for granted
from the outset; viz, the quantitative descriptions
\cite[p.~178]{fuchs4}. If they arise not as numerical
interpretations of something, but out of an experiment, then
observation should be the beginning, and the `manufacture of
numbers'"!---the end. In other words, the model `theory with boxes'
other than \eqref{box}--\eqref{Box} implicitly implies the logical
sequence \lrceil{mo\-del of process} $\rightarrowtail$
\lrceil{numerical interpretation}, in which empiricism holds a role
other than primary. It is clear that, regardless of the model, such
a situation will always remain unsatisfactory in the physical
respect.

\subsection{Observation}\label{observ}

The sequences addressed above lead to the following outcome.
\begin{itemize}
\item Any meaningful micro-act $\TRANS[1]{\sss\scr A}$ either saves a
 state ($\state\alpha \TRANS[1]{\sss\scr A} \state\alpha$) or turns
 it into a conserved one ($\statePsi \TRANS[1]{\sss\scr A}
 \state\alpha$).
\end{itemize}
The two extremes do not contradict this fact. The
first---""maximally rough observations""---is when all states are
destroyed into a certain one: $\statePsi \rightsquigarrow
\statePsi_0$ (`whatever and however we watch, all we see is one and
the same'). In this, the state $\statePsi_0$ is not destroyed:
$\statePsi_0 \rightsquigarrow \statePsi_0$. Another extreme is when
none of the states are destroyed: $\statePsi \rightsquigarrow
\statePsi$. This is the case of ideal (quantum) observation, but,
due to the absence of any changes, it is indistinguishable from the
case of if observations are entirely absent.

Situated in between these extremes lies a simplest case with two
distinctive states
\begin{equation}\label{a1122}
\state\alpha_1\TRANS{\scr A}\state\alpha_1\,,\qquad
\state\alpha_2\TRANS{\scr A}\state\alpha_2\;.
\end{equation}
Of course, these are prohibited from transitioning into each other.
Because there is still the free admissibility of transitions
$\statePsi\TRANS[1]{\sss\scr A}\state\alpha_1$,
$\statePsi\TRANS[1]{\sss\scr A}\state\alpha_2$, we can turn the
semantic sequence
\begin{equation*}
\big\lceil\text{arbitrariness}\quad\rightarrowtail\quad\text{preservation}\quad
\rightarrowtail\quad\text{distinctive $\state\alpha$'s}\big\rceil
\end{equation*}
into the more rigorous scheme
\begin{equation}\label{shema}
\big\lceil\boT=\{\statePsi\,,\,\statePhi\,,\,\ldots\}\big\rceil
\;\tplus\;\big\lceil\scr A
\text{-observations}\big\rceil\quad\rightarrowtail\quad
\{\state\alpha_1\,,\state\alpha_2\,,\ldots\}\FED \boT_{\!\sss\scr A}
\subset\boT\;,
\end{equation}
which gives, even though partially, rise to the concept of a
physical distinguishability (`distinguo'). It is formally defined
only on the subset $\boT_{\!\sss\scr A}$: the statement
$\state\alpha_1\not\approx\state\alpha_2$ is equivalent
to~\eqref{a1122}. To avoid overloading the further notation, we do
not use symbols like $\approx_{\!\!\!\sss\scr A}$ and
$\not\approx_{\!\!\!\sss\scr A}$; the context is always obvious.
\begin{itemize}
\item[\hypertarget{O}{{\red\textsf{O}}}] By a \emph{physical observation}
 $\scr A$ or, in short, \emph{observation} we will mean such
 interventions $\GOTO{\sss\scr A}$, in which the `never"=ending'
 chaos \eqref{chaos} is replaced by chaos with the notion of
 preservation, \ie, `chaos with rule~\eqref{Box}':
 \begin{equation}\label{rule}
 \statePsi\GOTO[3]{\scr A}\state\alpha\,,\quad\text{where~ }\quad
 \state\alpha\GOTO[3]{\scr A}\state\alpha\;.
 \end{equation}
 The set of $\state\alpha$"=objects $\boT_{\!\sss\scr A}$ with the
 property
 \begin{equation}\label{akk}
 \state\alpha_1\GOTO[3]{\scr A}\state\alpha_1\,,\qquad
 \state\alpha_2\GOTO[3]{\scr A}\state\alpha_2\,,\qquad\ldots
 \end{equation}
 is discrete, and the $\state\alpha_s$ themselves are termed
 \emph{the eigen \emph{(proper)} for observation $\scr A$}. They
 defines $\scr A$ and do not depend on $\cal S$. No logical
 connection between $\statePsi$ (the left of \eqref{rule}), family
 $\boT_{\!\sss\scr A}$, and system $\cal S$ exists.
\end{itemize}
(The comprehensive terminology here is this: micro-act of
observation by instrument $\scr A$. The zig-zag arrow
$\smash{\TRANS[1]{}}$ is replaced with the straight one $\goto$.)
Expressed another way, introduction of the conception `the eigen' is
equivalent to the following informal, yet minimal motivation:
\emph{at least some} certainty instead of \emph{total}
arbitrariness.

Two instruments $\scr A$ and $\scr B$ may have arbitrarily different
eigen"=states $\{\state\alpha_1$, \ldots, $\state\alpha_n\}$ $\ne$
$\{\state\beta_1,\ldots,\state\beta_m\}$. Accordingly, as regards
observation $\scr B$, the (distinctive) states $\{\state\alpha_s\}$
do not differ, in general, from the `regular' $\statePsi$'s, \ie,
from those chaotically destroyable into the $\scr B$"=eigen states:
$\state\alpha_j\GOTO{\sss\scr B}\state\beta_k$. All kinds of
instruments \{$\scr A$, $\scr B$, \ldots\} are thus defined by
aggregates \{$\boT_{\!\sss\scr A}$, $\boT_{\!\sss\scr B}$, \ldots\}. The
number $|\boT_{\!\sss\scr A}|$ of corresponding
$\state\alpha$"=objects therein may be an arbitrary integer. There
are also no (logical) grounds for restricting/"!prescribing the
composition of $\boT_{\!\sss\scr A}$. Any element of $\boT$ may be
the conserved one for a certain instrument. Parenthetically, the
notion of an eigen"=state---in different forms---is sometimes present
in axiomatics of \qm\ \cite{piron, beltrametti, mittel}.

In a generic case, the chaos present in~\eqref{rule} leaves open the
problem of correlating the recognizability
$\statePsi\not\approx\statePhi$ (or $\statePsi\approx\statePhi$)
with physics. Clearly, the issue is linked to the ambiguity of the
term $\statePsi$"=state itself, which is used in
pt.~\hyperlink{S}{\red\textsf{S}}---an important point---due to the
need to start with something, since building the mathematical
description without some sort of a set is impossible.

\begin{comment}\label{two}
Informally, metalinguistic semantics""---""association of meanings
with texts \cite{ludwig5}---is in general as follows. Inasmuch as we
are receiving \emph{different} $\state\alpha$"=responses to each
micro-act $\GOTO{\sss\scr A}$, let us say that `on the other side
from us there is something that can also be different', and all of
that is to be described. This reflects our intuitive perception of
reality, which, both at the micro- and the macro"=level, boils down
to pt.~\hyperlink{S}{\red\textsf{S}} and to an ineradicable pair:
\lrceil{some\-thing out\-side} \tplus\ \lrceil{that which can be
different for us}. If we give up either of these semantic
premises""---``something outside'' ($\statePsi$) or ``can be different''
($\state\alpha_j$)---then, as above, we face a linguistic dead end,
as the possibility for reasoning disappears. There must be two sides
present. Because of this, the arrow $\GOTO{\sss\scr A}$ must be
accompanied by `some things' to the left and the right of it. The
low"=level set $\boT=\{\statePsi$, $\statePhi$, \ldots,
$\state\alpha_1$, $\state\alpha_2$, \ldots\} does arise. Then the
arbitrary elements $\statePsi\in\boT$ (unrestricted chaos) are
assigned to the left of this act instead of `some thing' and the
micro $\not\approx$"=distinguishable $\state\alpha$"=objects
($\state\alpha_j\not\approx\state\alpha_s$) to the right. Put
another way,
\par\larger 
\begin{itemize}
\item \smaller what's being abstracted is not ``concrete things''
 \cite[p.~27]{gottfried} or ``behaviour of things''
 \cite[p.~414]{ulfbeck2} but a primitive element of perception""---a
 micro"=event---the $\state\alpha$"=click. Other than `the click', no
 entities, such as very small objects/"!particles, fields, or, much
 less the knowledge, human psychology, ``personal judgments'', ``memory
 configuration'' \cite{everett, everett1}, ``mysterious interaction
 \ldots\ brain of the observer'' \cite[p.~645]{peres0}, \cite[p.~11,
 thesis~3]{popper}, agents, their belief/"!consciousness \cite{mensk,
 tegmark, stapp2}, \etc, may exist in empiricism. This is a kind of
 ``\textsl{Radical Empiricism} \ldots\ [by] William James'' \cite[pp.~289,
 \ldots]{fuchs3}. The ``click \ldots\ and nothing more'' \cite[p.~42;
 \v{C}.~Brukner]{schloss3} is a kind of experimental the
 zero"=principium of \qt. Therefore the initial math"=premises of \qt\
 should contain nothing but the
 $\not\approx_{\!\!\!\sss\scr A}$"=distinguishability and formalization
 \eqref{rule}--\eqref{akk}.
\end{itemize}

\smaller 
Ideas of ``a click (signal) in a counter'' \cite[p.~758]{ulfbeck}
have, time and again, already been expressed in the literature
\cite{ulfbeck2}, \cite[A.~Peres]{fuchs4}, and we draw attention to
answers of \v{C}.~Brukner on pp.~41--43 in \cite{schloss3}, his work
\cite[p.~98]{brukner}, and page~635 in \cite{zeilinger1}. ``Having
grown up collecting clicks \ldots\ I would start with ``clicks'' as the
only point of contact between observer and observed'', wrote
J.~Summhammer in \cite[p.~261]{fuchs3}. It may be added that the
micro"=observation, as such, is terminated at the eigen elements; one
and the same $\state\alpha_j$ has always remained on the right.
\end{comment}

As a result, the minimal entity $\statePsi \GOTO{\sss\scr A}
\state\alpha_j$ constitutes, mathematically, an ordered pair
$(\statePsi,\state\alpha_j)$ of elements of the set~$\boT$, which
are labeled by the symbol~$\scr A$ that is equivalent to the
$\boT_{\!\sss\scr A}$"=family~\eqref{shema}. Accordingly, the
customary physical notion of the observation is substituted for a
micro"=event, an act. `Physics should forget' about processes or time
of interaction when observing, about the interaction itself, and
about anything but $\statePsi\GOTO{\sss\scr A}\state\alpha$. This
object represents a completed formalization of the
empirical/"!laboratory notion of a quantum micro"=event---a detector
click. The click is sometimes considered from the information
viewpoint as an information bit \cite{zeilinger1}. However it cannot
be such a (classical) bit with a reified content because it is
completely unpredictable. The next (different) click does entirely
negate the previous one, and the information bit is in turn a
concrete thing---the bit. For the same reason, there cannot be any
information behind the single micro"=event. It is `too small and too
momentary' to possess or to carry information about something
inasmuch as even the `something' is composed from elementary clicks;
see below.

\subsection{Numerical realizations}\label{makenumb}

Is there a possibility to rely exclusively on the inflexibility of
the eigen"=type elements~\eqref{akk}? Or to define the sought-for
ultimate distinguishability~$\not\approx$ through the
$\scr A$"=(micro)""distinguishabilities
$\state\alpha_j\not\approx_{\!\!\!\sss\scr A}\state\alpha_s$? Let us
formulate a thesis.
\begin{itemize}
\item[\hypertarget{T}{{\red\textsf{T}}}] There is no (linguistic) means of
 recognizing the system $\cal S$ to be different
 (pt.~\hyperlink{S}{\red\textsf{S}}), other than through the results of
 its destructions into the $\{\state{\alpha}_1$, $\state{\alpha}_2$,
 \ldots\}"=objects of observational instruments $\scr A$.
\end{itemize}
Granted, the stringency of this linguistic taboo
(\hyperlink{T}{\red\textsf{T}}) must be accompanied by something
constructive, and we will adopt the following programme, which
reflects the fact that the unequivocal description may only take the
form of a quantitative mathematical theory.
\begin{itemize}
\item[\hypertarget{R}{{\red\textsf{R}}$\phantom{\bullet}\bullet$}]%
 Out of the primary (`proto')""elements
 $\{\statePsi,\state\alpha,\ldots\}\in\boT$, one constructs a new set
 $\bbH$, of which the elements
 \begin{equation}\label{Repr}
 \ket\Xi\DEF
 {\bo\Oplus}(\frak a_1,\ket{\alpha_1};\frak a_2,\ket{\alpha_2}; \ldots)
 \in \bbH
 \end{equation}
 are said to be (number) \emph{representations} in the `reference
 frame for instrument $\scr A$', and $\frak a_s$ are the numerical
 objects. The distinguishability relation
 $\not\approx_{\!\!\!\sss\scr A}$ is carried over to $\bbH$ and admits
 an $\frak a$"=coordinate realization there---symbol~$\not\approx$.

\item[$\bullet\bullet$]No preferential or preordained observational reference frame
 $\scr A\{\state\alpha_1,\state\alpha_2,\ldots\}$---""instrument the
 absolute""---exists.
\end{itemize}

Identification \eqref{Repr} is always tied to a certain family
$\boT_{\!\sss\scr A}$. Accordingly, images of
$\state\alpha_s$---""symbols $\ket{\alpha_s}$---are present in
\eqref{Repr}, and character $\bo\Oplus$ is also no more than a
symbol here. Even though coordinates $\frak a_s$ are declared to be
numbers or aggregates of numbers, there is no arithmetic stipulated
for them yet. The number is as yet a name for $\frak a_s$.
Distinguishability $\ketPsi\not\approx\ket{{}\tilde{\bo\Psi}}$ of
two representatives
\begin{equation*}
{\bo\Oplus}(\frak a_1,\ket{\alpha_1};\frak a_2,\ket{\alpha_2};
\ldots)\FED\ketPsi\,,\qquad{\bo\Oplus}(\tilde{\frak a}_1,\ket{\alpha_1};
\tilde{\frak a}_2,
\ket{\alpha_2};\ldots)\FED\ket{{}\tilde{\bo\Psi}}
\end{equation*}
by means of numbers $\frak a_k\ne\tilde{\frak a}_k$ and mathematical
implementation of \eqref{Repr} and of the $\bbH$"=space, \ie, a
`coordinatization' scheme have yet to be established. This will
comprise the meaning of the word `constructs'
(sects.~\ref{Numbers}--\ref{statespace}), which may not be even
linked to the mathematical term mapping yet, since \emph{no math of
\qm\ exists at the moment}. It immediately follows that the question
about number entities""---""specifically, about \eqref{Repr}---is
nontrivial in physics.
\begin{itemize}
\item[\hypertarget{II}{{\red\textbf{\textsf{II}}}}] To speak of an exact
 correspondence between experiment and mathematics
 (\lrceil{observation \tplus\ measurement}) makes no sense until
 there has been detailed a \emph{mechanism for the
 emergence} of what is understood by number.\\
 \phantom.\hfill(\hbox{\embf{The $\bo2$-nd principium of quantum
 theory}})
\end{itemize}
In other words, we wonder what an empiricist/"!observer understands
(semantics) by the word (syntax) `number'. The underlying message
here implies that the reliance upon the all"=too"=familiar arithmetic
elucidates nothing. \emph{There is no arithmetic in
interferometers/"!colliders}---there are only clicks there---and
empirical nature of the arising that construction (along with the
measurement) must be scrutinized.

From pts.~\hyperlink{T}{\red\textsf{T}}, \hyperlink{R}{\red\textsf{R}},
and \hyperlink{II}{\red\textbf{\textsf{II}}} it also follows that the
search for a description through hidden variables, over which
something is averaged, is indistinguishable from the utopian
attempts to find out an intrinsic content of boxes~\eqref{box}.

\subsection{Macro and micro}\label{MM}

The task becomes more precise at this point. Instead of nonphysical
identity/"!noncoincidence ($\statePsi=\statePhi$ or
$\statePsi\ne\statePhi$) of two abstract elements $\statePsi$,
$\statePhi$ of the abstract set $\boT$, we need the concept of a
physical $\Approx$"=equivalence ($\not\Approx$"=distinguishability) of
$\bbH$"=representatives $\{\ketPsi$, $\ketPhi$, \ldots\}. That is, there
must hold either relation $\ketPsi\Approx\ketPhi$ or its negation
$\ketPsi\not\Approx\ketPhi$ for all $\ketPsi,\ketPhi\in\bbH$. The
primitive set $\boT$, initially required by
point~\hyperlink{S}{\red\textsf{S}}, must disappear from the ultimate
mathematics of symbols $\ketPsi\in\bbH$. Therefore elements
$\statePsi\in\boT$ are henceforth named \emph{primitives}.

Let us sum up the fallaciousness of a metaphysical belief in the
meaningfulness of the wording `there is a quantum state', \ie, the
belief in that the existence of a state has some math"=numerical
form.
\begin{itemize}
\item There is no a~priori way to endow the term (quantum) state of
 system $\cal S$ with any meaning \cite[p.~419]{lipkin}. It may not
 have a definition and any predefined semantics. This term should be
 created. Meanwhile, one cannot get around the concept of the
 (micro)""observation $\scr A$ \cite[pp.~98--100]{brukner},
 \cite[p.~646]{peres0}, \cite{benioff, ballentine2}. Essentially, no
 one thing, including $\statePsi$, $\state\alpha$, or the $\boT$"=set
 itself, can be the primary bearer of data about~$\cal S$.

 ``There is an entirely new idea involved, \ldots\ in terms of which one
 must proceed to build up an exact mathematical theory'' (P.~Dirac
 \cite[p.~12]{dirac}).
\end{itemize}
There is no escape from quoting K.~Popper: ``\ldots\ language for the
theory; \ldots\ it remains (like every language) to some extent vague
and ambiguous. It cannot be made ``\,precise\,'': the meaning of
concepts cannot, essentially, be laid down by any definition,
whether formal, operational, or ostensive. Any attempt to make the
meaning of the conceptual system ``\,precise\,'' by way of definitions
must lead to an infinite regress, and to merely \emph{apparent}
precision, which is the worst form of imprecision because it is the
most deceptive form. (This holds even for pure mathematics.)''
\cite[p.~13]{popper}.

The notions of a physical observable and of its observable values
are also ambiguous at this point \cite[p.~5]{ludwig1}. Their
ambiguity is even greater than that of state due to questions such
as `what is being measured?\@' and even `what is a measurement?'.
Nonetheless, up until the end of this section, we will not discard
the term state within the context of pt.~\hyperlink{S}{\red\textsf{S}}.

Irreproducibility of outcomes, \ie, the `turnability of
$\statePsi$"=primitives into the various' leaves only one option: `to
take a look at $\cal S$ again, once again, \ldots'. In other words, to
seek the source of description in repeatability. It is necessary,
then, to move to the subject of macro- rather than
micro"=observation. This intention fits perfectly with the undefined
verb ``constructs'' in pt.~\hyperlink{R}{\red\textsf{R}}, and the
following paradigm should be understood as the macro.

\begin{itemize}
\item[\hypertarget{M}{{\red\textsf{M}}}] The only way of handling the
 uncontrollable micro"=level changes is the treatment of the results
 of repeated destructions, accompanied by what we shall call the
 common physical macro"=setting (experimental context):
 \begin{equation}\label{macro}
 \begin{array}[c]{@{}c@{}c@{}c@{}cc@{}c@{}cc@{}}
 &\statePsi &\;\;\cdots\;\;& \statePsi &\statePsi&\;\;
 \cdots\;\;&\statePsi&\cdots\cdots\\
 \scr A&\DOWN&\DOWN&\DOWN&\DOWN&\DOWN&\DOWN&\DOWN\;\scr A\\
 &\state\alpha_1 &\cdots& \state\alpha_1&\state\alpha_2
 &\cdots&\state\alpha_2&\cdots\cdots
 \end{array}\quad\tplus\quad
 \lceil\text{common macro-environment }\mbf M\,\rceil\;.
 \end{equation}
\end{itemize}
To be precise, we should have to (and we shall do) indicate the
different $\{\statePsi$, $\statePsi'$, \ldots\} here because the same
ingoing $\statePsi$'s in \eqref{macro} is a preassumption""---which
we eschew throughout the work. This point will be very fully
addressed further below (sects.~\ref{A+S}, \ref{2.6}, and
\ref{mixture}).

The importance of repetitions and distinguishability had long been
noted (Bohr, von~Neumann et~al \cite{jammer}) and, recently, it was
particularly emphasized in the work \cite{zurek0}. The words
`co\-py/"!re\-peat\,\ldots/distin\,\ldots' occur 90 times therein.

Thus empiricism of quantum statics forces us to operate exclusively
with such formations of copies $\state\alpha$, \ldots, $\statePsi$, and
this is the maximum amount of data provided by the
supra"=mathematical problem setup. \embf{All} further mathematical
structures may come only from constructions like \eqref{macro} and
from nothing else. Getting ahead of ourselves, let us once again
turn our attention to the fact that implementation of this idea
\emph{is not short"=length}---``the mathematization process (cor) is
not simple'' \cite[p.~24]{ludwig5}, and sects.~\ref{Ans}--\ref{minus}
are devoted specifically to this; see, \eg, the
chain~\eqref{stream}.

One can once more repeat (sect.~\ref{physmath}) that much of what
follows does not and cannot contain the mathematical definienda and
proofs as they are usually present in the literature on quantum
foundations. Instead, there appears the step-by-step
\emph{inference} of objects as they result themselves: numbers,
operations, groups, algebras, \etc. The only instrument that may be
applicable here is the empirical inference.

The common macro"=environment $\mbf M$ in \eqref{macro} is also
viewed as a supra"=mathematical notion \cite{raseva}, the
mathematical implementation of which is yet to be created. The same
considerations regarding qualitative adjectives are applicable as to
the physical convention $\mbf M$ as well as the transition acts in
sect.~\ref{S}. Representations \eqref{Repr} will be the
formalization of the meaning \lrceil{observation} \tplus\
\lrceil{data on system $\cal S$}, but now with no references to the
elementary acts in \eqref{macro}. The physical distinguishability
criteria $\ketPsi\not\Approx\ketPhi$ may not be formulated yet
because the physical attributes are not yet available, but
$\ket{\alpha_s}$"=elements have already appeared in \eqref{Repr} as
prototypes of explicitly distinguishable~$\state\alpha_s$.

\begin{comment}\label{3}
Dual form of the typical quantum statements like `$\cal S$ is a
\emph{micro}"=system and $\scr A$"=instrument is a
\emph{macro}"=object' (N.~Bohr) is identical to the initial premise
`observation does always destroy a system'. It follows that there is
actually little need for that terminology. Indeed, \qm"=micro has no
internal structure and, hence, an oft"=discussed issue about boundary
(and limit\footnote{According to A.~Zeilinger, ``\ldots\ no limit. The
limit is only a question of \ldots\ money and of experience''
\cite[$13'09''$]{zeilinger2}.}) between micro and macro
\cite{saunders, home, ballentine, auletta} is devoid of sense; ``The
notions of 'microscopic' and 'macroscopic' defy precise definition''
\cite[p.~215]{bell}. Therefore this may be a matter only of
`different macro'---the `smaller/"!bigger'; \ie, when they describe
certain models.

As a (partially philosophical) aside, we note that what is
understood by observational randomness does, in fact, boil down to
distinguishability, and more specifically, to postulating the
micro"=chaos \eqref{rule}. In considering the denial of \eqref{rule}
as an impossible proposition, we arrive at the
\hyperlink{M}{\red\textsf{M}}"=paradigm and conclude that the only way
to deal with that which is contemplated for the subject"=matter of a
physical description must be the treatment of micro-acts as
assemblages \cite[Lect.~6]{mises}. In other words, and in accordance
with the outline of the clicks' analysis set out below, the
determinism of micro"=processes (micro"=ontology)---much less the
microscopic time"=arrow---is meaningless as a concept, since they are
not processes but rather structureless acts that have not even any
relationship to each other. Since there are no physical phenomena as
of yet, the claim that ``phenomenon"""~1 appears to be the cause that
precedes phenomenon"""~2 as the effect'' is no more than a collection
of words. To attribute physical content and mathematical formulation
at the micro"=level to them is impossible in principle; the `problem
of boxes' noted above. Accordingly, the cause of (classical)
macro"=indeterminism is the absurdity of the notion of its twin
concept---micro"=determinism""---and the unavoidable repetition of
the arrows $\goto$ (\hyperlink{M}{\red\textsf{M}}). N.~Bohr puts the
point very definitely: ``there can be no question of causality in the
ordinary sense of the word'' \cite[p.~351]{jammer}, and Heisenberg
adds that ``l'indeterminismo, \ldots\ \`e necessario, e non solo
consistentemente possibile'' \cite[sect.~IX.4]{accardi}. See
also~\cite[p.~223]{mises}.
\end{comment}

\subsection{Quantum ensembles and statistics}\label{A+S}

Let us call the upper row in \eqref{macro}, as a collection of the
$\statePsi$"=copies, (quantum) homogeneous \emph{ensemble}
(Kollektiv, by von~Mises \cite{mises}). We will designate it,
simplifying when needed, by
\begin{equation*}
\{\underbrace{\statePsi\:\statePsi\cdots\statePsi}_\text{$N$
times}\}\equiv\{\statePsi\cdots\statePsi\}_N
\equiv\ansPsi_N\;,
\end{equation*}
where $N$ is understood to be an arbitrary large number. Scheme
\eqref{macro} dictates also to consider the generic ensembles
\begin{equation}\label{mix}
\big\{\{\state{\alpha}_1\cdots\state{\alpha}_1\}_{n_1^{}}\;
\{\state{\alpha}_2\cdots\state{\alpha}_2\}_{n_2^{}}\;
\cdots\cdots\big\}\,,\qquad
\{\cdots\statePsi\cdots\statePsi\:\statePhi\cdots\statePhi\:
\state\Theta\cdots\state\Theta\cdots\}
\end{equation}
as collections of homogeneous sub"=ensembles. Ensembles are
symbolized in the same manner as sets but, for typographical
convenience, without the numerous commas and internal parentheses
$\{\}$ in \eqref{mix}; for example,
\begin{equation*}
\big\{a\,b\cdots b\{b\,c\,a\}\cdots\big\}=
\{a,b,\ldots,b,b,c,a,\ldots\}=\quad\cdots\quad\FED
\{ab\cdots bbca\cdots\}\;.
\end{equation*}

Scheme \eqref{macro} is the first point in which numbers emerge in
theory, and conversion
\begin{equation*}
\lceil\state\alpha\text{-ensemble
\eqref{mix}}\rceil\quad\rightarrowtail\quad(n_1\,,n_2\,,\ldots)
\end{equation*}
into the integer collection anticipates a \emph{numerical
$\scr A$"=measurement of} $\cal S$. Quantities $n_s\in\bb Z^+$,
however, should not be associated with such, as they are potentially
infinite. The minimal way of creating the knowingly finite numbers
out of independent and potential infinities $n_s$ (without loss of
their independence) is to divide each of them by a greater infinity,
which is a `constant' $\Sigma$ for the entire ensemble \eqref{mix}.
It is clear that one should put
\begin{equation}\label{fr}
\Sigma\DEF
n_1+n_2+\cdots\quad\text{and}\quad\Big\{\fr_1\DEF\frac{n_1}{\Sigma}\,,\quad \fr_2
\DEF\frac{n_2}{\Sigma}\,,\quad\ldots\Big\}\quad(\Sigma\rightsquigarrow\infty)\;,
\end{equation}
and that ensemble \eqref{mix} does not provide any numerical data
besides the relative frequencies \eqref{fr}. All the other data are
functions of $\fr_s$. An independence of the theory from the
ensemble's $\Sigma$"=constant, \ie, the scheme
$\Sigma\rightsquigarrow\infty$, is also implied to be a principle,
and it can be only the semantic one. Without it---the
$\Sigma$"=postulate of infinity""---there can be no question of a
rational theory, \ie, empiricism will not turn into a mathematics
(sects.~\ref{inv}--\ref{semi}). In turn, the concepts ``closely,
limit, the limiting frequencies'', \thelike\ will arise later when we
obtain the state of space as a Hilbert one $\bbH$ and topology on it
\cite{br3}.

Thus the \hyperlink{M}{\red\textsf{M}}"=paradigm \eqref{macro} does not
only give birth to a concept of numerical data in the theory per~se,
but also converts their $\bb Z^+$"=discreteness into the
$\bbR$"=continuum of real measurements. Namely, numbers
$\fr_s\in\bbR$ are the statistics $(\fr_1,\fr_2,\ldots)$ of
destructions $\GOTO{\sss\scr A}$ into the ensemble of primitives
$\{\{\state{\alpha}_1\}_{n_1}$, $\{\state{\alpha}_2\}_{n_2}$, \ldots\}.

\subsection{Distinguishability and numbers}\label{2.6}

Distinguishability of the two ensembles now turns out to be the
$\bbR$"=numerical, \ie, it is determined by the difference between
$\fr$"=numbers. As a result, and according to
pt.~\hyperlink{R}{\red\textsf{R}}, the two elements
$\ketPsi\not\approx_{\!\!\!\sss\scr A}\ket{{}\tilde{\bo\Psi}}$ of
$\bbH$ will differ in the numbers $\frak a_s$ and
$\tilde{\frak a}_s$, if the latter turn out to be the bearers of
different statistics
\begin{equation}\label{stat}
\fr_j(\frak a_1,\frak a_2,\ldots)\ne\skew1\tilde\fr_j
(\tilde{\frak a}_1,\tilde{\frak a}_2,\ldots)\;.
\end{equation}
In consequence, distinguishability $\not\approx$ is carried over to
$\bbH$ with an extension to the non"=eigen objects, but it is
inherently incomplete, since it does not take into account the most
significant fact---""arbitrariness of transitions~\eqref{Box}.

The collection $(\fr_1,\ldots)$, as a final result of transitions
$\{\statePsi\goto\state\alpha_s\}$, actually `knows nothing' about
their left hand side, much less about its uniqueness $\statePsi$.
For instance, if under the equal $\state\alpha$"=statistics
$\{\fr_s\}$ for the two families
$\{\state{\bfCourier{?}}\goto\alpha_s\}_{\sss N}$ and
$\{\statePsi\goto\alpha_s\}_{\sss N}$ (collectivity of
$\state{\bfCourier{?}}$'s), we would claim
$\state{\bfCourier{?}}=\statePsi$, this would mean a mass control
over transitions \eqref{rule}. Instead of a `black box' above, we
find that prior to acts $\GOTO{\sss\scr A}$ all the undefined
$\state{\bfCourier{?}}$'s were equal to $\statePsi$. This, however,
is the declaration of a property: `prior to observation the system
$\cal S$ was/"!dwelled in \ldots'. With any continuation of this
sentence, it is pointless and prohibited if one theoretically
accepts that, prior to observation, nothing exists and there are no
properties (sect.~\ref{S}). The words ``initial state of $\cal S$''
thus make no sense. The indeterminacy of the ingoing
$\state{\bfCourier{?}}$'s is therefore mandatory, and numbers
$(\fr_1,\fr_2,\ldots)$ required for recognition are manifestly
insufficient. Considering that the micro"=changeability of single
primitives $\statePsi$ also means nothing \cite{alter},
\cite[p.~493]{auletta}, \cite[p.~419\,(!), left column]{lipkin}, only
a generic ensemble
\begin{equation}\label{A}
\{\state{\bfCourier{?}}\cdots\state{\bfCourier{?}}\}\quad\rightarrowtail\quad
\{\cdots\statePsi\cdots\statePsi\:\statePhi\cdots\statePhi\:
\state\Theta\cdots\state\Theta\cdots\}\FED\frak A
\end{equation}
can be an intermediary in the sought"=for translation of
$\statePsi$'s onto representations $\ket\Xi\in\bbH$ under
construction~\eqref{Repr}.

In the accustomed physical terminology, the above is expressed in
the sequence
\begin{Align}\label{quant}
\lceil\text{state}\rceil\;
\mathrel{\Over[1.2]{\scr A_\text{quant}}
{\REPEAT{\mathchar"439}{5} \mathchar"044B}}\;&\;
\lceil\text{state$'\rceil$}&&\;\longmapsto\;
\lceil\text{measurement}\rceil\;.
\intertext{The removal of the intermediate component here, \ie,
switch to the sequence}
\label{class}
\lceil\text{state}\rceil\;
\mathrel{\Over[1.2]{\scr A_\text{class}}
{\REPEAT{\mathchar"439}{5} \mathchar"044B}}\;&
\,[6]\cdots\cdots&&\;\longmapsto\;\lceil\text{measurement}\rceil
\end{Align}
amounts to the rejection of micro"=destructibility and of
unpredictability. Even with the classical framework, this
supposition is questionable, since the notion of a `change when
observed' disappears. The relationships between the dual
concepts---(micro/macro)"=scopicity, big/middle/small, \etc---do also
get lost. That is the reason why, developing Heisenberg's asking the
question ``\ldots\ is it \ldots\ I can only find in nature situations which
can be described by quantum mechanics?\@'' \cite[p.~325]{jammer}, we
conclude that, strictly speaking,
\begin{itemize}
\item \emph{all observations}\eLab{ClassQuant}, regardless of (the
 envisioned physical) macro/meso/micro characteristics, do have the
 structure \eqref{quant}, \ie, are quantum. \emph{No non"=quantal
 observations exist}.
\end{itemize}
With their idealized `roughening', the classical description appends
numerical $\fr$"=statistics to \eqref{class}, which is when the
left/right sides of \eqref{class} become indistinguishable with
respect to the arrow symbols. The arrows may then be replaced with
the equivalence
\begin{equation}\label{grubo}
\lceil\text{state}\rceil\quad
\Over[1.7]{\scr A_\text{class}}
{\rlap{\raise0.45ex\hbox{${\dashleftarrow}\mathchar"439
\mathchar"439\mathchar"439$}}
\lower0.45ex\hbox{$\mathchar"439\mathchar"439
\mathchar"439{\dashrightarrow}$}}
\quad\lceil\text{$\fr$"=statistics numbers}\rceil\;.
\end{equation}

Supplementing the right hand side here with the concept of numerical
values $\{\alpha_s\}$ for all of the observables $\scr A=A(q,p)$ (or
for phase variables $\{q_1,q_2,\ldots;p_1,p_2,\ldots\}$), this side will
turn into an exhaustive numerical realization of the left hand side.
Criterion $\approx$, then, turn into the $\bbR$"=number equality $=$
of all the $\scr A$"=statistics or into an equality of phase
distributions $\varrho(q_1,\ldots;p_1,\ldots)$. This is a situation of the
classical (statistical) physics
({\smaller[1]C}lass{\smaller[1]P}hys), \ie, when `the physics is
initially identified' with quantities being numerical in character:
the particle coordinates/"!numbers, the number values of field
functions, \etc. The ill"=posedness of such a paradigm""---the core
motive of \qt---is discussed further below at greater length in
sects.~\ref{phys}, \ref{2slit}, and~\ref{Numbers}. Consequently,
`classicality' is not and cannot be regarded as a primitive in the
logical construct. In both these cases,
distinguishability~$\not\approx$ depends on the concept of
$\state\alpha$"=states.

\begin{comment}\label{rem4}
From this point onward, by state we will strictly mean
representations \eqref{Repr}. So it makes \emph{no sense to speak of
transitions between states}, much less of ``transition from possible
to actual'' \cite[p.~189; Everett]{barrett}, \cite{schloss, joos,
zurek}. The writing $\ket\Xi\goto\ket{\alpha}$ and its typical the
wave"=function collapse interpretation are not correct. Indeed, in
treating transition $\ket\Xi\goto\ket{\alpha_1}$ as a state-to"=state
destruction, its left hand side cannot carry any information about
$\fr_{\sss(\scr A)}$"=frequencies for other events
$\ket\Xi\goto\ket{\alpha_s}$, \emph{much less} about amount of
destruction from envisioned $\scr B$"=observations
$\ket\Xi\GOTO{\sss\scr B}\ket{\beta}$. Such
`$\fr_{\sss(\scr B)}$"=amounts' are always present at the
experimental interpretation of the $\ket\Xi$"=symbol. For this
reason, the concept of a state should not be used as a correct term
at all \cite{ludwig5}; the terminology, however, has been settled.
\end{comment}

The motivation given above---\hyperlink{S}{\red\textsf{S}} (system,
primitives), \hyperlink{O}{\red\textsf{O}} (observation) and
\hyperlink{R}{\red\textsf{R}} (representations),
\hyperlink{T}{\red\textsf{T}} (taboo), the semantic principia
\hyperlink{I}{\red\textbf{\textsf{I}}} (\qm"=statics) and
\hyperlink{II}{\red\textbf{\textsf{II}}} (numbers) complemented below with
the principium~\hyperlink{III}{\red\textbf{\textsf{III}}}---is sufficient
for further creating the basis of mathematical formalism of \qm.
These tenets should hardly be regarded as postulates, at least in
the common meaning of the phrase `postulates of a physical theory',
since they are of natural"=language nature and are, as we believe,
the points of departure for reasoning whatever the approach to the
micro-world. It is clear that they are directly concerned with the
familiar dialogs, which reflect, in the words by Bohr, ``[Einstein's]
feeling of disquietude as regards the apparent lack of firmly laid
down principles \ldots, in which all could agree''
\cite[p.~228]{schilpp}.

The underpinning of \qt\ must thus begin, at least to a large
extent, with a simplification/"!reducing the terminology in use and
putting the language and the semantics of observations/"!numbers in
order, rather than in giving the `improved' postulates or
definitions.
\begin{quote}
``The task is not to make sense of the quantum axioms by heaping more
structure, more definitions, \ldots, but to throw them away wholesale''
\\\phantom.\hfill C.~Fuchs \cite[p.~989]{fuchs2}
\\[1ex]
``Simplicity is implicit in the basic goals of scientific inquiry.
\ldots\ only simple theories can attain a rich explanatory depth. \ldots\
the Basic Propert[ies] should indeed be very simple''
\\\phantom.\hfill N.~Chomsky \cite[pp.~4--5]{chomsky+}
\end{quote}

As was underscored above, these (organizing) principles do not
stipulate for pre"=determined mathematics and physics, with the
exception of the linguistic/"!metamathematical understanding
\cite{kleene, raseva} of how to look at the mathematical axioms,
structures, rational theories, and their interpretations altogether.
See also Remarks~\ref{seven} and \ref{meta}, and
sects.~\ref{EmpMath}, \ref{interpret}.

\vbox{
\section{Ensemble formations}\label{Ans}

\flushright\tiny
\textsl{Your acquaintance with reality grows literally by buds or\\
drops of perceptions. \ldots\ they come totally or not at all}\\
\textsc{--- W.~James (1911)}
\smallskip\nopagebreak

\textsl{Are billions upon billions of acts of observer-\\
participancy the foundation of everything?}\\
\textsc{--- J.~Wheeler \cite[p.~199]{wheeler}}}
\smallskip\nopagebreak

The key corollary of the macro"=paradigm \eqref{macro} is not merely
the appearance of numerical data in the theory, but also the fact
that the further construct must rely not on isolated primitives but
on their aggregates being considered as an integrated whole, \ie, as
a set. This causes a choice for the ensemble notation.

\subsection{Mixtures of ensembles}\label{mixture}

Returning to the analysis of transitions
$\state{\bfCourier{?}}\goto\alpha_s$, one gets that the lower row in
\eqref{macro} comes actually from indeterminacy
\begin{equation*}
\begin{array}{c}
\{\state\alpha_1\}_{n_1}\{\state\alpha_2\}_{n_2}
\cdots\cdots\\[0.5ex]
\UP\quad\UP\quad\UP\quad\UP\quad\UP\\[0.5ex]
\;\state{\bfCourier{?}} \cdots \state{\bfCourier{?}} \cdots
\state{\bfCourier{?}}\,\,\cdots\cdots
\end{array}
\end{equation*}
and thus \eqref{macro}, by virtue of \eqref{A}, should be replaced
with the scheme
\begin{equation}\label{Macro}
\array{c}
\{\cdots\statePsi\cdots\statePhi\cdots\state\Theta\cdots\}\\[0.5ex]
\scr A\;\DOWN\quad\DOWN\quad\DOWN\quad\DOWN\quad\DOWN\;\scr A\\[0.5ex]
\big\{\cdots\state\alpha_1\cdots\state\alpha_2\cdots
\big\}
\endarray\;,
\end{equation}
wherein the composition of the upper ingoing row may not be
predetermined. Fundamentally, according to \eqref{quant}, it may not
be withdrawn from \eqref{Macro}, yet at the same time the meaning of
the row can in no way be aligned with the adjective `observable' via
typical empirical/"!physical words: properties, readings,
quantities/"!amounts, and other `observable' characteristics. Such a
non"=detectability is the equivalent of a box that may be prepended
to the scheme \eqref{Box}:
\begin{equation}\label{chain}
\def\fboxsep{1.5\dimen36}
\boxed{\cdots\cdots}\TRANS{}\statePsi\;
\smash{\ds\TRANS{\scr A}}\;\state\alpha\;.
\end{equation}
If $\state\beta$'s serve as $\statePsi$ in \eqref{chain}, then we
have the schemes of precedence and of continuation:
\begin{equation*}
\def\fboxsep{1.5\dimen36}
\boxed{\cdots\cdots}\;\smash{\ds\TRANS{\scr A}}\;
\state\alpha\;\smash{\ds\TRANS{\scr B}}\;
\state\beta\qquad\text{or}\qquad \boxed{\cdots\cdots}\;
\smash{\ds\TRANS{\scr B}}\;\state\beta\;
\smash{\ds\TRANS{\scr A}}\;\state\alpha\;.
\end{equation*}

Let an observer capture the fact of any distinguishability in the
penultimate $\scr A$. Section~\ref{S} tells us that this may only be
the distinguishability of objects
$\{\state\alpha_1,\state\alpha_2,\ldots\}$; hence, this very $\scr A$
turns into an observation (pt.~\hyperlink{O}{\red\textsf{O}}). The
\hyperlink{M}{\red\textsf{M}}"=paradigm then gives rise to the numbers
of $\state\alpha$"=events $(n_1,n_2,\ldots)$ and, thereupon, their
relative frequencies $(\varrho_1,\varrho_2,\ldots)$ by the rule
\eqref{fr}. If subsequently micro"=observations $\scr B$ are to
follow, then a composite macro"=observation $\scr B\circ\scr A$ has
been formed, and frequencies $\{\varrho_j\}$ cannot but impact on
statistical characteristics of these later $\scr B$"=observation's
micro"=events. However, being an ingoing ensemble for $\scr B$, each
homogeneous $\{\state\alpha_s\cdots\state\alpha_s\}_{n_s}$ is
indistinguishable from an indefinite ensemble
$\{\cdots\statePsi\cdots\statePhi \cdots\}_{n_s}$, since the concept
of `$\approx_{\!\!\!\sss\scr A}$"=sameness' is unknown for $\scr B$.
Instrument $\scr B$ is `aware of only its own
$\approx_{\!\!\sss\scr B}$ and cannot know \emph{what} it destroys',
or that the source"=object consists of one and the same
$\state\alpha_s$. Rejection of this point brings us once again
(p.~\pageref{bbox}) to attempts at `penetrating the black box' of
transitions \eqref{box}, \ie, to attempts at creating the physics of
a more primary level. According to pts.~\hyperlink{O}{\red\textsf{O}}
and \hyperlink{M}{\red\textsf{M}}, an instrument produces nothing more
than its own `destruction list'; in this case,
$(\{\state\beta_1\}_{m_1},\{\state\beta_2\}_{m_2},\ldots)$. This list
is completely independent of the preceding one since, according to
pt.~\hyperlink{R}{\red\textsf{R}}$^{\bullet\bullet}$, there cannot be restrictions
on $\boT_{\!\sss\scr A}$ and $\boT_{\!\sss\scr B}$. In case the set
$\{\state\alpha_s\cdots\state\alpha_s\}_{n_s}$ transits into
collection $\{\state\beta_k\cdots\state\beta_k\}_{n_s}$, this means
that $\state\alpha_s$ has always transited into one and the same
$\state\beta_k$ every time (under the convention
$\Sigma\rightsquigarrow\infty$), and there takes place merely a
coincidence $\state\alpha_s=\state\beta_k$ of eigen"=primitives in
the lists $\boT_{\!\sss\scr A}$ and $\boT_{\!\sss\scr B}$.

If $\scr B\circ\scr A$ is proceeded with a third observation
$\scr C$, the preceding analysis is repeated recursively with the
same result; only the values $\{\varrho_j\}$ will be changed. In
consequence, only the following two ingoing types for macro"=scheme
\eqref{Macro} are conceivable:
\begin{Align}
\big\{\cdots\statePsi\cdots\statePhi\cdots\state\Theta\cdots\big\}\quad&
\begin{array}{l}
\text{indefinite ensemble}\\\lceil\text{no statistics}\rceil\;,
\end{array}\label{m1}\\
\big\{\{{\cdot}{\cdot}{\cdot} \statePsi {\cdot}{\cdot}{\cdot}
\statePhi{\cdot}{\cdot}{\cdot}\state\Theta{\cdot}{\cdot}{\cdot}\}
^{{\sss(}\varrho_1^{}\sss)}\,
\{{\cdot}{\cdot}{\cdot} \statePsi {\cdot}{\cdot}{\cdot}
\statePhi{\cdot}{\cdot}{\cdot}\state\Theta{\cdot}{\cdot}{\cdot}\}
^{{\sss(}\varrho_2^{}\sss)}
\cdots\big\} \quad&
\begin{array}{l}
\text{ensemble mixture}\\\lceil\text{with statistics
$(\varrho_1,\varrho_2,\ldots)$}\rceil\;.
\end{array}\label{m2}
\end{Align}
It is reasonable to regard case \eqref{m2} as a `non"=interfering'
mixture of the system's $\scr A$"=preparations
\begin{equation*}
\{\cal S^{{\sss(}\varrho_1^{}\sss)},\:
\cal S^{{\sss(}\varrho_2^{}\sss)},\: \ldots\}\quad\hhence\quad
\{\state\alpha_1^{{\sss(}\varrho_1^{}\sss)},\:
\state\alpha_2^{{\sss(}\varrho_2^{}\sss)},\: \ldots\}\;,
\end{equation*}
to each of which one assigns the positive number $\varrho_s<1$
referred to as its statistical weight. These weights---``an element
of reality'' \cite[p.~649]{peres0}---are all that is inherited from
the preparation $\scr A$, and subsequent micro"=observation acts
$\scr B$ are performed again on indefinite ensembles \eqref{m1}.

It is clear that in the view of transitions $\goto$ in scheme
\eqref{Macro}, this situation is a derivative of \eqref{m1} and this
very type \eqref{m1} is crucial \cite[p.~53]{ballentine2}. In other
words, if preparation is regarded a concept as essential as
observation (pt.~\hyperlink{O}{\red\textsf{O}}), we still remain within
the framework of the binate essence of the transition:
\begin{equation*}
\statePsi\GOTO[3]{\scr B}\state\beta\,,\qquad
\state\alpha\GOTO[3]{\scr B}\state\beta\;.
\end{equation*}
Its left hand side should always be seen as an undetermined
primitive, even though we treat/call it the preparatory
(micro)""observation. See also ``preparation"=measurement reciprocity''
in \cite{ballentine4}.

\subsection{Ensemble brace}\label{brace}

According to pts.~\hyperlink{R}{\red\textsf{R}} and
\hyperlink{M}{\red\textsf{M}}, representations \eqref{Repr} must
reflect all information about physics of the problem:
primitives/"!incomes, transitions (`arrows' $\goto$), and outgoing
statistics. All that data are contained in scheme \eqref{Macro},
which is why the maximum that the model of a future mathematical
object---it characterizes everything we get while watching the
$\cal S$---can rely on is the \emph{ensemble brace}:
\begin{equation}\label{obj}
\obj{\state\Xi}\DEF
\mbig[13]\lgroup
\array{@{}c@{}}
\{\cdots\statePsi\cdots\statePhi\cdots\state\Theta\cdots\}\\[0.5ex]
{\scs\scr A}\;\DOWN\quad\DOWN\quad\DOWN\quad\DOWN\quad\DOWN\;{\scs\scr A}\\[0.5ex]
\big\{\cdots\state\alpha_1\cdots\state\alpha_2\cdots\big\}
\endarray
\mbig[13]\rgroup
\end{equation}
(or a couple of ensemble bunches).

It is immediately seen that \eqref{obj} carries the radical
difference between situation \eqref{quant} and its `roughening'
\eqref{grubo}; because of the upper row. The enormous arbitrariness
within the brace and arrows $\GOTO{\sss\scr A}$ is `programmed' to
give birth to the different processing rules of statistics and to
effects being typical for \qm. Thanks to the maximality of
\eqref{obj}, it is only this row that encodes all the sought-for
cases of distinguishability~$\not\approx$. In particular, by varying
the upper row, while the lower one remains unchanged, we get into a
situation when $\state\alpha$"=statistics $(\fr_1,\fr_2,\ldots)$ is
found to be the same for $\obj{\state\Xi}$ and
$\obj{\state{\tilde\Xi}}$, and meanwhile,
$\obj{\state\Xi}\not\approx\obj{\state{\tilde\Xi}}$.

The further problem is thus as follows. With the indefinite
$\frak A$"=ensemble \eqref{A} in hand, \ie, with the upper row of
\eqref{obj}, is it possible, based on the principles described
above, to bring the still incomplete relation $\not\approx$ to the
maximal quantum"=physical distinguishability of states?

\vbox{
\section{Why does domain $\bbC$ come into being?}%
\label{whyC}

\flushright\tiny
\textsl{\ldots\ quod ideo sint imaginariae, \ldots\ quod ideo sint \ldots\\
tum certe forent reales ideoque non imaginariae\footnotemark}\\
\textsc{--- L.~Euler} (1736)%
\medskip

\textsl{\ldots\ denn die imagin\"aren Gr\"o\ss{}en
existierten doch nicht\,?}\\
\textsc{--- D.~Hilbert} (1926)}%
\footnotetext{\ldots\ this is why they are imaginary. Were they \ldots,
they would certainly be real and therefore not imaginary.}
\smallskip\nopagebreak


The first priority, in the $\not\approx$"=distinguishability of
objects \eqref{obj}, is to separate the closest and unconditional
criterion""---the outgoing $\state\alpha$"=statistics. To do this,
let us split the lower row into families
$\big\{\{\state\alpha_1\}_{\infty_1^{}}
\{\state\alpha_2\}_{\infty_2^{}}\cdots\big\}$, where
\begin{equation}\label{infty}
\infty_1+\infty_2+\cdots=\infty\;,
\end{equation}
and, subsequently (rather than the reverse, otherwise \eqref{m2}),
taking into account the `arbitrariness of arrows', we also split the
upper row:
\begin{equation}\label{split}
\obj{\state\Xi}=
\mbig[13]\lgroup
\begin{array}{@{}r@{}c@{}c@{}c@{}l@{}}
\big\{&\{{\cdot}{\cdot}{\cdot} \statePsi {\cdot}{\cdot}{\cdot}
\statePhi{\cdot}{\cdot}{\cdot}\state\Theta{\cdot}{\cdot}{\cdot}\}
_{\hbox to0ex{$\scs\infty_1^{}$}}&\qquad
\{{\cdot}{\cdot}{\cdot} \statePsi {\cdot}{\cdot}{\cdot}
\statePhi{\cdot}{\cdot}{\cdot}\state\Theta{\cdot}{\cdot}{\cdot}\}
_{\hbox to0ex{$\scs\infty_2^{}$}}\qquad
&\cdots&\big\}\\
&\DOWN\quad\DOWN\quad\DOWN\quad\DOWN&\DOWN\quad\DOWN\quad\DOWN\quad\DOWN&\\
\big\{&\{\state\alpha_1\cdots\cdots\state\alpha_1\}
_{\hbox to0ex{$\scs\infty_1^{}$}}&
\{\state\alpha_2\cdots\cdots\state\alpha_2\}
_{\hbox to0ex{$\scs\infty_2^{}$}}&\cdots&\big\}
\end{array}
\mbig[13]\rgroup
\end{equation}
(the indication of observation $\GOTO{\sss\scr A}$ is omitted
further below, since it has been mirrored in primitives
$\state\alpha$). Hereafter, infinities $\infty_j$ stand for cardinal
numbers (a number of elements, possibly finite) of their own
ensembles. Therefore, the extension of distinguishability
\eqref{stat} should be produced by comparing the sub"=objects such as
\begin{equation}\label{tmp0}
\begin{array}{c}
\{{\cdot}{\cdot}{\cdot} \statePsi {\cdot}{\cdot}{\cdot}
\statePhi{\cdot}{\cdot}{\cdot}\state\Theta{\cdot}{\cdot}{\cdot}\}
_{\hbox to0ex{$\scs\infty_1^{}$}}\\
\DOWN\quad\DOWN\quad\DOWN\quad\DOWN\\
\{\state\alpha_1\cdots\cdots\state\alpha_1\}
_{\hbox to0ex{$\scs\infty_1^{}$}}
\end{array}\quad,
\end{equation}
that differ from each other in the upper-row composition.

\subsection{Continuum of quantum phases}

Cardinality of the $\boT$"=set cannot be finite. This would entail
finitely many $\state\alpha$"=primitives for all kinds of
instruments. But finiteness of this number $n_{\sss\boT}$ would mean
an exclusivity of its value that does not follow from anywhere. At
the same time, all the $\frak A$"=ensembles \eqref{A} are subsets of
the set $\boT$ (boolean\eLab{bulean} $2^\boT$); any finite portion
of it is ruled out. Hence, the endless variety of upper rows in
\eqref{tmp0} is uncountable.

Aside from the numerical $\fr$"=statistics,
program~\hyperlink{R}{\red\textsf{R}} does also require an association
of the numerical objects with each row
\begin{equation*}
\frak A=\{\cdots\statePsi\cdots\statePsi\:\statePhi\cdots\statePhi\:
\state\Theta\cdots\state\Theta\cdots\}
_{\scs\infty}\quad\hhence\quad\cdots\;,
\end{equation*}
because primitive's symbols must disappear in the ultimate
description. To avoid introducing the structures ad~hoc, we will
produce numbers here---the upper row---in the same manner, in which
statistics were producing in sect.~\ref{A+S}---the lower row.
Indeed, the genesis of the concept of the number must be single in
theory. That is, we should again take into account the presence of
copies of primitives and write
\begin{equation}\label{tmp}
\cdots\quad\hhence\quad\big\{
\underbrace{\{\statePsi'\}_{\infty'}\{\statePsi’’\}_{\infty’’}\:
\cdots}_{K\text{ times}}\big\}\;,
\end{equation}
and numbers per~se will come into being by the $\Sigma$"=convention
like \eqref{fr}, \ie, through the cardinal ratios
\begin{equation}\label{infty'}
\varkappa'\DEF\frac{\infty'}{\infty}\,,\qquad
\varkappa’’\DEF\frac{\infty’’}{\infty}\,,\qquad\ldots\;.
\end{equation}

Now, the discreteness of micro"=transition acts is embodied in
\eqref{tmp} with the sequence $(\statePsi'$, $\statePsi’’$, \ldots),
and the uncountability of micro"=arbitrariness is inherited by
attaching the symbolic `quantities'"!---`countless' characters
$(\infty'$, $\infty’’$, \ldots)---to elements of this sequence. The
global discreteness says that there are no grounds to assume a more
than countable infinity $\aleph_\textsf{o}$ for the set $\boT$, \ie,
$|\boT|=\aleph_\textsf{o}$. The infinity of the family \eqref{tmp},
hence, has the type
\begin{equation*}
2^{\aleph_\textsf{o}}=\aleph\;,
\end{equation*}
\ie, it is continual \cite{kurat}. Parenthetically, the
$2^{\aleph_\textsf{o}}$ is the only known way of introducing the
continual (more than discrete) mathematical infinity. Which
possibilities exist for the form of row~\eqref{tmp}?

The trivial case $\frak A=\big\{{\{\statePsi'\}_{\infty'}}\big\}$,
\ie, $K=1$ in \eqref{tmp} drops out at once, since element
$\statePsi'$ would always go into the same primitive:
\begin{equation}\label{N1}
\begin{array}{c}
\{\statePsi'\cdots\cdots
\statePsi'\}_{\hbox to0ex{$\scs\infty_1^{}$}}\\
\DOWN\quad\DOWN\quad\DOWN\quad\DOWN\\
\{\state\alpha_1\cdots\cdots\state\alpha_1\}
_{\hbox to0ex{$\scs\infty_1^{}$}}
\end{array}\quad=\quad
\begin{array}{c}
\{\statePsi'\}_{\hbox to0ex{$\scs\infty_1^{}$}}
\\\!\!{\scs\scr A}
\DOWN{\!\cdot{\cdot}\cdot\!}\DOWN{\scs\scr A}\\
\{\state\alpha_1\}_{\hbox to0ex{$\scs\infty_1^{}$}}
\end{array}\;.
\end{equation}
But this is tantamount to the identity
$\statePsi'\equiv\state\alpha_1$, which robs of any meaning the
concept of the transition $\statePsi\GOTO{\sss\scr A}\state\alpha$
whatsoever. We get a single number here---the number of
$\state\alpha_1$"=clicks---and arrive thereby at classical
statistics, the physics of which is inadequate with respect to the
interference patterns. Hence the following options are admissible
for the formations \eqref{tmp}:
\begin{multline}\label{N3}
\big\{\underbrace{\{\statePsi'\}_{\infty'}
\{\statePsi’’\}_{\infty’’}}_{K=2}\big\}\,,\quad\ldots,\quad
\big\{\underbrace{\{\statePsi'\}_{\infty'}\{\statePsi’’\}_{\infty’’}\,
\{\statePsi'\!\!'\!\!'\}_{\infty'\!\!\!'\!\!\!'}\:\cdots}_
{3\leqslant K<\infty}\big\}\,,\quad\ldots\quad \\
\ldots,\quad\big\{\underbrace{\{\statePsi'\}_{\infty'}
\{\statePsi’’\}_{\infty’’}\,
\{\statePsi'\!\!'\!\!'\}_{\infty'\!\!\!'\!\!\!'}\:\cdots}_{K=\infty}\big\}
\end{multline}
with minimal $K=2$. If some of the infinities
$(\infty',\infty’’,\ldots)$ are finite here or countable, this does not
change the total continuality $\aleph$. The extreme case
$K=\infty$---a countable infinity of continuums""---does also change
no this count because of $\aleph+\aleph+\cdots=\aleph$ \cite{haus}.
All of these infinities may be even countably duplicated without
augmenting the continuum since
$\aleph\cdot\aleph\cdots=\aleph^{\aleph_\textsf{o}}=\aleph$.

What can one say about relationship of cases \eqref{N3} between each
other? Do we have to deal with their total arbitrariness or with
only one of these schemes? The latter case---the
sameness/"!indistinguishability of upper rows in \eqref{tmp0}---would
correspond to the structural staticity of theory. Or, whether one
(unrecognizable upper) row should differ (why?) from another in the
number (what?) of defining primitives
$\{\statePsi',\statePsi’’,\ldots\}$ (which ones?)?

Suppose the variability of $K$. That is, consider the simultaneous
existence of, say, the $K=\{2,3\}$ rows
\begin{equation*}
\big\{\{\state\Psi'\}_{\infty'}\{\state\Psi’’\}_{\infty’’}\big\}\,,\qquad
\big\{\{\state\Psi'\}_{\infty'}\{\state\Psi’’\}_{\infty’’}
\{\state\Psi'\!\!'\!\!'\}_{\infty'\!\!'\!\!'}\big\}\;.
\end{equation*}
However each of the $2$"=row is a particular case of the $3$"=row with
a cardinal number $\infty'\!\!'\!\!'=""0$:
\begin{equation*}
\big\{\{\state\Psi'\}_{\infty'}\{\state\Psi’’\}_{\infty’’}\big\}=
\big\{\{\state\Psi'\}_{\infty'}\{\state\Psi’’\}_{\infty’’}
\{\state\Psi'\!\!'\!\!'\}_{\{\infty'\!\!'\!\!'=0\}}\big\}\subset
\big\{\{\state\Psi'\}_{\infty'}\{\state\Psi’’\}_{\infty’’}
\{\state\Psi'\!\!'\!\!'\}_{\infty'\!\!'\!\!'}\big\}
\end{equation*}
(the case in point is sets). Therefore these situations are
structurally indistinguishable from each other, and the $K=2$ theory
is a \emph{sub}theory for $K=3$. So, the cases $K=\{2,3\}$ are
actually not mutually exclusive; rather, they form an embedding. We
thus have arrived at the one cumbersome and common construct akin to
the Russian dolls $2\supset3\supset4\supset\cdots$. Hence the
minimal 2-theory will always be present inside of all the higher
orders $K>2$ as an `independent (sub)world'. For this reason, the
$K=2$ theory must be created in any way; incidentally, it will
enclose the $K=1$ case.

On the other part, we have no criteria to terminate the sequence
$2\rightarrowtail 3\rightarrowtail\cdots$ at some intermediate
$K<\infty$. Such a cut-off does immediately arise an issue of the
questionable empirical exclusivity of a certain `world integer
$K\geqslant3$' that defines the number of `physically inaccessible'
$\statePsi$"=objects. Besides, these options would be related to a
certain topological dimension $K\geqslant3$ that has an unmotivated origin.
We thus conclude that the non"=minimal options $K=3$, $K=4$, \ldots\ in
\eqref{N3} should be dismissed.

\begin{comment}
A few remarks may be made in connection with case $K=\infty$. It is
related to a conglomerate of infinities, which has the form of a
discretely infinite family of continual infinities $\{\varkappa'$,
$\varkappa’’$, \ldots\}, and things would have been even `worse' had
the staticity of the schemes \eqref{N3} been changed to variability.
Such formations would need to be equipped with topology and with
associated concepts of convergence and of limit. But all this
touches on principally unobservable numerical entities, for which it
is not clear how to motivate the further reductions to `finite
mathematics' as required: dimensions, finite approximations, finite
numbers (which ones?), \thelike. More to the point, all of that
would pertain to the global structural parameters of the theory
prior to constructing it per~se, not to mention the physical models.
To put it plainly, such an assumption would result not in a theory
but in a theory of theories, and so on ad~infinitum which should be
somewhere terminated in some way. For these reasons, we leave the
case $K=\infty$ aside, though it might be worth elaborating on it.
However, in sect.~\ref{naturalC}, we will give a further
justification of that the number domain of the theory is what it has
already been known in~\qm.
\end{comment}

As a result, one has a choice: the structural staticity $K=2$ or
entirely non"=structured/"!undetermined set of outgoing primitives
$\{\statePsi_j,\statePsi_k,\ldots\}$, \ie, extremely complex case
$K=\infty$. We do choose $K=2$. This option might have been adopted
even before on the ground that the most minimalistic construction,
which set"=theoretically gives rise, as a minimum, to the minimal
numerical object---a single number""---""corresponds to the minimal
$K=2$ in \eqref{N3}. The maximal case is problematic, while the
mid-ones are ruled out. That is to say, all possible assumptions
regarding the upper row structure in \eqref{tmp0} are
indistinguishable from a case just as if the row contained two
primitives only $\{\statePsi'$,
$\statePsi’’\}\FED\{\statePsi,\statePhi\}$. Functionality of the
symbol $\cup$, having regard to inclusion of the $\{\statePsi,
\statePhi\}$'s copies, is unchanged as is; see sect.~\ref{inv}
further below.

We establish also that, in the following writing
\begin{equation*}
\{\statePsi\cdots\statePsi\}_{\infty_1'}\cup
\{\statePhi\cdots\statePhi\}_{\infty’’_1}\quad
\setbox0=\hbox{$\vcenter{\hbox{\large$\scs\dashrightarrow$}}$}
\rlap{\copy0} \rlap{\raise0.9ex\copy0} \lower0.9ex\box0 \quad
\{\state\alpha_1\cdots\state\alpha_1\}_{\infty_1^{}}
\end{equation*}

\noindent
of the scheme \eqref{tmp0}, none of the primitives $\{\statePsi$,
$\statePhi\}$ coincide with $\state\alpha_1$. Otherwise, the
unrestricted adjunction of identical transitions
$\state\alpha_1\goto\state\alpha_1$ to \eqref{tmp0} would mean
indeterminacy of both the number $\varkappa_1$ and the actual
statistics $(\fr_1,\fr_2,\ldots)$.

Let us take into account that numbers \eqref{infty'} are
mathematically generated by the standard scheme: \lrceil{(ordered)
integers} $\rightarrowtail$ \lrceil{(or\-der\-ed) rationals}
$\rightarrowtail$ \lrceil{(or\-der\-ed) continuum}. The natural
ordering $<$ is always present here and, as is well known
\cite[p.~52]{stoll}, can be isomorphically represented by the
set"=theoretic inclusion~$\subset$ on a certain system of sets. That
inclusion (\,$=$~`to be contained in'), in turn, is directly
concerned with the semantics of sect.~\ref{S0}. The natural"=language
term `accumulating'"!---`the old is being nested into the new'---is
formalized to create sets by the cumulative ensembles (see
sect.~\ref{inv}).

We now conclude that all kinds of schemes \eqref{tmp0} form an
$\aleph$"=continuum, for which there is no reasonable rationale for
equipping it with a topology other than the standard order topology
of the 1"=dimensional real $\bbR$"=axis or its equivalents. Call the
quantity $\varkappa\in\bbR$ \emph{quantum phase}.

It should be added that in considering some two upper rows in
\eqref{tmp0} as infinite sets
\begin{equation*}
\{\statePsi\cdots\statePsi\}_{\infty'}\cup
\{\statePhi\cdots\statePhi\}_{\infty’’}\quad\text{and}\quad
\{\statePsi\cdots\statePsi\}_{\tilde\infty'}\cup
\{\statePhi\cdots\statePhi\}_{\tilde\infty’’}\;,
\end{equation*}
one can always establish their formal identity. However, physics
requires distinguishing the rows, which is what the numerical part
of pt.~\hyperlink{R}{\red\textsf{R}} and comparison of cardinals
$(\infty'$, $\infty’’)$ do `serve'.

\subsection{Statistics \tplus\ phases}\label{StatPh}

Thus the closest reconciliation of scheme \eqref{split} with the
\hyperlink{R}{\red\textsf{R}}$^\bullet$"=postulate is an ensemble brace of the
form
\begin{equation}\label{split'}
\obj{\state\Xi}=
\mbig[13]\lgroup
\begin{array}{@{}c@{}}
\big\{\{\statePsi\}_{\infty_1'}\{\statePhi\}_{\infty’’_1}\big\}\\
\DOWN\quad\DOWN\quad\DOWN\quad\DOWN\\
\{\state\alpha_1\cdots\cdots\state\alpha_1\}
_{\hbox to0ex{$\scs\infty_1^{}$}}
\end{array}
\quad
\begin{array}{@{}c@{}}
\big\{\{\statePsi\}_{\infty_2'}\{\statePhi\}_{\infty’’_2}\big\}\\
\DOWN\quad\DOWN\quad\DOWN\quad\DOWN\\
\{\state\alpha_2\cdots\cdots\state\alpha_2\}
_{\hbox to0ex{$\scs\infty_2^{}$}}
\end{array}\quad
\begin{array}{c}\cdots\cdots\\\\\cdots\cdots\end{array}
\mbig[13]\rgroup
\end{equation}
followed by the (upper) continual numeration through $\bbR$"=numbers
\begin{equation}\label{phase}
\varkappa_s\DEF\frac{\infty_s'}{\infty_s}\qquad
(\infty_s\DEF\infty'_s+\infty’’_s)\;.
\end{equation}
In other words, quantitative description in the theory is created on
the basis of the minimal building bricks
\begin{equation}\label{min}
\mbig[13]\lgroup
\begin{array}{@{}c@{}}
\big\{\{\statePsi\}_{\infty'}\{\statePhi\}_{\infty’’}\big\}\\
\DOWN\;\;\;\DOWN\;\;\;\DOWN\;\;\;\DOWN\\
\{\state\alpha\cdots\cdots\state\alpha\}_{\hbox to0ex{$\scs\infty$}}
\end{array}
\mbig[13]\rgroup\qquad(\emph{unitary brace\/})
\end{equation}
with \emph{two} abstract ingoing primitives.

Now, we've had cardinals connected by relation \eqref{infty} and the
structure \eqref{split'}--\eqref{phase}. In the above"=described
context, parentheses $\{\;\}$ and symbols
$\statePsi,\,\statePhi,\,\goto$ no longer carry meaning at this point.
Therefore, we may omit them as `extraneous' and write \eqref{split'}
as
\begin{equation*}
\obj{\state\Xi}\quad\hhence\quad\bigg\lgroup
\array{@{}c}\varkappa_1\\\infty_1\endarray\bigg|
\array{c}\varkappa_2\\\infty_2\endarray\bigg|
\array{c@{}}\cdots\\\cdots\endarray\bigg\rgroup=\cdots,
\end{equation*}
where $\state\alpha_s$ are well represented by a subscripted
numerals; observation $\scr A$ has been fixed so far. Let us now
introduce a statistics from the `embracing infinity'~\eqref{infty}:
\begin{equation*}
\cdots=\bigg\lgroup
\array{@{}c}\varkappa_1\\\fr_1\cdot\infty\endarray\bigg|
\array{c}\varkappa_2\\\fr_2\cdot\infty\endarray\bigg|
\array{c@{}}\cdots\\\cdots\endarray\bigg\rgroup=
\bigg\lgroup
\array{@{}c}\varkappa_1\\\fr_1\endarray\bigg|
\array{c}\varkappa_2\\\fr_2\endarray\bigg|
\array{c@{}}\cdots\\\cdots\endarray\bigg\rgroup
\array{@{}c@{}}\\\!\!\!{}\cdot\infty\endarray\;,\qquad
\fr_s\DEF\frac{\infty_s}{\infty}\;.
\end{equation*}
Then, by $\Sigma$"=postulate, one arrives at a continually numeral
labeling\eLab{labeling} of objects~\eqref{split'}:
\begin{equation*}
\obj{\state\Xi}\quad\hhence\quad\mbig[5]\{\mbig(
\smallmatrix \ds\varkappa_1\\\ds\fr_1\endsmallmatrix
\mbig),\;\mbig(
\smallmatrix \ds\varkappa_2\\\ds\fr_2\endsmallmatrix
\mbig),\;\ldots\mbig[5]\}\;.
\end{equation*}

Recall that the arithmetical operations on the emergent pairs
$(\fr,\varkappa)$ are still out of the question, and $\Sigma$"=limit
does not care the `innards' of $\obj{\state\Xi}$. Only one of all
the potentially infinite quantities tends to the
$\infty$"=infinity""---the total cardinality \eqref{infty} of brace
\eqref{split'}. What remains `non"=extraneous' in \eqref{split'} is
$\state\alpha$'s, and we return them to their place. Hence, from the
viewpoint of observation $\scr A$, the aggregate of all the possible
brace \eqref{obj} is indistinguishable from an order"=indifferent
2"=parametric family of data
\begin{equation}\label{set}
\obj\Xi=\mbig[5]\{\mbig(
\smallmatrix \ds\varkappa_1\\\ds\fr_1\endsmallmatrix
\mbig)\state\alpha_1,\;\mbig(
\smallmatrix \ds\varkappa_2\\\ds\fr_2\endsmallmatrix
\mbig)\state\alpha_2,\;\ldots\mbig[5]\}\;.
\end{equation}
We drop a lower bar in the symbolic designation $\obj{\state\Xi}$,
highlighting the fact that the meaning of the $\obj\Xi$"=object
becomes increasingly divorced from primitives in
pt.~\hyperlink{S}{\red\textsf{S}} and gets into the number domain to
match the programme~\hyperlink{R}{\red\textsf{R}}.

As an outcome, despite the freedom of ingoing collection in
\eqref{split} and quantum micro"=arbitrariness, the
distinguishability
$\obj{\state\Xi}\not\approx\obj{\state{\tilde\Xi}}$ is \emph{indeed
determinable}, it is determinable not only by statistics, and is the
$(\bbR\Times\bbR)$"=numerical:
\begin{equation}\label{not}
\obj{\Xi}\not\approx\obj{\tilde\Xi}\,,\quad\text{if}\quad
(\fr_s,\,\varkappa_s)\ne
(\skew1\tilde\fr_s,\,\skew2\tilde\varkappa_s)\;.
\end{equation}
What is more, the preliminary (classical) $\not\approx$"=criterium
\eqref{stat} fits in \eqref{set}--\eqref{not} as a particular case
by omitting the $\varkappa$"=numbers and middle link from
\eqref{quant}. That is to say, the ignoring of quantum
`$\varkappa$"=effects' is only possible via the
$(3\mapsto2)$"=reduction \eqref{quant} into \eqref{class}, with an
automatic imposition of the ClassPhys description. A simplified and
hypothetical version of \qm\ over $\bbR^1$ is also ruled out. It
would mean a reduction of the two numbers $(\fr,\varkappa)$ to a
single one. But they have fundamentally different origins. The
construct and reasoning in sect.~\ref{S} also tell us that the
attempt at a greater `quantum specification' to \eqref{box} and
\eqref{quant} is impossible by virtue of the 2-row
structure""---""ingoing/"!outgoing""---of the object
$\obj{\state\Xi}$, and distinguishability by numerical pairs
\eqref{not} is the highest possible.

The $\obj\Xi$"=objects \eqref{set} remain and they, as a family,
exhaustively inherit the problem's physics. The quantities $\fr_s$
are the really observable (unitless) numbers""---the percentage
quantity of events---which are declared by instrument/"!observer to be
the distinguishable $\state\alpha$"=objects. The quantities
$\varkappa_s$ are the internal and unremovable degrees of freedom.
Figuratively speaking, the $\varkappa$'s may be speculatively
referred to as phases, but they may not be associated with an actual
quantity of something. Not only is any material or the classical
treatment of these `amounts' impossible, but it is fundamentally
\emph{prohibited} since the converse would have meant endowing the
nonexistent boxes \eqref{box}--\eqref{Box} with a notional content
or asserting the nature of their origin. Justification is only
allowed here for the fact of their existence, which is mirrored by
the presence of the left hand side in the concept of the transition
$\statePsi\GOTO{\sss\scr A}\state\alpha$ (Remark~\ref{two}).

In view of numerous and ongoing discussions of the
meaning\eLab{smysl} to the quantum state \cite{nielsen}, note that,
for the same reason, any its (even merely similar)
classical/"!ontological and causal `visualization mechanisms'
\cite[p.~137]{silverman} as the wave function of a certain real
matter, of a hypothetical observable, of an `objective knowledge',
or of the classical data (whatever this all means) are---and this we
stress with emphasis""---""pointless. This is why, strictly
speaking, without further theoretical conventions,
\begin{itemize}
\item it\eLab{provo} is impossible \cite[p.~13]{englert} to
 make/"!prepare, observe/"!read-off, transmit or measure/"!approximate a
 state, or to endow it with the property of being known/"!unknown, or
 physically recognize/"!compare/"!distinguish it from the other.
\end{itemize}

We will be repeatedly turning back to this matter in
sects.~\ref{cat}, \ref{phys}, \ref{2slit}, and \ref{SM}. The present
thesis has no undergone a change even with regard to the word
`statistics' in the Born rule \cite{br2}, if only because the rule
is a substantial""---two-to-one---""reduction of the
$(\fr,\varkappa)$"=data. The state will itself, when created as a
mathematical object, determine the meaning\eLab{povtor} of all of
these words (see sect.~\ref{izmer}) with an appropriate concept of
the physical distinguishability (sect.~\ref{MM}). Cf.~the works
\cite{lundeen, pusey} and the ``methods to directly measure general
quantum states \ldots\ by weak measurements'' in \cite{lundeen2} and, on
the other hand, the statements in sect.~15.5 of the book
\cite{auletta}.

All the $\varkappa_s$ and $\fr_s$ are independent of each other,
except for relation $\fr_1+\fr_2+\cdots=1$. Taking into account the
admissible renormalization of both $\bbR$"=numbers, the pair
$(\fr,\varkappa)$ can be topologically identified with a point on
the complex plane:
\begin{equation*}
\mbig(\smallmatrix \ds\varkappa_{}\\[0.3ex]
\ds\fr\endsmallmatrix\mbig)\;\rightleftarrows\;
(\lambda,\mu)\in\bbR^2\FED\bbC\;.
\end{equation*}
That is, the domain $\bbC$ is at the moment just a 2"=dimensional
numeric continuum without algebra of complex numbers. Notice that
the pairs of $\bbR$"=numbers is a starting point---""different from
ours---to the \qt\ in ref.~\cite{goyal}. More than that,
impossibility of the real"=number \qm\ became a subject of the direct
experimental test to distinguish between the complex"=number and
real"=number representations of \qt: on photonic systems \cite{zheng}
and the superconducting qubits \cite{ming}.

The issue of the numerical domain over which the quantum description
is being conducted""---the real $\bbR$, the complex $\bbC$, the
quaternions $\msf Q$, or whatever""---is non"=trivial and continues
to be the subject of study \cite{hardy, aaronson, goyal, baez,
ludwig3}. The complexity $\bbC$ is often motivated by quantum
dynamics (Schr\"odinger's equation) \cite[p.~132;
Stueckelberg]{jauch1}, \cite{stueck}, however, such a motivation is
inconsistent, and as we have seen, there is no need for it. The
rigidity of the $\bbC$"=domain points to the fact that, in
particular, the quaternion \qm\ also has no place to originate from
\cite[sect.~10.1]{auletta}, although it was the object of
theoretical constructs in the 1960--70's \cite{jauch2}. Note that
even the most comprehensive works \cite[p.~217]{ludwig3},
\cite[p.~131]{jauch1}, \cite[p.~234]{beltrametti}, \cite{benioff,
goyal}, and \cite{baez}\,(!\@) observe a difficulty in the full
substantiation of the $\bbC$"=domain in \qt. Within the last decade,
this theme had also attracted the particular attention of the
information"=theoretic approach to \qt\ \cite{muller, goyal,
dAriano+}.

The above"=outlined emergence of the numerical quantities in theory
is as yet draft at the moment, and will be refined further below in
sect.~\ref{Numbers}.

\vbox{
\section{Empiricism and mathematics}\label{EmpMath}

\flushright\tiny
\textsl{Set theory does not seem today to have\\
\ldots\ organic interrelationship with physics}\\
\textsc{--- P.~Cohen \& R.~Hersh} \cite[p.~116]{cohen}%
\medskip

\textsl{\ldots\ physics has \ldots\ to say about the foundations\\
of mathematics \ldots ``if we believe in ZF there\\
is nothing for physics to say'' is not right}\\
\textsc{--- P.~Benioff} \cite[p.~31]{fuchs4}}
\smallskip\nopagebreak

Up to this point we have dealt, roughly speaking, with a single
abstract aggregate $\obj{\state\Xi}$ isolated from the others.
However, the constructional nature of the ensemble brace
\eqref{split'} entails the following closedness relation between
them. Every brace $\obj{\state\Xi}$ is composed of some others in
infinitely many ways (for remote analogies, see
\cite[sect.~11.2]{allah}), \ie, it is a union
\begin{equation}\label{union}
\obj{\state\Xi}=\obj{\state\Xi'}\cup\obj{\state\Xi’’}\;,
\end{equation}
and, to put it in reverse, any union of two brace is a third
object"=brace. In assemblages \eqref{union}, the operation $\cup$,
which generates them, is commutative and associative:
\begin{equation}\label{ComAss}
A\cup B = B\cup A\,,\qquad (A\cup B)\cup C=A\cup(B\cup C)\;,
\end{equation}
and these 2- and 3-term relations not only are not a formal
supplement, but should be read as the \emph{structural} properties
in general. Let us address the matter more closely.

\subsection{Union of ensembles}\label{inv}

 Consider the lower $\state\alpha$"=rows of brace
\eqref{split} and experimental forming the new real
$\state\alpha$"=ensembles from them. Let the procedure of that
forming be denoted by $\msf U(A,B,\ldots)$, where $(A, B,\ldots)$ are the
ensembles per~se. Its essence is such that it is comprehensively
determined by the following minimum. A rule that involves the
\emph{fewest} (\ie, two) number of arguments $\msf U(A,
B)=\bfCourier{?}$ and a rule of the \emph{repeated} applying
$\msf U$ to itself: $\msf U(\msf U(\ldots),\ldots)=\bfCourier{?}$.
Obviously, we should write
\begin{equation}\label{U}
\msf U(A, B)=\msf U(B, A)\,,\qquad
\msf U\big(\msf U(A,B),C\big)=\msf U\big(A,\msf U(B,C)\big)\;,
\end{equation}
which is of course merely the empirical rephrasing the standard
properties \eqref{ComAss} of operation $\cup$. However, the converse
is logically preferable: empiricism \eqref{U} is formalized into the
abstract properties \eqref{ComAss}. If we now attach the upper
`quantum' primitives to the low $\state\alpha$"=rows---a requirement
of sect.~\ref{S}---then the operationality of actions with the
resulting $\obj{\state\Xi}$"=braces would be just like that of
$\msf U$, \ie, \eqref{U}. In other words, we carry over (and had
already used everywhere) properties \eqref{U} to the general
operation on $\obj{\state\Xi}$"=brace, without distinguishing between
the essences of symbols $\cup$ and $\msf U$. `Micro"=operationality'
of empiricism and its formalization are confined, at most, by the
rules \eqref{ComAss}--\eqref{U}.

Let us temporarily discontinue using the numerical terminology as
applied to $\obj\Xi$"=objects. They differ from each other due to
relationships between their `innards', rather than because of our
assignment of differed symbols $(\lambda,\mu)$ to them. The brace
are comprised of elements that are combined into sets and are added
to them. In the language of the abstract logic, we are dealing with
the fact that transitions $x$ form the brace $A,\,B$, \ldots, \ie, they
are in the membership relationships $x\in A$, $x\in B$, \ldots\ or,
when accumulated as micro-acts, `get belonged to them'. That is to
say, the brace themselves and their formation (accumulation of
statistics for the $\Sigma$"=limit) are equivalent to a huge number
of propositional `micro"=sentences $x\in A$ \Courier{or} $x\in B$
\Courier{or} \ldots'. However, again, this is nothing but a logically
formal equivalent of the union operation~$\cup$:
\begin{equation}\label{cup}
A\cup B=\big\{x\,\big|\:(x\in A)\Vee(x\in B)\big\}\;,
\end{equation}
which is already being constantly exploited above.

\begin{comment}
As is well known \cite{stoll, kurat}, due to properties of logical
atoms $\in$ (\Courier{member\-ship}) and $\Vee$ (\Courier{or}), the
properties of sentences such as \eqref{cup} are determined precisely
by rules \eqref{ComAss} for $\cup$. Technically, we should also take
an idempotence $A\cup A=A$ into account, however. At the same time,
the need to have a number requires that the duplicates in ensembles
have to be taken into consideration. Nevertheless, this situation is
easily simulated by the set theory itself. Indeed, look first the
lower row in \eqref{obj} as a strictly abstract set
$\{\state\alpha',\state\alpha’’,\ldots\}\subset\boT$. Then, instrument
$\scr A$ `asserts' the distinguishable elements $\{\state\alpha_1$,
$\state\alpha_2,\ldots\}$ and those that should be thought of as their
equivalents:
\begin{equation*}
\state\alpha_1'\approx\state\alpha’’_1\approx\cdots
\FED\state\alpha_1\,, \qquad
\state\alpha_2'\approx\state\alpha’’_2\approx\cdots
\FED\state\alpha_2\,, \qquad\ldots\;.
\end{equation*}
This equivalence can be characterized, say, by words `a detector
click at one and the same place $\state\alpha_1$'. Upon such a
formalization, one obtains the formation
$\{\state\alpha_1'\,\state\alpha’’_1\cdots\}\,
\{\state\alpha_2'\,\state\alpha’’_2\cdots\}$ $\cdots$ $\approx$
$\{\state\alpha_1\cdots\}\, \{\state\alpha_2\cdots\}$ $\cdots$, \ie,
the very lower row in \eqref{split}. It is within this context that
we think of the union operation without running into
inconsistencies. Accordingly, $\obj{\state\Xi}\cup
\obj{\state\Xi}\ne\obj{\state\Xi}$ but the standard symbol $\cup$
continues to be used for simplicity.
\end{comment}

Therefore if we get back to the numeral labels \eqref{set}, but
ignore the `inner composition` of $\obj{\state\Xi}$, \ie, the
\hyperlink{M}{\red\textsf{M}}"=paradigm, thus excluding $\cup$ and
\eqref{ComAss} from the reasoning, then all possible
$\obj{\Xi}$"=objects would turn into the semantically `segregated
ideograms'. Micro"=transitions, their mass nature, arbitrariness,
$\not\approx$"=distinguishability, and the `quantumness' of the task
simply disappear. To take an illustration, the obvious statement
\begin{gather*}
\text{the brace $\{\statePsi\GOTO{\scr A}\state\alpha\}\FED
\obj{\state\Xi}$ has an empirical `kindred'}\\[-1ex]
\text{with its duplication
$\{\statePsi\GOTO{\scr A}\state\alpha,\,\statePsi
\GOTO{\scr A}\state\alpha\}\FED \obj{\state\Xi'}$}
\end{gather*}
becomes pointless, because the property $\obj{\state\Xi}\cup
\obj{\state\Xi}=\obj{\state\Xi'}$ is missing. And this is despite
the fact that the creating the transition copies in
$\obj{\state\Xi'}$ is a primary operation for generating the objects
and reasoning at all. Construction of the theory would then become
possible only with the interpretative introduction of the vanished
concepts anew. Therefore macro"=empiricism necessitates that the
relationships \eqref{ComAss} be operative rules, and with that the
quantumness or classicality of consideration is of no significance.

\begin{comment}\label{seven}
Let us take a closer look at the situation on the
opposite""---""mathematical""---side. The union of sets $\cup$ is a
fundamental operation already at the level of the set"=theoretic
formalization; \eg, the Zermelo--Fraenkel (\zf) axioms \cite{kurat}.
This is one of the first ways on how to create sets---the axiom of
union. So if we believe in the set"=theoretic mode of
explaining/"!creating the quantum rudiments, the quantitative
description will inevitably invoke operationality of the
mathematical primitive $\cup$ through rules \eqref{ComAss}. This
would be suffice to declare,
\par\larger 
\begin{itemize}
\item \smaller inasmuch as \emph{we have nothing but $\cup$ and
 $\obj{\state\Xi}$} \textup{(taboo~\hyperlink{T}{\red\textsf{T}})},
 commutativity/"!associativity of theory is then postulated from the
 outset by \eqref{ComAss}, with the subsequent carrying these
 structures over to numerical representations, \ie, to $\bbR$ or
 $\bbC$.
\end{itemize}
\smaller 
It is preferable, however, to adhere to the sequence order in
ideology more stringently""---\lrceil{observation} $\rightarrowtail$
\lrceil{mathematics}, \lrceil{empiricism} $\rightarrowtail$
\lrceil{numerical representation}---without substitution it for the
opposite. At least, if we rely upon the comprehension of the
empiricism as a formalization of the zero"=principium of \qt\
(Remark~\ref{two}):
\par\larger 
\begin{itemize}
\item \smaller our primordial perceptions are formalized only into sets
 and set"=theoretic $\cup$"=abstraction \eqref{cup}.
\end{itemize}
\smaller 
See also \cite[Ch.~3]{ludwig5}, \cite[p.~178]{fuchs4},
\cite[p.~323]{jammer}, \cite{bueno}, \cite[pp.~12, 86,
Ch.~4]{chomsky}, \cite[sect.~V.9]{frenkel}, sect.~\ref{phil}, and
\cite{benioff}.
\end{comment}

Summing up, we detect a kind of a junction point: the physical and
mathematical fundamentality of operation $\cup$ for describing the
elementary acts. That is to say, the mathematics of
$\obj{\state\Xi}$"=brace \eqref{split'} and of objects \eqref{set}
may not inherently be exhausted by them as `bare' sets without
structures.

Recalling now pr.~\hyperlink{II}{\red\textbf{\textsf{II}}}, we draw a
conclusion regarding the very construction of the theory.
\begin{itemize}
\item Reconciliation of the \hyperlink{R}{\red\textsf{R}}"=paradigm with
 empiricism must transform itself into \emph{rewriting} the primary
 ensemble $\cup$"=constructions \eqref{split}, \eqref{split'},
 \eqref{min}, and relationships between them into the language of
 numerical symbols.
\end{itemize}
More formally, we have the following continuation of
pt.~\hyperlink{R}{\red\textsf{R}}$^\bullet$.
\begin{itemize}
\item[\hypertarget{R+}{{\red$\msf R^\textsf{+}$}}]
 \emph{Homomorphism\eLab{R+} of the ensemble-brace properties `onto
 numbers'}: mutual $\cup$"=relationships \eqref{ComAss} between the
 $\obj{\state\Xi}$"=brace should be carried over to relations between
 their numerical $\obj\Xi$"=representations \eqref{set}.
\end{itemize}
Thereby we once again fix the maximum that is available for the
building up of quantum mathematics. One may handle only the
$\cup$"=aggregates of transitions""---""constructions
$\eqref{split'}$, \eqref{set}---and the minimal modules~\eqref{min}.

\subsection{Semigroup}\label{semi}

In line with \eqref{union}, let us split the unitary brace
\eqref{min} into two or combine two brace into one. Delete also the
symbols of primitives $\statePsi$ and $\statePhi$ from there. As was
pointed out above, they are not necessary at this stage. Replacing
the notation of upper cardinals \eqref{min} with pairs
$(\infty'_{\sss1},\infty'_{\sss2})$ and
$(\infty’’_{\sss1},\infty’’_{\sss2})$, upon the union one obtains
\begin{equation}\label{comp}
(\infty'_{\sss1},\infty'_{\sss2}) \cup
(\infty’’_{\sss1},\infty’’_{\sss2})=
(\infty'_{\sss1}+ \infty’’_{\sss1},
\infty'_{\sss2}+\infty’’_{\sss2})\;.
\end{equation}
Here, addition~$+$ obviously satisfies the properties
\eqref{ComAss}. If the cardinal `$\infty$"=coordinates' are replaced
with the `finite percentages' $(\varkappa,\frak S)$ introduced
above, \ie, if one puts
\begin{equation}\label{birat}
\Big\{\varkappa=\frac{\infty_{\sss1}}
{\infty_{\sss1}+\infty_{\sss2}}\,,\quad
\frak S=\infty_{\sss1}+\infty_{\sss2}\Big\}\,,\qquad
\Big\{\infty_{\sss1}=\varkappa\,\frak S\,,\quad
\infty_{\sss2}=(1-\varkappa)\,\frak S \Big\}
\end{equation}
as in \eqref{phase}, then rule \eqref{comp} acquires the form of a
number composition:
\begin{equation}\label{s}
(\varkappa',\frak S')\bo\circ (\varkappa’’,\frak S’’)=
\bigg(\frac{\ds\varkappa'\frak S'+\varkappa’’\frak S’’}{\ds\frak S'+
\frak S’’}\,,\;\frak S'+\frak S’’\bigg)\;.
\end{equation}
The commutativity/"!associativity properties of operation $\bo\circ$
hold here due to birationality of \eqref{birat}. Then, the formal
application of $\Sigma$"=postulate $\frak S'+\frak S’’\to\infty$
breaks, however, the symmetry $(')\leftrightarrow('')$ and
associativity of $\bo\circ$, since
\begin{equation}\label{s'}
(\varkappa',\frak S')\bo\circ (\varkappa’’,\frak S’’)\quad
\rightarrowtail\quad \varkappa'\bo\circ\varkappa’’=
s\cdot\varkappa'+(1-s)\cdot\varkappa’’\,, \qquad
s\DEF\frac{\ds\frak S'}{\ds\frak S'+\frak S’’}
\end{equation}
and $s$ is an undefined parameter. Consequence of the same kind
holds also true for the $\fr$"=components of pairs \eqref{set}, for
which a convex $w$"=combination of the statistical weights does
arise:
\begin{equation}\label{NU}
(\fr'_1,\fr'_2,\ldots)\bo\circ(\fr’’_1,\fr’’_2,\ldots)\FED
(\overline\fr\,'\bo\circ\overline\fr\,’’)=
w\cdot\overline\fr\,'+(1-w)\cdot\overline\fr\,’’\,,\qquad
w\DEF\frac{\ds\Sigma'}{\ds\Sigma'+\Sigma’’}\;.
\end{equation}

At the same time, the splitting \eqref{comp} is no more than an
`intrinsic reshuffle' of one and the same $\obj{\state\Xi}$"=brace,
which `knows nothing' about concept of a number (numbers $s$, $w$),
much less about the concept of observation or its numerical form.
Therefore mathematics of the ensemble structures should be
independent of any representation for \eqref{union} by such
operations as \eqref{s}. Composition
$\obj{\Xi'}\bo\Circ\obj{\Xi’’}=\obj\Xi$ should be determined solely
by its constituents $(\fr',\varkappa')$ and $(\fr’’,\varkappa’’)$,
\ie, such numbers as $(s,w)$ must not appear here.

\begin{comment}
In classical statistics, the foregoing has an analog as indifference
of data on events to the way of gathering and layout thereof. For
example, $(2,3)+(1,4)\equiv(0,6)+(3,1)\equiv\cdots\FED
\Courier{data}$. Then, the observation proper is being created by
the scheme $\Courier{data}\rightarrowtail(3,7)\mapsto
\big(\frac{3}{3+7},\frac{7}{3+7}\big)=(0.3,0.7)=(\fr_1,\fr_2)\FED
\Courier{observ}$. Parameters like $w$ can appear in $\obj\Xi$ only
if, prior to any of the $\cup$"=unions \eqref{union}, a construction
similar to \eqref{m2} has been fixed. That is, the invariantly
number-free brace \eqref{union} \emph{has been supplemented by an
external number} $w$ and ratio $w:(1-w)$. The correction
$\obj\Xi\rightarrowtail\big\{\obj{\Xi'}^{{\sss(}w\sss)},
\obj{\Xi’’}^{{\sss(}1-w\sss)}\big\}$ of the theory, related to this
number and to arrays \eqref{m2}, is very well known. This is a
$w$"=statistical mixture $\{(w;\psi'), (1-w;\psi’’)\}$ of wave
functions, accompanied by a formalization in terms of the
statistical operator $w\cdot\ket{{}\psi'}\bra{{}\psi'}+(1-w)\cdot
\ket{{}\psi’’}\bra{{}\psi’’}$.
\end{comment}

Now, to ensure that numerical $(\fr,\varkappa)$"=realization
\eqref{set} of ensemble brace \eqref{split'} inherits quantum
empiricism (\hyperlink{O}{\red\textsf{O}}, \hyperlink{M}{\red\textsf{M}})
and structural properties \eqref{union}--\eqref{ComAss} properly, we
reassign the quantities $(\fr,\varkappa)$ with a `percentage
meaning' and replace them by different numbers~$[\lambda,\mu]$:
\begin{equation}\label{set'}
\obj\Xi=\mbig\{\big[\smallmatrix \mu_1\\
\lambda_1\endsmallmatrix\big]\,\state\alpha_1\,,\;
\big[\smallmatrix\mu_2\\
\lambda_2\endsmallmatrix\big]\,\state\alpha_2,\;\ldots\mbig\}
\end{equation}
(this important move will be touched upon once again in
sect.~\ref{Dupl}). In so doing, each pair
$[\smallmatrix\mu{\sss'}\\[-0.25ex]\lambda{\sss'}\endsmallmatrix]$,
$[\smallmatrix\mu{\sss’’}\\[-0.25ex]\lambda{\sss’’}\endsmallmatrix]$
behaves as a whole, and, under coinciding $\state\alpha_s$, the
pairs are endowed with a composition
$[\smallmatrix\mu{\sss'}\\[-0.25ex]\lambda{\sss'}\endsmallmatrix]
\oplus
[\smallmatrix\mu{\sss’’}\\[-0.25ex]\lambda{\sss’’}\endsmallmatrix]$
that is to be commutative. Along with this, if symbol $\uplus$
denotes a composition of objects \eqref{set'} then it should
obviously copy properties~\eqref{ComAss}:
\begin{equation*}
\obj\Xi\uplus\obj{\Psi}=\obj{\Psi}\uplus\obj\Xi\,,\qquad
\big(\obj\Xi\uplus\obj{\Psi}\big)\uplus\obj{\Phi}=
\obj\Xi\uplus\big(\obj{\Psi}\uplus\obj{\Phi}\big)\;.
\end{equation*}

The finite ensembles are vanishingly small in their contribution
into infinite ones ($\Sigma$"=postulate), \ie, elements of the
$\obj{\state\Xi}$"=family, as infinite sets, are considered modulo
finite ensembles. Once again, the `finitely many' is forbidden in
theory. As soon as we put the numbers of $\state\alpha_1$, of
$\state\alpha_2$, \ldots\ to be finite, we get immediately the
numerical distinguishability $n_1\ne n_2$, \ldots, \ie, the act of
macro"=observation.

Let us designate the image of finite ensembles as $\obj{0}$, and,
due to property $\obj\Xi\uplus\obj0=\obj\Xi$, it is naturally
referred to as zero. The collection \eqref{set'} itself has also
been formed by the $\cup$"=combining the ingredients
\begin{equation*}
\mbig\{\big[\smallmatrix \mu_1\\
\lambda_1\endsmallmatrix\big]\,\state\alpha_1\,,\;
\big[\smallmatrix\mu_2\\
\lambda_2\endsmallmatrix\big]\,\state\alpha_2\,,\;\ldots \mbig\}\equiv
\mbig\{\big[\smallmatrix \mu_1\\
\lambda_1\endsmallmatrix\big]\,\state\alpha_1\mbig\}\cup
\mbig\{\big[\smallmatrix\mu_2\\\lambda_2\endsmallmatrix\big]\,
\state\alpha_2\mbig\}\cup\cdots=\cdots\;,
\end{equation*}
and which is why the same symbol $\uplus$ may be freely used between
objects with different~$\state\alpha_s$:
\begin{equation*}
\cdots=\mbig\{\big[\smallmatrix \mu_1\\
\lambda_1\endsmallmatrix\big]\,\state\alpha_1\mbig\}\uplus
\mbig\{\big[\smallmatrix\mu_2\\\lambda_2\endsmallmatrix\big]\,
\state\alpha_2\mbig\}\uplus\cdots\;.
\end{equation*}
For the sake of brevity, we omit the redundant curly brackets
further, redefining
\begin{equation}\label{set''}
\obj\Xi_{\!\sss\scr A}\DEF\big[\smallmatrix \mu_1\\
\lambda_1\endsmallmatrix\big]\,\state\alpha_1\uplus
\big[\smallmatrix\mu_2\\
\lambda_2\endsmallmatrix\big]\,\state\alpha_2\uplus\cdots\;.
\end{equation}

As a result, we have had that the set"=theoretic prototypes
\eqref{split}--\eqref{tmp0}, \eqref{split'} of states \eqref{Repr}
do invariantly exist in form of every possible
$\cup$"=decompositions. Thus, in dealing with the only instrument
$\scr A$, one reveals the following property.
\begin{itemize}
\item For each observation $\scr A$, the set of
 $\obj\Xi_{\!\sss\scr A}$"=objects forms an infinite commutative
 semigroup $\frak G$ with respect to operation~$\uplus$.

 An internal (beyond the observation) nature of
 $\obj\Xi_{\!\sss\scr A}$"=objects \eqref{set''} is characterized by
 commutative superpositions
 $\obj{\Xi'}_{\!\sss\scr A}\uplus\obj{\Xi’’}_{\!\sss\scr A}$ thereof,
 which are independent of the classical composition of observational
 $\fr$"=statistics.
\end{itemize}

\subsection{Measurement}\label{izmer}

The described above numerical $\obj\Xi$"=version of the
$\obj{\state\Xi}$"=brace `$\cup$"=phenomenology' makes it possible now
to preliminarily formalize a concept, the absence of which deprives
the theory of its basis. Namely, \emph{measuring statistics by
observation $\scr A$ over~$\cal S$}:
\begin{equation}\label{red}
\text{\qm-measurement:}\qquad\big([\lambda_1,\mu_1],
[\lambda_2,\mu_2],\ldots\big)\;\longmapsto\;(\fr_1,\fr_2,\ldots)\;.
\end{equation}
That is, the $[\lambda,\mu]$"=collection gets mathematically mapped
into the $\fr$"=statistics. This is a maximum of information provided
by observation $\scr A$. The mapping \eqref{red} annihilates the
pairs $[\lambda,\mu]$. Therefore the inheritance/"!homomorphism of
operations $\cup$ and $\uplus$ onto anything at all is eliminated.
Upon operation \eqref{red}, both the $(\fr,\varkappa)$"=sets and
$\cup$"=unions thereof, $\uplus$"=operations, and the semigroup
$\frak G$ per~se disappear. As a well-known result, the distinctive
feature subsequently referred to as superposition will also
disappear after measurement. The new numbers $\{\fr_s\}$ may be
`added up' only as required by the different, \ie, the classical
rule: forming the convex combinations \eqref{NU}. We note that the
formalization of measurement does not now depend on how the
mathematical map $[\lambda,\mu]\rightarrowtail(\fr\,)$ would be
further implemented""---it is a separate job \cite{br2}---or how the
$t$"=dynamics would be introduced.

\begin{comment}
Incorporation of $t$"=dynamics into the theory is still impossible
due to the absence of mathematics to be applied to instants $t_1$,
$t_2$. Accordingly, no physical $t$"=process, a temporal imitation of
the measuring, or its dynamical description may correspond to the
mathematical mapping \eqref{red}. The known `conceptual' problems
with the collapse' dynamics \cite{gottfried, weinberg, Hooft,
london, mittel} are actually non"=existent \cite{lipkin, ludwig1,
ludwig3}. More precisely, they stem from the blurring of meaning
that we typically give to the words `states' (what is that?),
`ensembles' (what are they comprised of?), and `dynamics/"!collapse'
(of what?). In regard to the latter, the authors of the book
\cite{ludwig5} speak out in the most definitive manner---the ``fairy
tales''. See also sect.~\ref{SM+} further below.
\end{comment}

In sect.~\ref{S}, the fundamental premise of the
$\state\alpha$"=symbol"=based distinguishability~$\not\approx$ was the
foundation of the entire subsequent language; ``two clicks are never
identical'' \cite[p.~761]{ulfbeck}. One then observes that the
measurement or its outcome will essentially remain a vacuous
term---``For microsystems nothing can be directly measured''
\cite[p.~304]{ludwig2}---up until it invokes the concept of a
\qm"=state, \ie, the $\obj\Xi$- and $\state\alpha$"=objects. In the
following, we shall see that, as a rough guide, everything that is
observable whatsoever is a function of the state and of the state
space.

Once again it is stressed""---the concept of the state must precede
the notion of measurement, rather than the reverse. ``[J.]~Bell
fulminated against the use of the word `measurement' as a primary
term when discussing quantum foundations'' \cite[p.~262]{home}. See
also the entire chapter~23---``Against `measurement'\,''---in
\cite[pp.~213--231]{bell}.

\subsection{Covariance with respect to observations
(`the same')}\label{SS}

Up to this point we had had no need for the matching of observation
$\scr A$ with observation $\scr B$, although it is clear that a
description based on a certain specified $\scr A$ will inevitably be
non"=invariant with respect to the tool $\{\scr A,
\scr B,\ldots\}$---`observation space'---and unacceptable
(pt.~\hyperlink{R}{\red\textsf{R}}) due to the impermissible
exclusivity of the set $\{\state\alpha_1,\ldots\}$. At the same time,
we do not have anything but $\{\scr A, \scr B,\ldots\}$ and micro-acts
\eqref{macro} (pts.~\hyperlink{T}{\red\textsf{T}} and
\hyperlink{M}{\red\textsf{M}}). In the brace, this fact has already
been present---""transitions $\GOTO{\sss\scr A}$ are combined into
integrities \eqref{obj}. Logically, however, the
$\obj\Xi_{\!\sss\scr A}$-, $\obj\Xi_{\!\sss\scr B}$"=objects are
incomparable and isolated from each other as carriers of statistics
of different origin.

On the other hand, `\emph{the same} is observed by instrument
$\scr B$, as by instrument $\scr A$'. Although this context has not
yet been invoked, without it the application of set"=theoretic
constructs to physics is devoid of meaning, just like the union of
the speeds of an electron and of the Moon into a set
$\{\bo v_\text{e},\bo v_{\sss\text{M}}\}$, with the subsequent
creating a certain physical characteristic of this `two"=body
system'; say, the mean velocity
$\frac12(\bo v_\text{e}+\bo v_{\sss\text{M}})$. Indeed, ``The
statements of quantum mechanics are meaningful and can be logically
combined \emph{only} if one can imagine a \emph{unique experimental
context}'' \cite[p.~115]{allah}.

Thus the global structuredness is required in the set of various
$\obj\Xi$"=data according to the context `the same, identical' or its
negation. Apparently, this addition implies such entities as `the
same particle', `in the same preparation/state', `under the same
temperature and $\mbf M$"=environment \eqref{macro}', `the same
closed system $\cal S$', `in the same external field', `in the same
interferometer' with `the same detectors/"!solenoids', \thelike\
\cite{allah, auletta, greenberg}; the short and generalized notation
\SM. All the notions here, including the state, are physical
conventions, yet their formalization and modeling are called for the
creation of a theory (sect.~\ref{MM}).

The notion `with the same initial data' falls under the same
category, if the intention is to use the term time $t$. Again, the
very creating the $\obj\Xi$"=brace as a set `by the piece' is from
the outset thought of as (sect.~\ref{MM}) a creation on the
assumption of common \SM. For instance, the $\scr A$"=statistics
$\obj\Xi_{\!\sss\scr A}$ is gathered within `the same' \SM\ as the
$\scr B$"=statistics $\obj\Xi_{\!\sss\scr B}$. On its part, any
variation is sufficient to obtain `not the same', even if we
`\emph{envision it as null}' in the spirit of the widely known
``without in any way disturbing a system'' \cite[p.~234]{schilpp}. To
take an illustration, equipment of interferometer
(sect.~\ref{2slit}) with additional `which-slit'\eLab{which}
detectors is already at variance with the notion of `the same
$\cal S$'. In similar cases, we end up in situations of
type~\eqref{m2}, since the detectors cause an
$\state\alpha$"=distinguishability.

Notice that the notions ``the same'' and ``distinguishable''
(Remark~\ref{two}), while antonymous, mutually exclude each other.
Semantically, one without the other makes no sense, which closely
resembles Bohr's conception of complementarity \cite{jammer}.

It follows from the above that in order to match $\scr A$ and
$\scr B$, the \emph{metatheoretical} \cite{frenkel} category \SM\ is
required, however, we are only in possession of the ensemble brace
$\obj\Xi_{\!\sss\scr A}$ and $\obj\Xi_{\!\sss\scr B}$
(pt.~\hyperlink{T}{\red\textsf{T}}). On the other hand, without joint
consideration of \emph{the two instruments}, \ie, without
introducing a mechanism for the mathematical matching
$\obj{\Xi'}_{\!\sss\scr A}\rightleftarrows\obj{\Xi'}_{\!\sss\scr B}$,
$\obj{\Xi’’}_{\!\sss\scr A}\rightleftarrows
\obj{\Xi’’}_{\!\sss\scr B}$, \ldots, the segregation of the
$\obj\Xi$"=objects is absolute. (It is clear that the matching of
single micro"=events $\statePsi\GOTO{\sss\scr A}\state\alpha_s$ and
$\statePsi\GOTO{\sss\scr B}\state\beta_j$ is also futile.) It is
impossible to associate physics with the abstractly segregated
$\obj\Xi_{\!\sss\scr A}$"=brace. Otherwise, the solitary object
$\obj\Xi_{\!\sss\scr A}$, generating nothing more than a statistics
provided by the single instrument $\scr A$, would yield a
description of everything, which is absurd by
pt.~\hyperlink{R}{\red\textsf{R}}$^{\bullet\bullet}$. The physical contents (to
come) arise precisely through the above"=mentioned matching; see
sect.~\ref{phys} below.

As a result, we adopt a kind of the relativity"=principle
analogue""---a tenet on the quantum observational covariance.
\begin{itemize}
\item[\hypertarget{III}{\red\textbf{\textsf{III}}}] Theory\eLab{III} should
 introduce a means of equating the macro"=observations
 (pts.~\hyperlink{O}{\red\textsf{O}} \tplus\
 \hyperlink{M}{\red\textsf{M}}) by differing instruments
 $\{\state\alpha_1,\ldots\}_{\!\sss\scr A}\ne
 \{\state\beta_1,\ldots\}_{\!\sss\scr B}$ under a common (the same)
 experimental environment \SM.\\
 \phantom.\hfill(\hbox{\embf{The $\bo3$-rd principium of quantum
 theory}})
\end{itemize}
Cf.~\cite[p.~632]{zeilinger1} and mathematical analogies
\cite{baez+, mazur}.

\subsubsection{Semantic closedness and the equal sign $=$}%
\addcontentsline{toc}{subsection}{\qquad\qquad Semantic closedness and the
equal sign $=$}

We are currently returning once again to sect.~\ref{S}, falling into
a situation when the case in hand does not just entail fundamental
theory in the form of \lrceil{math} \tplus\ \lrceil{phy\-si\-cal
`bla-bla-bla'}, while, continuing on an informal note, the
mathematics of physics---""quantum mathematics""---is being created
`from scratch'. When building up this math, it is impossible to
forego the physical conventions \SM, meanwhile, any preliminary and
the formal characterization for the \SM\ is ruled out.

Indeed, the attempts to mathematically formalize the physical
context of observation, rather than observation itself, will not
logically manage without another""---`more
fundamental'""---""observation, in this case, of the very
experimental environment. The semantic cycling is apparent here, and
any of its mathematization will lead to a retrogression of
definitions into infinity, which is known as the ``von~Neumann
catastrophe'' \cite[pp.~158--\ldots]{dewitt}) or as the ``trying to
swallow itself by the tail'' \cite[p.~220]{bell}. Which is why, once
again, the `box \eqref{Box} method' prohibitions are required. See
also a paragraph containing the capitalized emphasize ``\textsc{cannot
in principle}'' on p.~418 of the work \cite{lipkin}. Sooner or later,
it will have to be declared that mathematics will be created for
\emph{the} convention \SM, and that this mathematics will be a
mathematical model for this \SM. The analogous
argument""---``mathematics is there to serve physics, and not the
other way round'' \cite[p.~242; L.~Hardy]{schloss3}---has already
long been met in the literature \cite{allah, auletta, fuchs3}. In
connection with the ``general contextual models'', see the books
\cite{hren1, hren9} (the V\"axj\"o"=model, ``quantum contextuality'')
and bibliography therein.

\begin{comment}[semantic]\label{meta}
To avoid the just mentioned linguistic closedness""---a kind of
mathematical `pathology' of the physical and natural languages""---a
description that lays claim to the role of an unambiguous/"!rigorous
theory requires a careful separation of the object- and
meta"=languages. For more detail, see \cite[sects.~14--16]{kleene},
\cite[sect.~V.1]{raseva}, \cite[sect.~3.9]{stoll}. For this reason,
the constructs should track the blending of the object \qm"=domain
(syntax) and the meta"=domain (semantics) and, more generally, the
penetration of extra"=linguistic elements of thinking \cite{chomsky}
into \qm. The notion of `the same \SM', which is intuitive in the
natural language, should explicitly be indicated as an external and
fundamental category (pr.~\hyperlink{III}{\red\textbf{\textsf{III}}}), and
its circular re"=interpretations/"!re"=translations within the theory
should be banned. That is, re"=stating `the sameness \SM, the
identical \SM' by way of word or of the equality symbol~$=$ between
some other entities is forbidden.
\par\larger%
\begin{itemize}
\item \smaller ``The same'' may no have a definition in terms of anything
 else. It exists prior to theory and has only a meaning (\,$=$~verbal
 context), though its natural"=language descriptions may be of great
 variety and be ``presented to us in wildly different ways''
 \cite[p.~2 and the whole of sect.~``The awkwardness of
 equality'']{mazur}.
\end{itemize}
\smaller%
One could, \eg, accept the typical verbal vehicle ``a complex of
conditions, which allows of any number of repetitions'' (quotation
from the literature). It is clear that the words ``complex \ldots\
allows \ldots\ repetitions'' here are just another semantic equivalent
to the word ``the same \SM''. The physics terminology per~se
(sects.~\ref{phys}, \ref{2slit}, and \ref{divis}) will become
accessible when physical concepts are introduced via the
originating""---and obligatorily very
ascetic""---""quantum"=mechanics language. See also the selected
thesis on page~\pageref{forbiden}.
\end{comment}

It is crucial to immediately note that, in the same manner, the
classical description contains the cited arguments in their
entirety. It is easy to convince that such a description also
implies implicitly that which is designated above as \SM; otherwise,
the physical reasoning would be entirely impossible. ``[W]e often
prefer to regard a number of outcomes of distinct physical
operations as registering the same property, \ldots\ representing the
same measurement. \ldots\ permitting an unrestricted identification of
outcomes would lead to \texttt{"}grammatical chaos\texttt{"}''
(Foulis--Randall \cite[p.~232]{foulis}). More to the point, the
physics and mathematics not merely have been closely interwoven with
each other. Any recursive procedure of definitions will inevitable
result in either a cyclic definition at some level, or a definition
which refers outside not only of the physics but even of the math.
Hence, the hierarchical arrangement of notions/"!\ldots/"!definitions""---a
property that is frequently uncontrolled and violated in the human
thinking""---can only be meaningful if at least one knot in the
definition network is externally defined. In this work, that basic
points are, as a rough guide, the brace
$\statePsi\GOTO{\sss\scr A}\state\alpha$ and the notion of `the same
\SM'; motivated in Remarks~\ref{two} and \ref{meta}, respectively.

\begin{comment}
Here, the situation is similar to the role of the axiom of choice in
the \zf"=system \cite{cohen, kurat}. It has been well known for a
long time that the axiom is often subconsciously implied
\cite[Chs.~II, IV]{frenkel}; it can also not be either circumvented
or ignored. Another counterexample to `infinite retrogression and
circularity' in logic comes from the very same system. This is a ban
on infinite chain of set memberships $\in$ on the left
\begin{equation*}
\bo\|\cdots\in X_n\in\cdots\in X_2\in X_1\in X_0
\end{equation*}
(the regularity axiom $[\forall x\in X,\: x\cap
X\ne\varnothing]\Rightarrow [X= \varnothing]$) under the
permissibility of the infinite $(\in)$"=continuing to the right:
\begin{equation*}
X_0\in X_1\in X_2\in\cdots\in X_n\in\cdots\in\cdots
\end{equation*}
(not rigorously, the infinity axiom) \cite{shenfield, kurat}.

The obvious parallels here are the famous Russell paradox
\cite{frenkel} or a chaos in the computer file system when the `hard
links' from a folder to the parent folder are allowed. Thus the
relations $\in$ `downwards' to the left and necessarily terminates
in something, \ie, in a set $X_0$ that contains nothing:
$\varnothing=X_0\in\cdots \in X_n\in\cdots$. Therefore one needs to
give `meaning' to the only set---the empty one $\varnothing$.
Incidentally, it is these axioms that guarantee the existence of
infinitely many ordinal numbers \eqref{ordinal} and uniqueness of
this structure. The ordinals, and numbers at all, have yet to be
dealt with further below in more detail.
\end{comment}

All that remains is to add that no theory in physics is feasible
without re"=calculations of physical units, of vectors/"!tensors,
without transformations in the fibre superstructures over manifolds,
\etc. Accordingly, the considerations on invariance and on
transformations should be present in the quantum case as well, but
it---which is its principal difference from the classical
case---still lacks the concepts of physical quantities/"!properties
(see sect.~\ref{phys}). Therefore such argumentation may only be
applied to those objects that we have at our disposal, \ie, to the
$\obj\Xi$"=brace. The renunciation of
pr.~\hyperlink{III}{\red\textbf{\textsf{III}}} would actually be
tantamount to the inability to make the physics theories whatsoever.

Now, pr.~\hyperlink{III}{\red\textbf{\textsf{III}}} and the `quantum
diversity of the reference frames' $\{\scr A$, $\scr B$, \ldots\}
require a kind of factorization of the entire family
$\{\obj\Xi_{\!\sss\scr A}, \obj\Xi_{\!\sss\scr B},\ldots,
\obj\Xi_{\!\sss\scr A}',\obj\Xi_{\!\sss\scr B}',\ldots\}$ with respect to
the conception \SM, \ie, the introduction of an operation of
equating the results $\obj\Xi_{\!\sss\scr A}$,
$\obj\Xi_{\!\sss\scr B}$ that came when observing $\cal S$. The
immediate putting $\obj\Xi_{\!\sss\scr A}\mathbin{\Over[1]{?}{=}}
\obj\Xi_{\!\sss\scr B}$ should not be made, since these brace are
simply different sets. That is why, with isolated semigroups
\begin{equation*}
\big\{\underbrace{%
\Over[1.5]{\sss\scr A\!\!}{\uplus}\;;\;
\obj{\Xi'}_{\!\sss\scr A}\,,\obj{\Xi’’}_{\!\sss\scr A}\,,\,\ldots}_
{\frak G_{\!\!\!\sss\scr A}}\big\}\,,\quad
\big\{\underbrace{
\Over[1.5]{\sss\scr B\!}{\uplus}\;;\;
\obj{\Xi'}_{\!\sss\scr B}\,,\obj{\Xi’’}_{\!\sss\scr B}\,,\,\ldots}_
{\frak G_{\!\!\!\sss\scr B}}\big\}\,,\quad\ldots
\end{equation*}
at our disposal, we have to conceive of them as elements of a new
set \itsf{H} of objects having a single nature and 1) to carry out
the mapping $\{\frak G_{\!\!\!\sss\scr A}\,, \frak G_{\!\!\!\sss\scr B}\,,
\,\ldots\}\mapsto\itsf{H}$, assigning new representatives
$\ket{\Xi_{\!\sss\scr A}}\in\itsf{H}$ to the $\obj\Xi$"=brace, and 2)
to equip \itsf{H} with an equivalence relation
$\ket{\Xi_{\!\sss\scr A}}\approx\ket{\Xi_{\!\sss\scr B}}$ (the concept
`the same' above). Let us implement all of that by the scheme
\begin{equation}\label{o+}
\begin{array}{r@{}l@{}c@{}r@{}l@{}r@{}ccc}
\obj\Xi_{\!\sss\scr A}&{}\DEF\big[\smallmatrix \mu_1\\
\lambda_1\endsmallmatrix\big]\,\state\alpha_1
\mathbin{\Over[1.5]{\sss\scr A\!\!}{\uplus}}
\big[\smallmatrix\mu_2\\
\lambda_2\endsmallmatrix\big]\,\state\alpha_2
\mathbin{\Over[1.5]{\sss\scr A\!\!}{\uplus}}
\cdots&\quad\rightarrowtail\quad
&\big[\smallmatrix \mu_1\\
\lambda_1\endsmallmatrix\big]\,\bet{\alpha_1}\Uplus
\big[\smallmatrix\mu_2\\
\lambda_2\endsmallmatrix\big]\,\bet{\alpha_2}\Uplus\cdots{}
\FED{}&\ket{\Xi_{\!\sss\scr A}}&{}\in\itsf{H}\;,\\[2ex]
\obj\Xi_{\!\sss\scr B}&{}\DEF\big[\smallmatrix \mu_1\\
\lambda_1\endsmallmatrix\big]\,\state\beta_1
\mathbin{\Over[1.5]{\sss\scr B\!}{\uplus}}
\big[\smallmatrix\mu_2\\
\lambda_2\endsmallmatrix\big]\,\state\beta_2
\mathbin{\Over[1.5]{\sss\scr B\!}{\uplus}}
\cdots&\quad\rightarrowtail\quad
&\big[\smallmatrix \mu_1\\
\lambda_1\endsmallmatrix\big]\,\bet{\beta_1}\Uplus
\big[\smallmatrix\mu_2\\
\lambda_2\endsmallmatrix\big]\,\bet{\beta_2}\Uplus\cdots{}
\FED{}&\ket{\Xi_{\!\sss\scr B}}&{}\in\itsf{H}\;,\\[1ex]
&\mathrel{}\ldots\ldots\ldots&\ldots\ldots&\ldots\ldots\ldots\mathrel{}
\end{array}
\end{equation}
In this, the new addition $\Uplus$ must of course homomorphically
inherit operations $\Over[1.5]{\sss\scr A\!\!}{\uplus}$,
$\Over[1.5]{\sss\scr B\!}{\uplus}$, \ldots, and extension of this
definition throughout \itsf{H} is then made with the aid of the very
equivalence~$\approx$:
\begin{equation*}
\ket{\Xi'_{\!\sss\scr A}}\Uplus\ket{\Xi’’_{\!\sss\scr B}}=
\Big| \ket{\Xi’’_{\!\sss\scr B}}\approx
\ket{\Xi’’_{\!\sss\scr A}} \hence\Big|=
\ket{\Xi'_{\!\sss\scr A}}\Uplus\ket{\Xi’’_{\!\sss\scr A}}=
\ket{\Xi'_{\!\sss\scr B}}\Uplus\ket{\Xi’’_{\!\sss\scr B}}\;.
\end{equation*}
Negation $\not\approx$ of the relation $\approx$, \eg,
$\ket{\Xi'_{\!\sss\scr A}}\not\approx \ket{\Xi’’_{\!\sss\scr A}}$, is
exactly the very same distinguishability that was discussed in
sects.~\ref{S0}--\ref{Ans}.

For the sake of convenience, we adopt the regular sign $=$ for
$\approx$, in order not to introduce yet a further homomorphism,
which are already numerous, with more to come. In other words, the
physics \SM\ is `concentrated' in the sign $=$, turning the
empirical structures \eqref{o+} into the $\scr A$-,
$\scr B$"=implementations of the object
$\ket\Xi\equiv\ket{\Xi_{\!\sss\scr A}}= \ket{\Xi_{\!\sss\scr B}}$
under construction. The adequate term for it---the
\Courier{Info/"!Data}-\Courier{Source}\eLab{datainfo} or
``representative of information'' (\v{C}.~Brukner
(2014))---""corresponds to the preliminary prototype of the concept
of a state, but we will remain within the standard term,
disregarding its variance.

\vbox{
\section{Quantum superposition}\label{SP}

\flushright\tiny
\textsl{How come the quantum?}\\
\ldots\ No space, no time\\
\textsc{--- J.~Wheeler (1989)}
\medskip

\textsl{\ldots\ postulation of something as a Primary\\
Observable is itself a sort of theoretical\\
act and may turn out to be wrong}\\
\textsc{--- T.~Maudlin} \cite[p.~142]{maudlin}}
\smallskip\nopagebreak

\subsection{Representations of states}

Let us simplify notation according to the rule
$[\smallmatrix\mu\\[-0.25ex]\lambda\endsmallmatrix]\FED\frak a$.
The sought-for relationships between $\scr A$, $\scr B$, \ldots\ then
turn into the key point of further construct""---the equalities
\begin{equation}\label{g1}
\mbig[7]\lceil\![5]
\begin{array}{c}\text{representations}\\\text{of $\ket\Xi$"=state}
\end{array}\![5]\mbig[7]\rceil:\qquad
\frak a_1\,\bet{\alpha_1}\Uplus\frak a_2\,\bet{\alpha_2}\Uplus\cdots
\mathbin{\Over{\sss\text{`the same'}}{\scalebox{2}[1]{$=$}}}
\frak b_1\,\bet{\beta_1}\Uplus\frak b_2\,\bet{\beta_2}\Uplus\cdots=
\cdots\;.
\end{equation}
They furnish \emph{representations $\ket{\Xi_{\!\sss\scr A}}$,
$\ket{\Xi_{\!\sss\scr B}}$, \ldots\ of quantum state $\ket\Xi$ of
system~$\cal S$}. By design, this \Courier{Data\-Source} object
$\ket\Xi$ carries data $\obj\Xi_{\!\sss\scr A}$,
$\obj\Xi_{\!\sss\scr B}$ and, more generally, $\obj\Xi$"=data
\eqref{set''} from arrays of \emph{any} observations, including the
imaginary ones. That is what eliminates the initial need for the
$\obj\Xi_{\!\sss\scr A}$"=brace \eqref{obj} to come from the
observation $\scr A$, which is reflected in the shortening of the
term ``representation of state'' to simply ``state'' $\ket\Xi$. It
should be added that the straightforward storing of objects
$\{\ket{\Xi_{\!\sss\scr A}}$, $\ket{\Xi_{\!\sss\scr B}}$, \ldots\} in a
certain set \itsf{H}, but with the independence of operations
\{$\Uplus^{\!\!\!\sss(\scr A)}$, $\Uplus^{\!\!\!\sss(\scr B)}$, \ldots\}
preserved, would not differ from the tautological substitution of
symbols. Accordingly, the semantic autonomy of
$\obj{\state\Xi}$"=brace would also be inherited, whereas
covariance~\hyperlink{III}{\red\textbf{\textsf{III}}} requires an
elimination of precisely this autonomy. What is more, the
set"=theoretic original copy for operations
\{$\Over[1.5]{\sss\scr A\!\!}{\uplus}$,
$\Over[1.5]{\sss\scr B\!}{\uplus}$, \ldots\} and $\Uplus$ is one and the
same---the union~$\cup$.

The symbols $\bet{\alpha_s}$ and $\bet{\beta{}_s}$ in \eqref{g1} are
no more than symbols. Hence the objects' property \eqref{g1} of
being identical must be reflected in terms of their coordinate
$\frak a,\frak b$"=components (pt.~\hyperlink{R}{\red\textsf{R}}$^\bullet$).
This means that any aggregate $(\frak a_1$, $\frak a_2$, \ldots) is
unambiguously calculated by means of a certain transformation
$\widehat U$ into any other $(\frak b_1$, $\frak b_2$, \ldots) when the
two aggregates represent a common~$\ket\Xi$:
\begin{equation*}
(\frak a_1,\frak a_2,\ldots)=\widehat U (\frak b_1,\frak b_2,\ldots)\;.
\end{equation*}
The $\widehat U$ then becomes an isomorphism between these
aggregates (a preimage of the future unitary transformation
\cite[p.~14]{br2}) and, accordingly, their lengths must coincide.
This length---a certain single constant""---will be symbolized
as~\textsf{D}.

\subsection{Representations of devices. Spectra}

Naturally, instrument is converted to the \itsf{H}"=structure
language along with $\obj\Xi$"=objects. It is a set of symbols
$\{\bet{\gamma_1}$, $\bet{\gamma_2}$, \ldots\} in place of the previous
$\{\state\gamma_1$, $\state\gamma_2$, \ldots\}. As has just been shown,
their number for any $\scr C$"=instrument should be equal to \textsf{D}.
However, generally speaking,
$|\boT_{\sss\scr A}|\ne|\boT_{\sss\scr B}|$, since
$\boT_{\sss\scr A}$ and $\boT_{\sss\scr B}$ are assigned in an
arbitrary way (pt.~\hyperlink{O}{\red\textsf{O}}). Therefore if we take
an illustration $\scr A\{\state\alpha_1$, $\state\alpha_2\}$ and
$\scr B\{\state\beta_1$, $\state\beta_2$, $\state\beta_3\}$, then
\itsf{H}"=representation of instrument $\scr A$ should appear at
least as $\{\bet{\alpha_1}$, $\bet{\alpha_2}$, $\bet{\alpha_3}\}$.
Clearly, the already present distinguishability
$\state\alpha_1\not\approx\state\alpha_2$ (sect.~\ref{observ}) is
automatically converted into an abstract distinguishability of new
symbols $\bet{\alpha_1}\ne\bet{\alpha_2}$, and empirical
$\scr A$"=distinguishability is confined exclusively by these two
symbols. In that case, for the purpose of noncontradiction, the
added third symbol $\bet{\alpha_3}$, as an adjunction to the
abstract relations $\bet{\alpha_3}\ne\bet{\alpha_1}$ and
$\bet{\alpha_3}\ne\bet{\alpha_2}$, should be complemented with the
notion of its physical \emph{in}discernibility from $\bet{\alpha_1}$
or $\bet{\alpha_2}$. By an extension of this argument, one gets that
every $\scr A$"=instrument should be endowed with the
(non)""equivalence relation (${\Bumpeq}/{\not\Bumpeq}$) in terms of
the \itsf{H}"=structure by its formal
$\{\bet{\alpha_1},\ldots\}$"=representations. How do we do this?

Let us proceed further from a self"=suggested extension of
pt.~\hyperlink{R}{\red\textsf{R}}. Let us declare---and it is more than
natural""---that the number representations $\alpha_s$ are linked
not only to observations but to instruments as well. Each $\alpha_s$
is the new object of a numerical type: a number or a collection of
numbers. Then, indiscernibility, say
$\bet{\alpha_3}\bumpeq\bet{\alpha_1}$, is recorded by coincidence of
the numeral labels $\alpha_3=\alpha_1$ attached to the symbols
$\bet{\alpha_3}$ and $\bet{\alpha_1}$ respectively. The abstract
(`old') distinguishability $\bet{\alpha_3}\ne\bet{\alpha_1}$,
meanwhile, remains as it is. From these we have the following
formalization of the relationship between $\bumpeq$ and $=$ by means
of dropping/"!adding the brackets~$\bet{\ }$:
\begin{equation}\label{spectr}
\left.
\begin{aligned}
\bet{\alpha_s}\not
\bumpeq\bet{\alpha_k}\quad&\hhence\quad\alpha_s\ne\alpha_k\\
\bet{\alpha_s} \bumpeq\bet{\alpha_k}\quad&\hhence\quad\alpha_s=\alpha_k
\end{aligned}
\right\}\quad\text{under}\quad \bet{\alpha_s}\ne\bet{\alpha_k}\;.
\end{equation}
Call the quantity $\alpha_s$ (numerical) \emph{spectral
label/"!marker} of eigen"=element $\bet{\alpha_s}$. Then, by the
\itsf{H}-\emph{representation $\bo[\scr A\bo]$ of instrument}
$\scr A$ we will mean the set of objects $\big\{\bet{\alpha_1}$,
\ldots, $\bet{\alpha_{\textsf{D}}}\big\}$ supplemented with the spectral
structure~\eqref{spectr}:
\begin{equation}\label{spectr-}
\bo[\scr A\bo]\DEF\big\{\bet{\alpha_1}_{{\sss\lfloor\!}\alpha_1^{}},\,
\bet{\alpha_2}_{{\sss\lfloor\!}\alpha_2^{}},\, \ldots\big\}\;.
\end{equation}
It is not difficult to see that if
$\bet{\alpha_1}\not\bumpeq\bet{\alpha_2}$ then either
$\bet{\alpha_3}\bumpeq\bet{\alpha_1}$ or
$\bet{\alpha_3}\bumpeq\bet{\alpha_2}$. Otherwise, spectral markers
$\smallc[2]\lfloor\!\alpha_1=\smallc[2]\lfloor\! \alpha_2$ should
coincide, and primary primitives
$\state\alpha_1\not\approx\state\alpha_2$ lose their empirical
distinguishability in contrast with \eqref{a1122}. Multiple
coincidence of $\smallc[2]\lfloor\!\alpha_s$"=markers is admissible.

In the presence of relations \eqref{spectr}, it is natural to state
that instrument $\scr A$ is coarser (more symmetrical) than $\scr B$
and, terminologically, to declare that the degeneration of the
spectral"=label values takes place. In cases of embeddability like
$\scr A_2\{\state\alpha_1$, $\state\alpha_2\}\subset
\scr A_3\{\state\alpha_1$, $\state\alpha_2$, $\state\alpha_3\}$,
instrument $\scr A_2$ can even be called the same as (coinciding
with) $\scr A_3$, but with a more rough scale. Or conversely,
$\scr A_3$ is a more precise extension of $\scr A_2$. In particular,
the natural notion of a device resolution fits here.

All instruments may then be mathematically imagined as having the
same resolution, but, perhaps, with degeneration of spectra. The
non"=coinciding instruments may be interpreted as non"=equivalent
reference frames $\scr A\ne\scr B$ in an observation space.
According to pts.~\hyperlink{R}{\red\textsf{R}}$^{\bullet\bullet}$ and
\hyperlink{III}{\red\textbf{\textsf{III}}}, they are mandatorily present
in the description. The spectral degenerations are also always
present, since element $\state\alpha_1$ can always be removed from
$\boT_{\!\sss\scr A}$, and there are no logical foundations to
prohibit an observational instrument with family
$\boT_{\!\sss\scr A}-\{\state\alpha_1\}$. Hence it follows that
introducing the spectra---""instrumental readings""---is required
even formally, without physics. It is of course implied here that
spectral (in)""discernibility is realized in the same manner as its
statistical counterpart in sects.~\ref{2.6} and \ref{whyC}, \ie, by
numbers. Incidentally, such a property of
$\smallc[2]\lfloor\!\alpha_s$---\ie, of being a numerical object---is
not at all necessary at the moment. Spectrum
$\{\smallc[2]\lfloor\!\alpha_1,\smallc[2]\lfloor\! \alpha_2,\ldots\}$ may
be thought of as an abstract set of labels attached to the
eigen"=elements. As numbers, it is introduced for the subsequent
creation of models to classical/"!macroscopic dynamic, and they are
numerical.

Returning to \textsf{D}, we note that, in any case, the toolkit
$\{\scr A,\scr B,\ldots\}\FED\cal O$ in real use has always been
defined, fixed, and is finite. Consequently, the constant
\begin{equation}\label{D}
\textsf{D}\geqslant2
\end{equation}
has also been defined and fixed, and it becomes the globally static
observable characteristic""---an empirically external parameter.
Meanwhile, the entire scheme internally contains the natural method
of its own extension $\textsf{D}\mapsto\textsf{D}+1$, and the potentially
all"=encompassing choice $\textsf{D}=\infty$ may be considered the
universally preferable one in \qt. By freezing the different
$\textsf{D}<\infty$, the theory makes it possible to create models, and
they are not only admissible but also well-known. Their efficiency
is examined in experiments. Once again,
\begin{itemize}
\item the \textsf{D}"=constant, the concept of spectra, and their
 degenerations are created by the ($\scr A, \scr B$)"=covariance
 requirement, \ie, by
 principium~\hyperlink{III}{\red\textbf{\textsf{III}}}.
\end{itemize}

As a result, the structure of \itsf{H}"=representations of states and
of instruments are liberated from the arbitrariness in assigning the
subsets $\boT_{\!\sss\scr A}$ in \eqref{shema}. The statistical
unitary pre"=images \eqref{min} and \itsf{H}"=elements of the form
$\frak c\,\bet{\gamma_s}$ can be associated with any `eigen symbol'
$\bet{\gamma_s}$. They are always available because every possible
brace \eqref{split'} are known to contain subfamilies when ingoing
$\statePsi,\statePhi$"=primitives get to a single one, \eg, to
$\state\gamma_1$. Therefore every representation
$\frak a_1\,\bet{\alpha_1}\Uplus\cdots$ is always equivalent to a
$\obj{\state\Xi}_{\!\sss\scr C}$"=brace for some observation $\scr C$
with a homogeneous outgoing ensemble
$\{\state\gamma_1\cdots\state\gamma_1\}$. That is, one may always
write
\begin{equation}\label{eigen}
\frak a_1\,\bet{\alpha_1}\Uplus\frak a_2\,\bet{\alpha_2}\Uplus\cdots=
\frak c_1\,\bet{\gamma_1}\Uplus0\,\bet{\gamma_2}\Uplus\cdots
\FED\frak c_1\,\bet{\gamma_1}\;,
\end{equation}
while naturally referring to $\frak c_1\,\bet{\gamma_1}$ as one of
the \emph{eigen"=states of instrument} $\scr C$, with an appropriate
adjusting the similar definition in pt.~\hyperlink{O}{\red\textsf{O}}.
The construction of the representation"=state space is far from being
complete, since it is still a `bare' semigroup~\itsf{H}.

\subsection{Superposition of states}\label{ssuper}

Since writings \eqref{g1} exist for any ensemble $\obj\Xi$"=brace,
let us consider the following two representations:
\begin{equation}\label{ab}
\begin{aligned} \frak a_1\,\bet{\alpha_1}\Uplus
\frak a_2\,\bet{\alpha_2}&=\frak b_1\,\bet{\beta_1}\Uplus\frak b_2\,
\bet{\beta_2}\Uplus\cdots\;,\\
\frak a_2\,\bet{\alpha_2}&=
\frak b_1'\,\bet{\beta_1}\Uplus\frak b_2'\,\bet{\beta_2}\Uplus\cdots\;.
\end{aligned}
\end{equation}
Comparison of these equalities tells us that the second one is a
solution of the first one with respect to
$\frak a_2\,\bet{\alpha_2}$. Hence, the semigroup operation $\Uplus$
admits a cancellation of element $\frak a_1\,\bet{\alpha_1}$. This
means that there exists an \itsf{H}"=element
$\tilde{\frak a}_1\,\bet{\alpha_1}$ such that
\begin{Align*}
\big\{\tilde{\frak a}_1\,\bet{\alpha_1}\Uplus
\frak a_1\,\bet{\alpha_1}\big\}\Uplus\frak a_2\,\bet{\alpha_2}&=
\tilde{\frak a}_1\,\bet{\alpha_1}\Uplus\big\{\frak b_1\,\bet{\beta_1}
\Uplus\frak b_2\,\bet{\beta_2}\Uplus\cdots\big\}
\\[-1ex]&\,\,\Downarrow\\[-1ex]
0\,\bet{\alpha_1}\Uplus \frak a_2\,\bet{\alpha_2}&=\tilde{\frak a}_1\,
\bet{\alpha_1}\Uplus\frak b_1\,\bet{\beta_1}\Uplus\frak b_2\,
\bet{\beta_2}\Uplus\cdots\\[-1ex]&\,\,\Downarrow\quad
\text{(due to \eqref{eigen})}\\[-1ex]
\frak a_2\,\bet{\alpha_2}&=\frak b_1'\,\bet{\beta_1}
\Uplus\frak b_2'\,\bet{\beta_2}\Uplus\cdots
\\\Downarrow\qquad\quad&\qquad\quad\Downarrow\\
\makebox[0em][r]{$\bet{0}\DEF0\,\bet{\alpha_1}=
\tilde{\frak a}_1\,\bet{\alpha_1}\Uplus
\frak a_1\,\bet{\alpha_1}$}\,,\quad&\qquad
\makebox[0em][l]{$\tilde{\frak a}_1\,\bet{\alpha_1}\Uplus
\frak b_1\,\bet{\beta_1}\Uplus\cdots=
\frak b_1'\,\bet{\beta_1}\Uplus\cdots\;,$}
\end{Align*}
where $\bet{0}$ stands for a zero in semigroup \itsf{H} (image
$\obj0$ of the finite-length brace $\obj{\state\Xi}$) and 0 in
$0\,\bet{\alpha_1}$ is a symbol of its $[\lambda,\mu]$"=coordinates.
By canceling out $\frak a_s\,\bet{\alpha_s}$, one by one, if
necessary, one deduces that any element of \itsf{H} does have an
inversion. That is, \itsf{H} is actually a group. We re"=denote
inverse elements $\tilde{\frak a}_s\,\bet{\alpha_s}$ by
$(-\frak a_s)\,\bet{\alpha_s}$ and inversions of sums are formed from
$(\Uplus)$"=sums thereof. Moreover, all the $[\lambda,\mu]$"=pairs
turn into a set $\{\frak a,\frak b, \ldots\}$ equipped with the above
mentioned composition $\oplus$, which follows from an obvious
property of unitary brace:
\begin{equation}\label{s20}
\frak a\,\bet{\alpha_1}\Uplus \frak b\,\bet{\alpha_1}=
(\frak a\oplus\frak b)\,\bet{\alpha_1}
\end{equation}
(inheritance of clossedness under the $\cup$"=operation). This
composition is also a $\oplus$"=operation of a group and of a
commutative one:
\begin{equation}\label{g+}
\frak a\oplus\frak b=\frak b\oplus\frak a\,,\qquad
(\frak a\oplus\frak b)\oplus\frak c=\frak a\oplus
(\frak b\oplus\frak c)\,,\qquad\frak a\oplus0=\frak a\,,\qquad
\frak a\oplus(-\frak a)=0\;.
\end{equation}
Therefore the group nature of semigroup \itsf{H} and the group
\eqref{g+} come from the scheme
\begin{equation*}
\boxed{\begin{array}{c}
\ds \bigg\lceil\!\!\!\begin{array}{c}
\text{single observations}\\
\text{$\scr A$, $\scr B$, \ldots}
\end{array}\!\!\!\bigg\rceil\quad\hence\quad
\bigg\lceil\!\!\!\begin{array}{c}
\text{semigroups}\\\text{$\frak G_{\!\!\!\sss\scr A}$,
$\frak G_{\!\!\!\sss\scr B}$, \ldots}\end{array}\!\!\!\bigg\rceil
\quad\rightarrowtail\hspace{2em}
\\\\
\hspace{3em}\ds \rightarrowtail\quad
\bigg\lceil\!\!\!\begin{array}{c}
(\scr A,\scr B)\text{-covariance},\\
\text{\SM\ and principium~\hyperlink{III}{\red\textbf{\textsf{III}}}}
\end{array}\!\!\!\bigg\rceil \quad\hence\quad
\big\lceil\text{\,group \itsf{H}\,\,}\big\rceil
\end{array}}
\end{equation*}
and, technically, from equatings/"!identifyings~\eqref{g1}, \ie, from
conception `the same' (sect.~\ref{SS}). For its part, it is this
very structure of algebraic operations""---the 2-,~3-term (and
nothing else) axioms of commutativity/"!associativity, \ie, the
group---that comes from properties~\eqref{U}. All of this provides
an answer to the key question: where do the (semi)group and the
minus sign come from and why?

Thus handling the $\ket\Xi$"=objects breaks free from its ties to the
notion of observation, and the objects admit the formal writings
$\frak a\,\bet\Psi\Uplus\frak b\,\bet\Phi\Uplus\cdots$. Call them
\emph{s\,\,u\,\,\-p\,\,e\,\,r\,\,\-p\,\,o\,\,\-s\,\,i\,\,\-t\,\,i\,\,o\,\,n\,\,s}. However, as
soon as they or the state are associated in meaning with the word
`readings' (this is discussed at greater length in sects.~\ref{phys}
and \ref{2slit}), this term should be replaced with a non"=truncated
one---""representation of the state with respect to a certain
observation. Specifically, the statistical weights $\fr_j$ are
extracted from such expressions only after their conversion into a
sum over eigen"=states of the form \eqref{g1}; a task of the
subsequent mathematical tool.

No superposition
$\frak a\,\bet\Psi\Uplus\frak b\,\bet\Phi\Uplus\cdots$, including
\eqref{eigen}, has any physical sense in and of itself
\cite[p.~137]{silverman}, \cite{englert} nor is it preferable to any
other one. It merely mirrors the closedness of states with respect
to operation $\Uplus$, since any $\ket\Xi$ is re"=recorded as a sum
of various \{$\frak a\,\bet\Psi$, $\frak b\,\bet\Phi$, \ldots\} in a
countless number of ways and is linked to any other such sum.
Without a system of $\bet{\alpha_s\!}$"=symbols for instrument
$\scr A$, nothing observable is extractable out of the aggregate of
coefficients $\{\pm\frak a$, $\pm\frak b$, \ldots\} (and, of course, of
the $\bet\Psi$"=letters themselves) in any imaginable way.
Accordingly, it is incorrect to speak of---a widespread
misconception""---the destruction of the superposition or of the
``relative"=phase information'' \cite[p.~253]{schloss}, associating the
word destruction with the physical/"!observational meanings or
processes.

As a result, \emph{even without having a numerical theory yet} and
without recourse to the concept of a physical
quantity""---""superposition may not address whatever physical
concepts, we arrive at the paramount property, which characterizes
the most general type of micro"=observation's
ensembles~\eqref{quant}.
\begin{itemize}
\item \embf{Superposition principle}\eLab{princ}\\
 A $(\Uplus)$"=composition of quantum states $\frak a\,\bet\Psi$ and
 $\frak b\,\bet\Phi$, which are admissible for system $\cal S$, is an
 admissible state
 \begin{equation}\label{s1}
 \frak a\,\bet\Psi\Uplus\frak b\,\bet\Phi=\frak c\,\bet\Xi
 \end{equation}
 and, with that, the set $\big\{\frak a\,\bet\Psi,\frak b\,\bet\Phi,
 \frak c\bet\Xi,\ldots\big\}\FED\itsf{H}$ forms a commutative group
 with respect to operation $\Uplus$. The family
 $\{\frak a,\frak b,\ldots\}$ of coordinate $\bbR^2$"=representatives of
 states \eqref{g1} is also equipped with the same group structure
 under the $\oplus$"=operation \eqref{g+} and with the rule of
 carrying the operation $\Uplus$ over to~$\oplus$:
 \begin{equation}\label{s2}
 \frak a\,\bet\Psi\Uplus\frak b\,\bet\Psi=
 (\frak a\oplus\frak b)\,\bet\Psi\;.
 \end{equation}
\end{itemize}

Let us clarify the transferring \eqref{s20} to \eqref{s2}. The union
of the state prototypes $\frak a\,\bet\Psi$,
$\frak b\,\bet\Psi\in\itsf{H}$ is known to belong to $\frak G$. So
the composition $\frak a\,\bet\Psi\Uplus\frak b\,\bet\Psi$ should be
identical to a certain element $\frak c\,\bet\Psi\in\itsf{H}$. It is
clear that $\frak c$ depends on $\frak a$, $\frak b$ and, hence,
$\frak a\,\bet\Psi\Uplus\frak b\,\bet\Psi=
\frak c(\frak a,\frak b)\,\bet\Psi$. The exhaustive properties of
dependence $\frak c(\frak a,\frak b)$ are given by formulas
\eqref{g+} and \eqref{s2} under notation
$\frak c(\frak a,\frak b)\FED(\frak a\oplus\frak b)$.

\subsubsection{`Physics' of superposition}\label{cat}
\addcontentsline{toc}{subsection}{\qquad\qquad `Physics' of superposition}

Besides the essentially unphysical nature of the
($\Uplus$)"=superpositions""---``we cannot recognize them''
\cite[p.~13]{englert}, the primary and salient property of quantum
addition is in the fact that, due to the group subtraction, it is
possible the experimental obtainment of a `quantum zero' in
statistics from `non"=zeroes'. With that, these `seem to be'
positive, but there are `negative non"=zeroes'---""negative numbers
(sect.~\ref{92}). Subtraction manifests by the typical obscurations
in interference pictures. S.~Aaronson adds to this: ``We've got minus
signs, and so we've got interference'' \cite[p.~220]{aaronson+}. No
classical composition
\begin{equation}\label{w1w2}
w\,\varrho_1^{}+(1-w)\,\varrho_2^{}
\end{equation}
of non"=zero statistics $\varrho_1^{}$, $\varrho_2^{}$ can provide a
zero value, since the zero will never be obtained via the
$\cup$"=unions. The same is true for the pre"=superposition in
isolated brace $\obj\Xi_{\!\sss\scr A}$, \ie, when one instrument is
in question.

\begin{comment}\label{Qcat}
One cannot but mention yet another counterexample to the
superposition's `physicality'"!---the (in)famous `quantum cat'. Any
combination of the dead and living animal is meaningless as a
statement about new/"!nonclassical entity such as a `(half-)dead/alive
cat' or such statements about particles as `their being neither here
nor there but everywhere'; especially with the stress on `at the
same point in time' (see pr.~\hyperlink{I}{\red\textbf{\textsf{I}}}). It
makes absolutely no sense to add (allegedly in accord with the
character~$+$) to each other the nature's phenomena and notions that
have not yet been created and are dynamical (`alive') at that.
\par\larger 
\begin{itemize}
\item \smaller\embf{What's being added is states, not their
 denominations} \emph{or verbal descriptions of envisioned
 \emph{(`fantasized', ``fantastic phantoms'' \cite[p.~15]{englert})}
 physical properties} such as spin up/"!down or dead/"!alive.
 Cf.~\cite[pp.~134\,(!), 135]{maudlin}.
\end{itemize}
\smaller 
The `cat-box open' is a click, not state; without a notion of ``a
cat''. Accordingly, the word combination ``the quantum objects exist
in ``superpositions'' of different possibilities'' (a representative
excerpt from the literature) is at most an interpretative allegory
(sect.~\ref{interpret}) without physical and mathematical content.
That is to say, strictly speaking,
\par\larger 
\begin{itemize}
\item \smaller no quantum (micro)system has ever been/"!dwelled in any
 state, much less in a superposition one, and much less at an
 instant $t$. Ludwig, on pp.~16 and 78 of the book \cite{ludwig5},
 insists that it is a ``myth'' and ``a fairy tale, \ldots\ the very
 widespread idea that each microsystem has a real state \ldots\
 represented by a vector in a Hilbert space'', and M.~Nielsen remarks
 in \cite{nielsen} that ``Saying $0.6\,\ket{{}0}+0.8\,\ket{{}1}$ is
 simultaneously $0$ and~$1$ makes about as much sense as Lewis
 Carroll's nonsense poem \emph{Jabberwocky}: \ldots''. K.~Svozil does
 also underscore that ``\,`coherent superpositions' just correspond to
 improper, misleading representations of non-existing aspects of
 physical reality. They are delusive because they confuse ontology
 with epistemology'' \cite[p.~26]{svozil}.
\end{itemize}
\smaller 
The meaning of the word `add' is still being created, including an
implementation at objects to be thought of as the `atomic
irreducible' entities""---the numbers (sect.~\ref{Numbers}).

T.~Maudlin notes on p.~133 of the work \cite{maudlin}: ``Our job \ldots\
is to invent mathematical representations \ldots, rather than merely
linguistic terms such as ``$z$-up.'' \ldots\ we are in some danger of
\emph{confusing physical items with mathematical items}'' (italics
supplied). Here is an example of confusion. If we are going to
measure the $z$"=spin in one of the $(\rightleftarrows_x)$"=beams in a
Stern--Gerlach device, then why and when does this
`\emph{observable}' certainty""---say
$\ket{{}{\rightarrow}_x}$---get turned into a
$(\uparrow\!\!\!\downarrow_z)$"=uncertainty $\ket{{}{\uparrow}}+
\ket{{}{\downarrow}}$? (see \cite[p.~232]{ballentine+}). But what if
we are about not to do this? We come up against the question:
\par\larger 
\begin{itemize}
\item \smaller What does one mean by an equal-sign $=$ in the orthodox
 notation $\ket{{}{\rightarrow}}=\ket{{}{\uparrow}}+
 \ket{{}{\downarrow}}$?
\end{itemize}
\smaller 
Which state does the system `intend' to fall into: the
$z$"=uncertainty or the $x$"=determinacy? Which of the states is it
in, after all? Examples to the `physicality of states' may be
continued endlessly \cite{barrett2}.
\end{comment}

A statement about \qm"=superposition (without $\bbC$"=numbers) as a
non"=independent axiom can be found in the book \cite[p.~108]{jauch1}
but arguments given there are circular: \lrceil{Hilbert space}
$\rightarrowtail$ \lrceil{quantum logic of propositions}
$\rightarrowtail$ \lrceil{superposition principle}. Similarly, in
the works \cite{piron} and \cite[p.~164]{beltrametti}, all of that
is `derived' from modular lattices \cite{birkh}. However, the
lattices are known to enter \qm\ from the Hilbert space structure
and, on the other hand, the purging quantum rudiments of such a
space' axiomatics constitutes Birkhoff's 110-th problem
\cite[p.~~286]{birkh}. Note also that, in connection with the formal
logic approaches to the theory construction \cite{beltrametti,
foulis, jauch1, laloe, piron, varadar, zierler}, the issue of
vindicating the \emph{matters} that this logic deals \emph{with}
should not be neglected. What me mean here is the questions: logic
of what? \cite{ludwig5, ludwig3, spekkens} of propositions?
\cite{kleene, raseva} of relations? of (math-logic) classes/sets?
\cite{shenfield} of phenomena/"!properties? (which ones?) of
quantum/"!classical events? \ldots? ``For example, would one have to
develop a quantum set theory?\@'' \cite[p.~17]{edwards}. ``If by
``logic'' we mean something like ``correct reasoning,'' then it would
make no sense to think of logic as ``just another theory.''\,''
\cite[p.~258]{stairs}. The more so as the abstract micro"=events and
Boolean logic we have used in metamathematical reasoning at the
moment \cite[pp.~189, 193]{ludwig1} contain nothing that depends on
classical physics. That is, quantum foundations do not require
\cite{ludwig5} a different""---""quantum/"!non"=classical""---logic.
See also \cite[p.~29]{foulis0}.

\subsubsection{What is non-commutativity? Whence this structure?}
\addcontentsline{toc}{subsection}{\qquad\qquad What is non-commutativity?
Whence this structure?}

Yet another fact that results from the above constructs is that the
availability of a superposition math"=structure \eqref{s1} reflects
the presence of at least \emph{two} $\scr A$, $\scr B$ with
\emph{non"=coinciding} families of eigen"=primitives
$\{\state\alpha_s\}$, $\{\state\beta_k\}$. This consequence of
pt.~\hyperlink{R}{\red\textsf{R}}$^{\bullet\bullet}$ should be particularly
emphasized, since it will manifest in the \emph{non"=commutativity}
of operators $\hA$ and $\hB$ in the future. Although the present
work does not get to operators as a mathematical structure, it is
clear that the emergent eigen"=states and spectra have a direct
bearing on them. In this context, the `commuting instruments'
$\{\bet{\alpha_1},\bet{\alpha_2},\ldots\}=
\{\bet{\beta_1},\bet{\beta_2},\ldots\}$ can be treated, roughly
speaking, as coinciding, because this fact is independent of the
specific spectrums $\{\smallc[2]\lfloor\! \alpha_1,
\smallc[2]\lfloor\! \alpha_2,\ldots\}$, $\{\smallc[2]\lfloor\! \beta_1,
\smallc[2]\lfloor\! \beta_2,\ldots\}$ assigned to them. If they differ,
this is merely a different (numerical) graduation of the spectrum
scale. It is the same for all instruments, and its length is the
parameter~\textsf{D}.

Notice that definition of an $\scr A$"=observation is not different
from the formal assignment of the family $\boT_{\!\sss\scr A}$
(pt.~\hyperlink{O}{\red\textsf{O}} and \eqref{shema}), which is why the
non"=coinciding sets $\boT_{\!\sss\scr A}$, $\boT_{\!\sss\scr B}$ do
always exist. This provides a kind of abstractly deductive
existence's proof for the non"=commutativity,
\qm"=interference""---see sect.~\ref{2slit} further below---and for
the utmost low"=level finality of \qm\ altogether \cite{ludwig3,
gottfried, auletta}. The whys and wherefores of theory do not
require invoking the physical conceptions; cf.~\cite[p.~2]{landau}.

Of no small importance is that this point entails an independence of
the (existence/"!presence of) classical physics or of its formal
deformation, which are yet to be created from the quantum one (cf.~a
selected thesis on page~\pageref{ClassQuant}). In particular, no use
is required of the notion of a certain pretty small---again the
classical/"!physical term---""quantity, \ie, the Plank
constant~$\hbar$ \cite[sect.~6.5]{ulfbeck2}. (Parenthetically, no
the numerical value of this constant matters here; the more so as it
is not dimensionless and its zero limit is not meaningful.) What is
more, the quantum paradigm \eqref{quant}--\eqref{grubo} tells us
that the classical description begins---\ie, we do
create/"!introduce""---with the notions of a micro"=event's average and
of time, whereas these conceptions are still absent at the moment
and in the present work. Similarly for the notions of locality,
causality, the classical event, and the classical object.

\subsection{Physical properties}\label{phys}

Now, the `general physics' \SM\ is mathematized into representations
\eqref{g1} of states $\ket\Xi$ of system $\cal S$. There is,
however, an ambiguity, the source of which is the fact that the
natural/"!classical language also lays claims to a similar
formulation. This refers to the belief in the existence of
mathematics (`bad habit' \cite{mermin}; see also \cite{zeilinger},
\cite[p.~122]{ludwig5}, and \cite{mermin2}) that describes $\cal S$
as an individual object with properties regardless of observation;
an observation that is not a \emph{functioning} attribute of the
mathematics itself. In classical description, it is specified by
definitions: point $\point P$ of a phase space,
$(q,p)$"=coordinatization of the point (manifold), and statistical
distribution~$\varrho(q,p)$.

On the other hand, quantum empiricism provides nothing more to us
besides the ensemble brace and $\ket\Xi$"=states
(pt.~\hyperlink{T}{\red\textsf{T}}). Preordained definienda with
physical contents are unacceptable, \ie, $\cal S$ should not be
conceived as `something with \emph{physical} properties' or as an
`\emph{individual} object' \cite[p.~645]{peres0}, \cite{ludwig3,
ludwig4}. However, since the observational data (in the broadest
sense of the word) may not originate from anywhere but a certain
$\ket\Xi$"=object, there should subsequently create
\begin{itemize}
\item[1)] the very concept of physical objects and properties
 \cite[pp.~211--230]{wart} and
\item[2)] their numerical values/"!characteristics, \ie, the ``physical
 attributes of objects'' \cite[p.~238; N.~Bohr]{schilpp}.
\end{itemize}
This is habitually referred to as elements/"!images of reality
\cite[pp.~194]{greenstein}, \cite[sect.~10.2]{allah},
\cite[sect.~XIII.4.8]{ludwig4}---Bell's ``beables'' \cite{bell}---or
what we have been calling attributes of a physical system.
\begin{itemize}
\item ``The very notion of 'phenomenon' or of 'the appearance of things,'
 \ldots\ is a cognitive and perceptual act of abstraction''
 \\\phantom.\hfill M.~Wartofsky \cite[p.~220]{wart}
\end{itemize}
That is to say, the physical phenomena per~se do not exist
\cite{brukner}, \cite[p.~310]{ludwig2}.

Indeed, the primary ideology of sects.~\ref{physmath} and \ref{S}
tells us that an invasion of physically self"=apparent images into
the theory should be avoided \cite[p.~69]{ludwig1}, because ``quantum
theory not only does not use---it does not even dare to
mention""---the notion of a ``real physical situation''\,''
\cite[p.~198; E.~Jaynes]{greenstein}. Continuing a quotation from
R.~Haag on page~\pageref{haag}, one requires ``the renunciation of
the absolute significance of conventional physical attributes of
objects'' \cite[p.~238; N.~Bohr]{schilpp} and of concomitant and
accustomed logic in reasoning. In fact, we are led to (re)""build
the language of the classical description. Therefore everything,
with no exceptions, should be created mathematically: coordinates,
momenta, energies, optical spectra, device readings,
lengths/"!distances and time, extension and lifetime of objects, the
language of particles, their number/"!numeration (Fock space),
(in)""discernibility/"!individuality (bosons/"!fermions), the notions of
a subsystem of system $\cal S$ (see~\eqref{m2}), and even a notion
of the physical rigor (in reasoning), \etc.

Degrees of freedom, the concepts of the
field/"!body/"!mass/"!inertia/"!interaction, Newtonian mechanics with its
equations and the concepts of the force, interaction, and the
causality of classical events, thermodynamics, the very term `the
classical state', the numerical labels of the space"=time continuum
and numerical forms of what is known as the classical reference
frames---""coordinates on manifolds""---need to also be created.
Once more to underscore, the numerical forms of the classical space
coordinates and the time (\eg, the metric tensor
$g_{\alpha\beta}(x)\,dx^\alpha dx^\beta$) have a quantum empirical
origin. The latter fact is required for carefully posing the
questions of quantum gravity, and it should be noted in passing that
the simultaneity is an ill"=defined term not only in the (general)
relativity theory; in \qt\ it is even worse. In common with the
simultaneous measurability, this term appears to have come from the
classical framework, which is why it is illegal as a
quantum"=theoretical primitive (pr.~\hyperlink{I}{\red\textbf{\textsf{I}}}
and \cite{ludwig1}).

\subsubsection{Waves/particles?}
\addcontentsline{toc}{subsection}{\qquad\qquad Waves/particles?}

The concept of a (non"=elementary) particle, which is conceptually
close to the notion of a subsystem/part, is also a physical
convention and can only arise from the $\ket\Xi$ or its models:
Bose"=condensates, deformation excitations in crystal lattices,
quasi"=particles in a superfluid phase, quantum theories of various
fields (relativistic or non), and more. By a particle is meant here
the classical kinematic conception. ``What do we detect? The presence
of a particle? Or the occurrence of a microscopic event?'', wondered
R.~Haag (2013). H.~Zeh and G.~Ludwig do answer: ``There are no
particles in reality'' \cite{zeh}, ``we must abandon the notion of a
microscopic ``object'', one to which we have been accustomed''
\cite[p.~69]{ludwig1}.

Clearly, the \qft{}s is a subclass of \qm\ rather than its
extension; not that we have yet given a definition of \qm. In
particular, it is common knowledge in \qft\ that there is no logical
way to distinguish a particle from a certain state---""normally, a
vacuum excitation. One word should therefore be used for both. To
this extent, the familiar ``dualism of \ldots\ the \emph{particle
picture} and the \emph{wave picture}'' \cite[p.~28]{popper},
\cite[sect.~7.2]{jammer}, \cite{jammer2} simply disappears.
K.~Popper is rather emphatic concerning this `problem' and puts it,
in his ``thirteen theses'' \cite{popper}, quite rightly in the
following terms: ``\!\emph{the great quantum muddle}'', ``alleged
``duality'' or ``complementarity'', \ldots\ \emph{this} kind of
``understanding'' is of little value'', ``has not the slightest bearing
on either physics, \ldots'', ``fashionable among quantum theorists, \ldots\
a vicious doctrine'', \thelike. As a matter of fact, both the
particles and waves are the classical terms \cite{henry} and, in
quantum language, they turn into the derivatives of the concepts of
state and mixture \eqref{m2}.
\begin{itemize}
\item Like waves, \emph{the particle is already an appearance}---an
 observable one (phenomenology, derivative)---rather than a logical
 primitive or a fundamental substance, which is why it may not exist
 \cite{zeh} prior to theory's principles \cite[p.~762\,(!)]{ulfbeck}.
 Paraphrasing Heisenberg, Haag remarks, in the context of his ``event
 theory'', that ``Particles are the roof of the theory, not its
 foundation'' \cite[p.~300]{haag2}.
\end{itemize}
Both these notions should be superseded by a mathematics of clicks.

The $\fr$"=statistics also falls under observable quantities, and
constant $\textsf{D}$, if declared finite, is an example of an already
created characteristic; the dimension of a state space to come. A
tensorial structure of this space---""compound systems""---also
pertains to the physical properties but we do not touch upon this
point here. As an aside, this compositional structure will provide
the means of distinguishing the aforementioned models under
$\textsf{D}=\infty$.

In other words, the logic of the above constructs prohibits not only
endowing the phraseology `internal state of an individual object
$\cal S$' and `the system is in a (definite) state'
\cite{ludwig3}--\cite{ludwig5}, \cite{muynck} with a meaning, but
also indirectly using its numerical forms. That would work in the
circumvention of empiricism, assuming the a~priori availability of
mathematical structures that do not rest on the state space.
L.~Ballentine remarks in this regard: ``the habit of considering an
individual particle to have its own wave function is hard to break''
\cite[p.~238]{ballentine2}; cf.~``To speak of a single possible
initial apparatus state is pure fantasy'' \cite[pp.~241--242;
N.~Graham]{dewitt}.

\subsection{Interference}\label{2slit}

Let us go on with comments as to involving the
\emph{physics}"=related argumentation to explicate of the quantal
behavior. We have already mentioned above that for this purpose
there is simply no language of physics (sects.~\ref{S},~\ref{phys})
and of mathematics yet (sects.~\ref{makenumb},~\ref{EmpMath}). That
is why analogies of this sort are not only deceptive, but must be
prohibited for exactly the same reasons that accompanied boxes
\eqref{box}. The typical examples in this connection are the
simultaneous measurability mentioned above and the 2-slit
interference \cite{silverman, accardi}.

First and foremost, the two cases---whether one or two slits are
open---are the utterly ``different experimental arrangements''
\cite[p.~236]{ballentine+}, \cite[p.~58]{hren1}:
\begin{equation*}
\SM\!'\ne \SM\!’’\;.
\end{equation*}
There is nowhere to seek a means of their comparison or the
transference of one into another \cite[p.~236]{ballentine+}.
Nonetheless, the classical approach, when opening another slit
\SM$\!_2'$ together with the first one \SM$\!_1'$, does literally
envision properties for \SM$\!’’$ (see sect.~\ref{phys}). In doing
so, the transference method itself---`addition of the two 1"=slit
\SM$\!'$"=phys\-i\-cae' by the rule of arithmetical addition of
statistics \eqref{w1w2}---is meanwhile considered self"=apparent.
Thus, natural questions arise, such as `why/where are the zeroes
coming from, they shouldn't be there'. In accordance with the
aforesaid, everything here is erroneous, including the `natural'
questions. There are no rules at the outset""---(non)""classical and
even quantum, just as there is no addition per~se. An a~priori
assumption that stem from the obvious images for \SM$\!_1'$ and
\SM$\!_2'$ is actually a declaration of the physical properties for
\SM$\!’’$, but they do not follow from anywhere \cite[p.~55]{hren1},
\cite{szabo}, \cite{accardi}. The (illegal) assumption of the
'negligible effect of the which-slit detectors' were mentioned on
p.~\pageref{which} is identical with a declaration of a physical
property; as well as a solenoid's switch-on/off in the
Aharonov--Bohm effect.

Taken alone, the $\varrho$"=distributions""---""separate for
\SM$\!_1'$ and \SM$\!_2'$---are entirely correct observational
pictures, but introducing the rule \eqref{w1w2} is indistinguishable
from `invention' of physics; a logically prohibited operation. Or,
as Slavnov had put it, ``to invent the physical exegesis of a \ldots\
mathematical scheme'' \cite[p.~304]{slavnov2}. ``Our custom of seeing
classical mechanics as a no"=nonsense description of 'reality as it
is' does not seem to be justified. This custom is actually based on
a confusion of categories \ldots'' (W.~de~Muynck \cite[p.~89]{muynck}).
In other words, the mere fact of non"=adherence to this rule means
that the grammatical conjunction of the verbs `to understand/"!deduce'
with the noun `micro"=phenomena' is unacceptable even linguistically.
It is the point~\hyperlink{T}{\red\textsf{T}} that prohibits predefined
(classical) semantics, and this was faithfully summarized by
C.~Fuchs: ``badly calibrated linguistics is the predominant reason
for quantum foundations continuing to exist as a field of research''
\cite[p.~xxxix]{fuchs4}. To figure out or deduce (from mathematics)
that quantal phenomena are unfeasible \cite[p.~111]{silverman} and
are ``\!\emph{absolutely} impossible, to explain in any classical way''
(quotation by Feynman). Just as with the elucidation of the nature
of the quantum state on p.~\pageref{smysl}, any (circum-)""classical
justification or even motivation are guaranteed to fail here, since
they are based upon significant and implicit assumptions.

The classical theory is a theory of \emph{observational} objects
with \emph{observational} properties expressed by
\emph{observational} numbers. We possess none of the three items
required to create the quantum (\,$=$~correct) description
(sect.~\ref{S0}). The adjective `observational' itself is a
linguistic notion of the classical vocabulary (sect.~\ref{observ}).
Accordingly, the description can only be changed into the `to
describe in newly created terms'. A.~Leggett notes \cite{leggett1}
that which is understood as common"=sense should also be changed (see
also \cite[p.~10]{englert}). The reason is clear.
\begin{itemize}
\item Common"=sense operates""---and that is perfectly normal""---with
 observational categories rather than with structureless
 `microscopy' \eqref{rule} and $\cup$"=abstractions of
 sect.~\ref{inv};
\end{itemize}
cf.~Bohr's correspondence principle \cite{jammer}. In effect, we
have dealt with a `fundamental chasm' between the right
description""---``what is really going on?\@''
\cite[p.~12]{englert}---and our ability to give a (naturally
speaking) explanation in terms of these categories:
\begin{quote}
``All our intuition, all our sense of what constitutes concreteness
are based upon our everyday experience, and the terms used to
describe a phenomenon concretely are necessarily drawn from that
experience. There is no indication that such a language could be
used without contradictions for phenomena which are as far removed
from it as those of microscopic physics'' (A.~Messiah. \emph{Quantum
mechanics}).
\end{quote}
The total dismissal of this very concrete has to be at the heart of
quantum reconstructing.

\subsubsection{Detector micro-events}
\addcontentsline{toc}{subsection}{\qquad\qquad Detector micro-events}

For similar reasons, we may not think or envision that a particle in
an interferometer `flies through the slit', `has (not) arrived', ``is
located somewhere in the region of space'' \cite[p.~7]{dirac}, `here,
not there', `now/later', that `the choice of a detector has been
delayed' \cite[Wheeler]{greenstein}, \cite{wheeler}, or that a
``photon \ldots\ interferes \ldots\ with itself'' \cite[p.~9]{dirac}, and
that, generally, `something is flying along a trajectory', and
`something' is a particle at an intuitive understanding. Cf.~Dirac's
description of ``the translational states of a photon'' in sect.~3 of
\cite{dirac}.
\begin{itemize}
\item ``Photons are just clicks in photon detectors; nothing real is
 traveling from the source to the detector'' (ascribed to
 A.~Zeilinger),
\end{itemize}
and this point is supported by all the known varieties of
interferometers. There has to be an amendment here.

The clicks themselves are the clicks not \emph{of
photons/"!particles}, just `merely clicks'. ``[T]he click is no \ldots\
produced by a particle. \ldots\ nothing takes place in the source that
could be a cause of the click \ldots, the genuinely fortuitous click
comes without a cause and has no precursor'' \cite[pp.~758,
765]{ulfbeck}. Nothing really interferes inside interferometers, nor
is anything superposed/"!reinforced. For example, that the path of
`photons' is not represented by trajectories was impressively
demonstrated with the nested Mach--Zehnder experimental setup in the
work \cite{danan}. Asking ``where the photons have been'' \cite{danan}
is also the matter of a certain $\state\alpha$"=distinguishability.
An interferometer""---the entire installation""---should be
perceived as nothing more than a black box \SM---the box
\eqref{box}---""outside of space and time. This is a kind of
irreducible element that produces the only
entity---""distinguishable $\state\alpha$"=events, and no other. The
box contains no `flying particles'. Exempli gratia, none of the
words in the typical sentence `photon propagates a definite path'
are well"=defined. Any assessment of the screen flashes observed
within the interferometer""---`is zero statistics possible in any
spot?'---lacks meaning until the theory's numerical apparatus is
presented.

\begin{comment}\label{13}
Thus Young's interference of the light beams (1803) is
\emph{inherently} the quantum, not the classical effect: a
micro"=events' accumulation is usually termed as the light intensity.
The classical electromagnetism and optics, in an exact sense, do not
explain, only describe, the phenomenon quantitatively with the use
of the numerical concepts of the positive and, which is important,
the negative values of observable strength"=fields $\vec{\bo E}$ and
$\vec{\bo H}$. (The negative numbers are specifically discussed
further in sect.~\ref{92}.) Accordingly, operations of their
addition/"!subtraction `rephrase' the effect in words ``superposing,
suppressing, waves, intensities'', and we call this `the
explanation'. In quantum way of looking at it, all of these concepts
are not yet available, and the phenomenon per~se is no more than
statistics of the `positively accumulative' quantal clicks;
\par\larger 
\begin{itemize}
\item \smaller there are no particles, waves, and subtractions there.
\end{itemize}
\smaller 
The same---visible with the naked eye and `explainable by
waves'---""macroscopic effect would take place, if we had a `laser'
of, say, mono"=energetic very slow electrons (a proposal for
experimentalists). To put it more precisely: a gun or emitter of
something we envision as the `tiny bodily formations' the electrons,
molecules, microbes, \thelike. It is self-evident that we would have
seen the wave"=like manifestation even from a single slit.
\end{comment}

Criticism of the typical (a common event"=space) examination of the
2-slit experiment \cite{feynman} is already abundant in the
literature. See, for example, the works \cite{szabo}\,(!),
\cite[pp.~55--58]{hren1}, \cite[Ch.~2]{hren0}, \cite[sects.~V.1,
VI.1--2]{accardi}, \cite[p.~93]{peres}, \cite{fine}.

By way of a continuation to the last sentence in Remark~\ref{3}, we
add the following. To force an electron-click to happen each time at
the same (or predictable) place is no different from ``completely
describing \emph{everything} that we have'', \ie, from the precise
setting of `the same' and of macro"=context \SM. It is amply evident
that this is a manifest absurdity. Hence it immediately follows that
\emph{the unpredictability of microscopic events must exist in
principle} and macro"=determinism may be only an idealization through
a (math) model: the model description of the \SM\ itself.

Summing up, it is not the quantum interference that requires
interpretative comprehension but its classic `roughening'. In other
words, a scheme that latently presumes the rule \eqref{w1w2} of
extrapolation of what is observed in macro and micro \cite[(!)~last
sentence on p.~101]{wart}. It is this scheme and not the quantum
approach that contradicts the logic and experience. Pauli
characterizes this as habits ``known as `ontology' or `realism'\,''.
More than that, the chief component of
constructs""---\lrceil{observation $\rightarrowtail$ sta\-te$'$}---is
cast out and replaced with \eqref{grubo} under such a
transformation. The \Courier{Data\-Source} object
(p.~\pageref{datainfo}) begins to be identified with
\emph{observational} and \emph{numerical} characteristics (see a
paragraph preceding Remark~\ref{rem4}), while the logic of the
micro-world requires precisely distancing these two concepts, with
no need for the characteristics themselves.

Thus we should not be deriving the physics of one phenomenon from
another \cite[p.~92]{ludwig5} and making (super)""generalizations,
as soon as the incorrectness of the previous derivation method was
established.
\begin{itemize}
\item Quantum\eLab{micro} mathematics is not a physical theory---and
 that is its distinguishing feature""---but rather a single
 syntactical \emph{\emph{(meta)}principle of forming the
 mathematical models being subsequently turned into} (the physical)
 \emph{theories}. This principle is not subject to any physical
 validation.
\end{itemize}
Scott Aaronson was likely the first to advance the line of thought
about non"=physicality. On page~110 of the book \cite{aaronson+}, he
writes that ``it's \textsl{not} a physical theory in the same sense as
electromagnetism or general relativity \ldots\ quantum mechanics sits
at a level \textsl{between} math and physics \ldots\ \textsl{is the operating
system} \ldots''. Fuchs--Peres provoke: ``quantum theory does \emph{not}
describe physical reality'' \cite[p.~70]{fuchs1}.

To create the models, we already have a good deal of
\emph{latitude}: the toolkit $\cal O=\{\scr A,\scr B,\ldots\}$, the
parameter \textsf{D}, the families $\{\boT_{\!\sss\scr A}$,
$\boT_{\!\sss\scr B}$, \ldots\}, numbers $\{\varrho_s\}$ of mixtures
\eqref{m2}, and---thanks to the notion of
covariance~\hyperlink{III}{\red\textbf{\textsf{III}}}---spectra, a
structure of a group, and the concept of (different) representations
of a mathematical structure. This liberty will be subsequently
augmented with the key notions of a mean and of time $t$, and also
with the composite systems, the classical Lagrangians/"!Hamiltonians,
their symmetries, gauge fields, and phenomenological constants. This
is what is currently termed the quantum phenomenology or a
\emph{quantization procedure} of the classical models: the
path"=integrals, $S$"=matrix, et similia.

All that remains is to examine the numerical constituent of quantum
mathematics. The further strategy
(sects.~\ref{Numbers}--\ref{minus}) lies in the fact that the
numbers need to be created at first as a theoretical
concept---""arithmetic""---and then as the `numerical values for
observable quantities'"!---the observations numbers. The sections
\ref{divis}--\ref{92} contain some more explanations along these
lines.

\vbox{
\section{Numeri}\label{Numbers}

\flushright\tiny
\textsl{By number we understand not so much a multitude of\\
unities, as the abstracted ratio of any quantity to another\\
quantity of the same kind, which we take for unity}\\
\textsc{--- I.~Newton} (1707)}
\smallskip\nopagebreak

\subsection{Replications of ensembles}\label{Dupl}

In connection with the emergence of a group, the numerical
representation of brace also undergoes a change, since the
`doubling' of a semigroup into a group through adjoining the
inversions deprives coordinate $\frak a$ of its distinction in
comparison with the inversion $-\frak a$. Given the involution
\begin{equation}\label{unar}
-(-\frak a)=\frak a\;,
\end{equation}
it makes no difference what to call an element and what to call its
inversion in the pair $\{\frak a,-\frak a\}$. This doubling is
formally known as a symmetrization of the commutative associative
law (monoid) \cite{burbaki}. Curiously, under commutativity and
associativity \cite[sect.~1.10]{clifford}, solution to the problem
of embedding is unique \cite[pp.~15--17]{burbaki}, and otherwise, no
solution, in general, exists. There exist the classes (Mal'cev
(1936)), which are not axiomatized by finitely many
$\forall$"=formulas \cite[pp.~216--217]{maltsev}.

The aforesaid is best demonstrated by another way of `numeralizing'
the empiricism, which is realized as the infinite replication of
finite ensembles
\begin{equation}\label{dupl}
\big\{\{\statePsi\}_{\msf n}\{\statePsi\}_{\msf n}\cdots
\big\}=\big\{\{\statePsi\}_{\msf n}\big\}_\infty
\FED\{\statePsi\}_{\msf n\infty}\;.
\end{equation}
That is, empirically, any infinite ensemble is thought of as created
by repetitions (copies) of the finite objects
$\{\statePsi\}_{\msf n}$. It is in this sense, and in this sense
alone, that one should read the writing
$\Sigma\rightsquigarrow\infty$ for the infinity postulate
\eqref{fr}, because at the moment we possess neither the mathematics
nor the topological concepts, such as a passage to the limit
$\Under[1.1]{\sss\Sigma\to\infty}{\text{lim}}$. For example,
expression $\Sigma\times\infty$ can be viewed as a conjunction of
the actual and potential infinity \cite{harin, kleene}. Simply put,
the case in point is not an \emph{axiomatic} act---an imposition of
the \emph{math}"=existence condition for numbers $\{\fr_j\}$ in
\eqref{fr}. The latter has been typically criticized as an idea of
the stable limiting frequencies in \qm\ \cite[pp.~15, 183, 211,
\ldots]{saunders}, \cite[pp.~97--99]{sklar}, \cite{wallace, fuchs3}.
Rather, we claim that the only way to consistently incorporate the
language notions of infinity and of the finite (observational)
numbers in theory---``to cross an abyss'' (Poincar\'e)---is the above
semantics and correspondence between symbols
$\{n_j,\fr_j;\Sigma,\rightsquigarrow,\times,\infty\}$. See also
subsection ``\SL\ and infinity'' in the work \cite{br2}.

In turn, the above mentioned copies $\{\statePsi\}_{\msf n}$ are
replications of the atomic primitive $\{\state\Psi\}_1$. Replication
is thus an operation of the same significance as $\cup$ and
$\uplus$. With this point, the $\obj\Xi$"=brace is characterized by
the `numerical' combination
\begin{equation*}
\eqref{set}\quad\rightarrowtail\quad\big\{[\msf n_1\infty,\,\msf m_1\infty],\;
[\msf n_2\infty,\,\msf m_2\infty],\;\ldots\big\}
\;\rightleftarrows\;\obj\Xi
\end{equation*}
(indices label the $\state\alpha_s$"=primitives), which has been
created from the unitary brace by the scheme
\begin{equation}\label{nm'}
\eqref{min}=\mbig[13]\lgroup\![11]
\begin{array}{c}
\big\{\{\statePsi\}_{\infty'}\{\statePhi\}_{\infty’’}\big\}\\
\DOWN\;\;\;\DOWN\;\;\;\DOWN\;\;\;\DOWN\\
\{\state\alpha\cdots\cdots\state\alpha\}_{\hbox to0ex{$\scs\infty$}}
\end{array}\![8] \mbig[13]\rgroup\quad\rightarrowtail\quad
\mbig[13]\lgroup\![11]
\begin{array}{c}
\big\{\{\statePsi\}_{\msf n\infty}
\{\statePhi\}_{\msf m\infty}\big\}\\
\DOWN\;\;\;\DOWN\;\;\;\DOWN\;\;\;\DOWN\\
\{\state\alpha\cdots\cdots\state\alpha\}
_{\hbox to0ex{$\scs(\msf n+\msf m)\infty$}}
\end{array}\hspace{1.3em} \mbig[13]\rgroup
\quad\rightarrowtail\quad[\msf n\infty,\msf m\infty]\,\state\alpha\;.
\end{equation}
The semigroup union $\obj{\Xi'}\uplus\obj{\Xi’’}$ is then conformed
with the writing
\begin{multline}\label{nm''}
\big\{[\msf n_1'\infty,\,\msf m_1'\infty],\;[\msf n_2'\infty,\,\msf m_2'
\infty],\;\ldots\big\}\uplus\big\{[\msf n’’_1\infty,\,
\msf m’’_1\infty],\; [\msf n’’_2\infty,\,
\msf m’’_2\infty],\;\ldots\big\}=\\
=\big\{[(\msf n_1'+\msf n’’_1)\,\infty,\,
(\msf m_1'+\msf m’’_1)\,\infty],\; [(\msf n_2'+\msf n’’_2)\,\infty,\,
(\msf m_2'+\msf m’’_2)\,\infty],\; \ldots\big\}\;.
\end{multline}
Moreover, the $\msf n$-, $\msf m$"=quantities may be freely thought
of as real ones due to the $\bbR^2$"=continual infinity of ensembles
proven above (sect.~\ref{whyC}). The empirical rationale of this is
apparent; namely, fractions of the arbitrarily large ensembles
$\{\statePsi\,\,\,\statePsi\cdots\}$.

This way of matching the infinity with $\Sigma$"=postulate
automatically inherits translation of associativity/"!commutativity,
because the `percentages', such as $s$ and $w$, just as the rules
\eqref{s'}--\eqref{NU} themselves, do not even emerge. There, these
numbers were originating from $\Sigma$"=postulate, but it, in turn,
was \emph{demolishing the pair $(\varkappa,\frak S)$ itself} in
\eqref{s}: $\frak S\to\infty$. It is clear that, according to
\eqref{nm''}, the semigroup structure $\frak G$ is also inherited,
turning into the addition of the numerical pairs
\begin{equation}\label{oplus}
(\msf n',\msf m')\oplus(\msf n’’,\msf m’’)=
(\msf n'+\msf n’’,\msf m'+\msf m’’)\;.
\end{equation}

Returning to the group, we observe that the `negative symbols'
$(-\msf n,-\msf m)$ might be initially taken as the semigroup
$\frak G$ being duplicated, with equal success and with the same
arithmetical addition $\oplus$, while the positive $(\msf n,\msf m)$
could be thought of as inversions thereof.

Summing up, let us specify the rules of passing to the numerical
representations
\begin{equation}\label{nmPsi}
\eqref{min}\quad\Leftarrow\![11]=\![11]\Rightarrow\quad
\{\pm{\Over[1.5]{\sss\statePsi}{\msf p}},
\pm{\Over[1.5]{\sss\statePhi}{\msf q}}\}\,{\state\alpha}\,,\qquad
(\msf p,\msf q)\in\bbR^2
\end{equation}
and, to avoid ambiguity, replace the binary"=composition symbols
$\{\uplus,\Uplus\}$ with a new symbol~$\Tplus$ for
objects~\eqref{nmPsi}:
\begin{equation*}
\{\Over[1.5]{\sss\statePsi}{\msf p},
\Over[1.5]{\sss\statePhi}{\msf q}\}\,{\state\alpha}\Tplus
\{\Over[1.5]{\sss\statePsi}{\msf n},
\Over[1.5]{\sss\statePhi}{\msf m}\}\,{\state\alpha}\;.
\end{equation*}
The previously dropped primitives $\statePsi$, $\statePhi$ have been
restored here, since they will be further needed for theory's
covariance (sects.~\ref{C},~\ref{C*}), although they are still
unnecessary at the moment.

It is not accidental that we spoke of ``numeral labeling'' the brace
(p.~\pageref{labeling}), since the question of arithmetic on them
had not yet arisen. Although $\fr$"=statistics""---the real
$\bbR$"=numbers""---are already involved, their use was based on an
accustomed perception of the number. In accordance with
pr.~\hyperlink{II}{\red\textbf{\textsf{II}}}, the numerical formalization
of ensemble empiricism should be considered in greater detail.

\subsection{The number as an operator}\label{arif}

Let us take up the `process of manufacturing' the numbers
(pr.~\hyperlink{II}{\red\textbf{\textsf{II}}}). We begin with the
classical simplification
\begin{equation}\label{setdupl}
\frak A=\big\{\{\statePsi\},\,\{\statePsi\,\statePsi\},\,
\{\statePsi\,\statePsi\,\statePsi\},\,
\{\statePsi\,\statePsi\,\statePsi\,\statePsi\},\;\ldots\big\}\;,
\end{equation}
and the notion of the number does not yet appear in any form.

The mathematical abstracting the observation micro-acts is an
employment of the operation $\cup$ and of its closedness (see
sect.~\ref{inv}). For example,
$\{\statePsi\}\cup\{\statePsi\,\statePsi\,\statePsi\}=
\{\statePsi\,\statePsi\,\statePsi\,\statePsi\}$. All the symbols in
\eqref{setdupl}, as well as the character $\cup$, is of course
merely a convention, and they may be changed. By writing
\eqref{setdupl} in symbols like $\{a\,,b\,,c\,,d\,,\ldots\}$ and $+$, this
set should be supplemented with identities as $a+b=c$, $b+b=d$, \ldots,
\ie, with a binary construction~$+$. Then (semi)group and
commutative superpositions arise. Though note that the introducing
the numbers at this point---even if only as symbols""---is not
necessary. It would reduce to re"=notating the set's elements, to be
precise. But the empirical description calls for their unification,
as manifested in the numerical notation like
$\{\statePsi\}\FED1\,\{\statePsi\}$, $\{\statePsi\,\statePsi\}
\FED2\,\{\statePsi\}$, \ldots. It is precisely this pattern that was
implicitly kept in mind in procedures \eqref{min}--\eqref{set} and
\eqref{dupl}--\eqref{nm'}, \ie, when introducing the numbers
$\msf n$ by means of replication of finite or infinite ensembles:
\begin{equation*}
\{\statePsi\cdots\statePsi\}_{\msf n}\;\hhence\;
\msf n\,\{\statePsi\}\,,\qquad
\big\{\{\statePsi\}_\infty\cdots\{\statePsi\}_\infty
\big\}_{\msf n}\;\hhence\;\msf n\,\{\statePsi\}\;.
\end{equation*}
The symbol $\hhence$ should read here as `the same thing as'.
Clearly, the very idea of conjunction of the two
entities""---""empirical brace \eqref{min} and the concept of a
(quantitative and ordinal) number
(sects.~\ref{inv},~\ref{semi})---is not otherwise implementable.
That is to say,
\begin{itemize}
\item we have no any means of translating the aggregates of micro-acts
 $\GOTO{\sss\scr A}$ (\ie,
 macro"=observations~\hyperlink{M}{\red\textsf{M}}) into the numerical
 language, other than through the counting of things \cite{koerner},
 \ie, through the natural"=language notion of the `quantity of
 something':
 \begin{equation}\label{shema'}
 \begin{array}{r@{}c@{}l}
 &\makebox[0em][c]{{\smaller[2]\DOWN$\cdots$\DOWN\;\DOWN}\;
 \text{$\scr A$"=transitions}\;
 {\smaller[2]\DOWN\;\DOWN$\cdots$\DOWN}}&\\&
 \makebox[0em][c]{$\Downarrow\qquad\Downarrow$}\\
 \makebox[0em][r]{$\lceil$quantity of$\rceil$}\;&&
\;\makebox[0em][l]{$\lceil$something$\rceil$\quad(replication)}\\&
 \makebox[0em][c]{$\Downarrow\qquad\Downarrow$}\\
 \lceil\text{numbers}\rceil\;&&\;
 \text{$\lceil\statePsi$"=primitives, ensembles$\rceil$}\\
 \searrow\quad&&\quad\swarrow\\
 &\makebox[0em][c]{$\msf n\,\{\statePsi\}$}
 \end{array}\;.
 \end{equation}
\end{itemize}

Heisenberg stresses an obligatory relationship with ``the natural
language because it is only there that we can be certain to touch
reality'' \cite[pp.~201--202]{heisenberg2}. Otherwise, the
quantitative theory would have nowhere to originate even at the
level of calculating the natural entities by the $\bb N$"=number
tokens. It may be added that arising the numbers is a permanently
present (innate) process of creating the thought objects by an
abstracting in the human brain: the mental suppressing/"!neglecting of
the inessential and identifying the distinguishable
entities""---""perceptual objects---""irrespective of their nature
\cite[``forming collections, \ldots\ putting objects together''; pp.~99,
251]{lakoff}. It is something that humans do all the time without
even realizing they are doing it. This process, say,
\begin{equation*}
\begin{aligned}
&\text{\lrceil{language, words}}\cdots\rightarrowtail
\text{\{\textsf{sheep, $\statePsi$, verb, $\statePsi$, theory, \ldots}}\}
\rightarrowtail\\
&\qquad \{\textsf{a sheep, a $\statePsi$, a verb, \ldots}\} \cdots
\rightarrowtail\lceil\text{something/thing/\ldots/St\"ucke}
\rceil\cdots\rightarrowtail\\
&\qquad\{\bullet\,\textsf{St\"uck}, \bullet\,\textsf{St\"uck}, \bullet\,\textsf{St\"uck},
\bullet\,\textsf{St\"uck}, \bullet\,\textsf{St\"uck},\ldots\} \rightarrowtail\{\bullet, \bullet, \bullet,
\bullet, \bullet;\textsf{St\"ucke}\}\rightarrowtail\\
&\qquad\{1, 1, 1, 1, 1;\textsf{St\"ucke}\}\rightarrowtail
5\,\textsf{St\"uck} \rightarrowtail 5\;\textsf{\sout{St\"uck}}
\rightarrowtail 5 \rightarrowtail \text{\lrceil{abstraction
\textsf{5}}}\;,
\end{aligned}
\end{equation*}
is akin to Cantor's introducing the concept of a~Menge \cite[Ch.~1,
sect.~1.1]{stoll} and has \emph{no} the mathematical (math"=logic)
nature. Rather, the math of numbers does originate from it
\cite[Ch.~3]{lakoff}; see also
\cite[sect.~2.4.5.1~\textsc{arithmetic}]{gray}.

Incidentally, the `inessential and identifying' just mentioned have
the nature just like the `the same' in sect.~\ref{SS}. It is with
these notions---a key feature of the natural/"!physical language and
of speech---that any abstracting begins: the `abstracting from \ldots'.

On the other hand, the numerical tokens are `affixed' not only to
the `atom' $\{\statePsi\}$, but also to other objects, \emph{any} at
that; for more details, see Remark~\ref{units} further below.
Therein lies the primary meaning of this still proto"=mathematical
concept \cite[``Psychologie du nombre'']{santerre}. One might even
say, a definition according to which this notion has been conceived
(`20~St\"uck', `half an hour', \ldots) and is being used universally.
Here are a few examples:
\begin{equation}\label{real}
\begin{array}{c}
\{\statePsi\,\statePsi\,\statePsi\}\equiv3\,\{\statePsi\}\,,\qquad
\{\state\Theta\,\state\Theta\,\statePhi\,\statePhi\}\equiv
2\,\big(\{\state\Theta\}\cup\{\statePhi\}\big)\\[1ex]
\{\statePhi\,\statePsi\}\mathrel{\Over{2}{\rightarrowtail}}
\{\statePhi\,\statePsi\,\statePhi\,\statePsi\}\,,\qquad
a\mathrel{\Over{3}{\rightarrowtail}}3\,a\,,\qquad
c\mathrel{\Over{1}{\rightarrowtail}}1\,c
\end{array}\;.
\end{equation}
Accordingly, in between the elements, there arise identities such as
$2\,b\equiv4\,a$, $3\,a\equiv c$, $1\,c\equiv c$. In other words, as we
complete simplification \eqref{setdupl},
\begin{itemize}
\item While abstracting the empirical contents of the number entities
 into math"=symbols, they should be defined as unary operations
 \{$\oper[-3]1$, $\oper[-3]2$, $\ldots$,
 $\widehat{\raisebox{0.4ex}{$\scs3$}\!\!\!\Smaller[2]/\!\!\!
 \raisebox{-0.4ex}{$\scs4$}}$, $\ldots$, $\oper[-1]\pi$, \ldots\} that
 take action at $\frak A$-set \eqref{setdupl} as automorphisms:
 $\{\oper[-3]2\,b=\oper[1]4\,a,\, \oper[-3]1\,c=c\,,\ldots\}$.
\end{itemize}

That said, replication is formalized as an operator $\oper[-1]n$
with its numerical symbol~$\msf n$:
\begin{equation}\label{real'}
\psi\mathrel{\Over{\skew1\widehat n}{\mapsto}}\msf n\,\psi\,,\qquad
\psi,\, \msf n\,\psi\in\frak A\,,\qquad\msf n\in\bbR\;,
\end{equation}
where $\psi$ is understood to be any (sub)""ensemble/(sub)set. In
the language of the \zf"=theory, $\msf n\,\psi$ would be formally
organized as an ordered pair $(\msf n,\psi)\DEF\{\{\msf n\},
\{\msf n,\psi\}\}$ \cite{kurat}, where $\msf n$ is a cardinality of
a set consisting of copies of the object/"!set~$\psi$. We will refer
to these facts as the implementation of a replication operator by
numbers.

Attention is drawn to the fact that the case in point at the moment
\emph{is not} a math"=logical \emph{definitio/"!formalization} of the
concept of a number, such as \eqref{ordinal}, but is an introducing
of what is understood by number in the empirical/"!physical theory
(\hyperlink{II}{\red\textbf{\textsf{II}}}). For example, Chomsky says with
regard to this point: ``When multiplying numbers in our heads, we
depend on many factors beyond our intrinsic knowledge of arithmetic''
\cite[p.~3]{chomsky+}.

\subsection{\qm\ and arithmetica}\label{qm+A}

We immediately observe the following properties.

The operators are applicable to each other. Being a family
$\{\oper[-1]n\,,\oper[-1]m,\,\oper p\,,\ldots\}$, they are closed with
respect to their composition
$\oper[-1]n\,(\oper[-1]m\,\psi)=(\oper[-1]n\circ
\oper[-1]m)\,\psi=\oper p\,\psi$, and among them, there is an
identical operator $\oper[-4]{\mathds1}\,\psi=\psi$. The empirical
meaning of the concept---a fractional portion of ensemble
(see~\eqref{dupl})---""requires that for each $\oper[-1]n$ there
exists its inversion $\oper[-1]n^{\sm1}$. Hence, the composition of
replications $\oper[-1]n\circ\oper[-1]n^{\sm1}$ must return the
former `quantity':
$(\oper[-1]n\circ\oper[-1]n^{\sm1})\,\psi=\oper[-4]{\mathds1}\,\psi$.
Therein lies ``the actual meaning of division. \ldots\ this [operator]
construction really corresponds to division'' \cite[p.~37]{baez+}. By
virtue of the fact that family
$\{\oper[-1]n\,,\oper[-1]m\,,\oper p\,,\ldots\}$ provides automorphisms of
the $\frak A$-set, these operators entail the associative identities
$((\oper[-1]n\circ\oper[-1]m)\circ\oper p)\,\psi=(\oper[-1]n\circ
(\oper[-1]m\circ\oper p))\,\psi$. This point is a property, and it
has a proof \cite[sect.~I.1.2]{kurosh}. The common nature of the
replication and of the $\cup$"=union also signifies that there are
relations in place that mix the actions of the unary $\oper[-1]n$'s
and the binary union of ensembles. At a minimum, suffice it to
define the action of the replicator on a `$\cup$-sum' of
replications. Clearly, the case in point is a distributive
coordination of $\circ$ and~$\cup$:
\begin{equation*}
\oper p\,(\oper[-1]n\,\psi\cup\oper[-1]m\,\psi)=
(\oper p\circ\oper[-1]n)\,\psi\cup(\oper p\circ\oper[-1]m)\,\psi\;.
\end{equation*}

We now observe that the indication of $\psi$ everywhere in the
identities above loses the necessity, and the $\psi$"=label becomes a
semblance of a dummy index or the unit symbol [\textsf{kg}], which can
be changed. As we omit it, the theory is freed of $\psi$ as a
`calculation unit'. Then the last relation, as an example, acquires
the form of a property between the operator $\msf n$"=symbols
\eqref{real'}, if $\{\cup,\circ\}$ are replaced with the symbols of
binary operations $\{+,\times\}$:
\begin{equation}\label{distr}
\msf p\times(\msf n+\msf m)=
(\msf p\times\msf n)+(\msf p\times\msf m)\;.
\end{equation}
Supplementing this relation with other empirically determining
properties, one infers that the \emph{unary} operationality of
$\oper[-1]n$"=replications \eqref{real'} is \emph{indistinguishable}
from the \emph{binary} operationality on their $\msf n$"=symbols. The
latter, in turn, acquires the multiplicative structure of a
commutative group
\begin{equation}\label{mult}
\msf n\times\msf m=\msf m\times\msf n\,,\qquad
(\msf n\times\msf m)\times\msf p=
\msf n\times(\msf m\times\msf p)\,,\qquad\msf n\times1=
\msf n\,,\qquad\msf n\times\msf n^{\sm1}=1\;,
\end{equation}
and, as for the addition~$+$, it is already binary and commutative
due to properties of~$\cup$ (sect.~\ref{inv}):
\begin{equation}\label{Ax}
\msf n+\msf m=\msf m+\msf n\,,\qquad(\msf n+\msf m)+\msf p=
\msf n+(\msf m+\msf p)\,,\qquad\msf n+0=\msf n\;.
\end{equation}
Incidentally, the 3-term multiplicative associativity relation in
\eqref{mult} has the same operatorial nature and origin as
operations $\{\cup,\bo\cup\}$ do in \eqref{U}. We have already
commented on the additive analog to this situation""---a
determinative structure of the binary operation""---after formula
\eqref{g+}.

It is also clear that rules \eqref{distr}--\eqref{Ax} must be
supplemented with the concept of a negative number
\begin{equation}\label{n-}
\msf n+(-\msf n)=0\;,
\end{equation}
for such numbers have been fully justified in the superposition
principle.

After having acquired properties \eqref{distr}--\eqref{n-}---call
them \emph{arithmetica}, symbols $\{\msf n,\msf m,\ldots\}$ turn into
abstract numbers, although their operator genesis does not go away
and is yet to be involved. This is where a full list of requirements
for the concept of a real number should be added, and which have to
do with ordering~$<$, completeness/"!continuality, \emph{and} their
relations with algebraic rules \eqref{distr}--\eqref{n-}. We will
take that this is conducted axiomatically \cite[pp.~35--38]{zorich},
although the algebraic constituent of this `axiomatics', as we have
seen, is not axiomatical but deducible from empiricism.
Multiplication~$\times$, and also the subsequent
$\odot$"=multiplication of $\bbC$"=numbers \eqref{riC}, is a most
nontrivial part in deriving the structure `the arithmetic'.

As an outcome, we reveal an essential asymmetry in the genesis of
the standard binary structures~$+$ and~$\times$
(cf.~\cite[p.~60]{lakoff}), and thereby a greater primacy of
\qm"=consideration even over the (seemingly
self"=evident)\eLab{comment} arithmetic. Indeed, binarity may come
only from operation $\cup$, which is primordially unique and,
thereby, is inherited only to the one natural
prototype""---""addition.
\begin{itemize}
\item Multiplication is not featured in the superposition principle, nor
 does it arise directly as a binary structure. The absence of a
 multiplication symbol in \eqref{s1}--\eqref{s2} is no accident.
\end{itemize}

The multiplication originates in the closedness of replications
$\oper[-1]n\circ\oper[-1]m$, and they are required according to the
\hyperlink{M}{\red\textsf{M}}"=paradigm \eqref{macro}. In effect, any
non"=operatorial way of introducing the $\msf n$"=numbers is not
self"=evidence for empiricism. An operator nature of the number is
precisely that which gives rise to the second binary operation.
Moreover, without such a comprehension of the number the `linear
nature' of \qm\ (sect.~\ref{lvs}) will remain axiomatic at all times
and, as will be seen below, quantum foundations will be doomed to
never"=ending interpreting the mathematical symbols. However, the
pure axiomatic declaration of arithmetic \eqref{distr}--\eqref{n-}
will, in one way or another, require a (reciprocal to
\eqref{shema'}) treatment of the number in a context of `the
quantity of what?', while its empirical pre"=image always appears in
the pair \lrceil{the quantity of} \tplus\ \lrceil{some\-thing}.
Another way to put it is that,
\begin{itemize}
\item in the foundations of theory, there arises a predecessor/"!analog to
 the notion of a physical unit,
\end{itemize}
though the ultimate description is a description in terms of binary
structures \eqref{distr}--\eqref{n-}. It is carried out by
dropping/"!attaching the symbols such as $\psi$, which is a quantum
generalization to the independence of a physical theory from the
measurement units.

Certainly, when formalized, the $\oper[-1]n$"=replication and its
binary $\msf n$-twin become universally abstract. For example, the
$\oper[-1]n$"=operator \eqref{real'} may be applied to the quantum
case in which the object $\psi$ has already an internal structure
associated with the presence of $\statePsi,\statePhi$"=primitives.
This changes no the essence of the matter\eLab{QMdim}. Another
example is when numbers $\msf n$ give birth to the really observable
quantities. See also sect.~\ref{izmer}, Remark~\ref{units}, and
additional discussion in sect.~\ref{minus}. Let us now proceed from
the fact that the comprehension/"!relation of the number and its
operator has been formalized as described above. This is the set of
`axioms' \eqref{distr}--\eqref{n-}.

As concerns the philosophical literature, the issue of numbers was
likely discussed \cite{russell, shapiro, knott} (see also
\cite{koerner} and non"=philosophical book \cite{santerre}), and it
would be appropriate to quote T.~Maudlin: ``\ldots\ numbers: they can be
added to one another, \emph{perhaps} multiplied by one another, \ldots.
But it is typically obscure what sort of \emph{physical} relation
these mathematical operations could possibly represent''
\cite[p.~138; first emphasis ours, second in original]{maudlin}.
Cf.~Einstein's remarks regarding the ``concepts and propositions'' and
``the series of integers'' on p.~287 in~\cite{einstein2}.

\subsection{`2-dimensional' numbers}\label{C}

A number in and of itself, as a replication operator, may be applied
to any ensemble and to anything at all. However, in quantum case,
the `upper' primitives are attached to every `lower'
$\state\alpha$"=event. These primitives, as was noted above, have to
be discarded. At the same time, the minimal structure associated
with the homogeneous array $\{\state\alpha_s\cdots\state\alpha_s\}$
as a whole is a unitary brace $\{\Over[1.5]{\sss\statePsi}{\msf n},
\Over[1.5]{\sss\statePhi}{\msf m}\}\,\state\alpha_s$ containing two
`upper' primitives $\statePsi$, $\statePhi$. Their order, however,
is arbitrary there. That is to say, given
$(\msf n,\msf m)\,\state\alpha$ there are two quite equal objects
$\{\Over[1.5]{\sss\statePsi}{\msf n},
\Over[1.5]{\sss\statePhi}{\msf m}\}\,\state\alpha$ and
$\{\Over[1.5]{\sss\statePhi}{\msf n},
\Over[1.5]{\sss\statePsi}{\msf m}\}\,\state\alpha$ that are subjected
to a replication. Each of them should be in a relationship (see
sect.~\ref{inv}) to any other brace \eqref{nm'}, which is already
apparent in the example of `$1$"=dimensional' versions
$(\msf n,0)\,\state\alpha$ and $(\msf n',0)\,\state\alpha$. We mean
that for each pair
$\{(\msf n,0)\,\state\alpha,(\msf n',0)\,\state\alpha\}$, there always
exists the number $\msf m$ such that $\oper m\,
(\msf n,0)\,\state\alpha=(\msf n',0)\,\state\alpha$, \ie,
$\msf m\times\msf n=\msf n'$.

As in the classical case \eqref{real}, the sought-for
generalizations of replicators are the transitive automorphisms on
\emph{unitary} $\state\alpha$"=brace \eqref{nmPsi}, but they
\emph{are not} abstract and \emph{not} arbitrary. They are strictly
bound to the declared meaning of the number: $\oper N$"=operation of
creating the copies. Therefore, by virtue of the equal rights of
$\statePsi$ and $\statePhi$, it is imperative to bring the two
1-fold copying acts $\oper N\{\Over[1.5]{\sss\statePsi}{\msf n}\,,
\Over[1.5]{\sss\statePhi}{\msf m}\}\,\state\alpha$ and
$\oper M\{\Over[1.5]{\sss\statePhi}{\msf n}\,,
\Over[1.5]{\sss\statePsi}{\msf m}\}\,\state\alpha$ into play, which
differ in the permutation of primitives
$\statePsi\rightleftarrows\statePhi$. This point will determine a
quantum extension of the replication.

As a result, since we have nothing but the copying $\oper N$ and
`union' $\Tplus$, the most general transformation of the brace
$\{\Over[1.5]{\sss\statePsi}{\msf n},
\Over[1.5]{\sss\statePhi}{\msf m}\}\, \state\alpha$ into (any) brace
$\{\Over[1.5]{\sss\statePsi}{\vbox to1.5ex{}\msf n}{}',
\Over[1.5]{\sss\statePhi}{\vbox to1.5ex{}\msf m}{}'\}\,
\state\alpha$, that has been in a quantum"=replication relation with
it, is determined by the rule
\begin{equation}\label{nmU}
\{\Over[1.5]{\sss\statePsi}{\msf n}\,,
\Over[1.5]{\sss\statePhi}{\msf m}\}\,\state\alpha\quad
\Over[1.5]{(\widehat{N\!\!\!,M})}
{\scalebox{2.5}[1]{$\rightarrowtail$}} \quad
\{\Over[1.5]{\sss\statePsi}{\vbox to1.3ex{}\msf n}{\cramp'},
\Over[1.5]{\sss\statePhi}{\vbox to1.3ex{}\msf m}{\cramp'}\}\,
\state\alpha \qquad\hence\qquad
\{\Over[1.5]{\sss\statePsi}{\vbox to1.3ex{}\msf n}{\cramp'},
\Over[1.5]{\sss\statePhi}{\vbox to1.3ex{}\msf m}{\cramp'}\}\,
\state\alpha= \oper N\{\Over[1.5]{\sss\statePsi}{\msf n},
\Over[1.5]{\sss\statePhi}{\msf m}\}\,\state\alpha\;\Tplus\;\oper M
\{\Over[1.5]{\sss\statePhi}{\msf n},
\Over[1.5]{\sss\statePsi}{\msf m}\}\,\state\alpha\;.
\end{equation}
This is the quantum version of operators
\eqref{real}--\eqref{real'}, and the foregoing ideology of
$\oper N$"=operators and of liberation from the $\statePsi$"=symbols
remains in force and entails the following. The numeral
implementation of replicating the unitary brace \eqref{nmPsi}, along
with the $(\msf n,\msf m)$"=representation of itself, is also
determined by a certain pair $(\msf N,\msf M)\in\bbR^2$, \ie, by an
operator symbol $(\widehat{N\!\!,\!M})$.

The aforesaid means that the numerical form $(\msf n,\msf m)\;
\Over{\sss(\widehat{N,M})}
{\scalebox{2.2}[1]{$\rightarrowtail$}}\;(\msf n',\msf m')$ of
transformation \eqref{nmU} is indistinguishable from a composition
of pairs
\begin{equation*}
(\msf N,\msf M)\odot(\msf n,\msf m)=(\msf n',\msf m')\;,
\end{equation*}
where $\odot$ is a designation for the new binary operation. Its
resultant structure is derived from the arithmetical nature
\eqref{mult} of the 1"=dimensional replication \eqref{real} described
above, \ie, from the rules
\begin{equation}\label{NM}
\oper N\{\Over[1.5]{\sss\statePsi}{\msf n},
\Over[1.5]{\sss\statePhi}{\msf m}\}\,\state\alpha=
\{\msf N{\,\Over[1.7]{\sss\statePsi}{\mathstrut\times}\,}\msf n,\;
\msf N{\,\Over[1.7]{\sss\statePhi}{\mathstrut\times}\,} \msf m\}\,
\state\alpha\,,\qquad\oper M\{\Over[1.5]{\sss\statePhi}{\msf n},
\Over[1.5]{\sss\statePsi}{\msf m}\}\,\state\alpha=
\{\msf M{\,\Over[1.7]{\sss\statePhi}{\mathstrut\times}\,}
\msf n,\;\msf M{\,\Over[1.7]{\sss\statePsi}{\mathstrut\times}\,}\msf m\}\,
\state\alpha\;.
\end{equation}
Here, a positivity/"!negativity of symbols $(\msf n,\msf m)$ in
\eqref{nmPsi} should also be taken into account. Having regard to
the foregoing, rules \eqref{nmU}--\eqref{NM} generate the Ansatz
\begin{equation}\label{ansatz}
(\msf N,\msf M)\odot(\msf n,\msf m)=(\pm\msf N\,\msf n
\pm\msf M\,\msf m,\;\pm\msf N\,\msf m\pm\msf M\,\msf n)\;,
\end{equation}
wherein all four signs~$\pm$ are independent of each other, and the
$\smallc({\times}\smallc)$"=multiplication of 1"=dimensional numbers
in \eqref{mult} and \eqref{NM} have been re"=denoted by the habitual
standard $\msf N\,\msf m\DEF\msf N\times\msf m$. What should the
pair"=composition rule \eqref{ansatz} be?

As was the case previously, the just emerged binarity for $\odot$
should inherit""---due to its operator origin""---""associativity,
existence of unity $\mathds1$, and of inversions. Namely, if the
$(\msf n,\msf m)$"=pairs are identified with the notation \eqref{g+}
according to the convention
\begin{equation}\label{nma}
(\msf n,\msf m)\FED\frak a\;,
\end{equation}
then the following properties should be declared:
\begin{equation}\label{m+}
(\frak a\odot\frak b)\odot\frak c=\frak a\odot(\frak b\odot\frak c)\,,
\qquad\frak a\odot \mathds1=\frak a\,,\qquad\frak a\odot{\frak a}^{\sm1}=
\mathds1\;.
\end{equation}
From \eqref{nmU}--\eqref{NM} it is not difficult to see that the
combining \eqref{m+} with \eqref{g+} leads to a distributive
coordination of operations $\oplus$ and $\odot$:
\begin{equation}\label{m++}
\frak c\odot(\frak a\oplus\frak b)=
(\frak c\odot\frak a)\oplus(\frak c\odot\frak b)\;.
\end{equation}
However, the direct examination of this property shows that Ansatz
\eqref{ansatz} satisfies it automatically. More than that, we can
consider this Ansatz even with parameters
$\{\alpha,\beta,\gamma,\delta\}$ instead of $(\pm)$"=signs:
\begin{equation*}
\frak a\odot\frak b=(\msf N,\msf M)\odot(\msf n,\msf m) =
(\alpha\,\msf N\,\msf n
+\beta\,\msf M\,\msf m,\;\gamma\,\msf N\,\msf m+\delta\,\msf M\,\msf n)\;.
\end{equation*}
Then the straightforward calculation shows that distributivity
\eqref{m++} holds under the arbitrary
$\{\alpha,\beta,\gamma,\delta\}$.

In turn, the examination of associativity""---the first equality in
\eqref{m+}---under the same meaning for
$\{\alpha,\beta,\gamma,\delta\}$ yields $\alpha=\gamma=\delta$ and
free $\beta$. Returning to the $(\pm)$"=values of these parameters,
this associativity particularizes \eqref{ansatz} into the expression
\begin{equation*}
(\msf N,\msf M)\odot(\msf n,\msf m)=
\pm(\msf N\,\msf n\pm\msf M\,\msf m, \msf N\,\msf m+\msf M\,\msf n)\,;
\end{equation*}
now, with two independent signs $\pm$. Moreover, in passing we
reveal the commutativity
\begin{equation}\label{m+++}
\frak a\odot\frak b=\frak b\odot\frak a\;,
\end{equation}
though it was not presumed prior to that.

The search for unity $\mathds1$ and subsequent finding an inversion
of the element $(\msf n,\msf m)$ yield:
\begin{equation*}
\mathds1=(\pm1,0)\,,\qquad(\msf n,\msf m)^{\sm1}=
\mbig[7](\frac{\msf n}{\msf\Delta},
-\frac{\msf m}{\msf\Delta}\mbig[7])\,,\qquad\msf\Delta\DEF
\msf n^2\pm\msf m^2\;.
\end{equation*}
Both the $(\pm)$"=symbols continue to be independent here. The choice
$\msf\Delta=\msf n^2-\msf m^2$ results in the absence of inversions
$(\msf n,\msf n)^{\sm1}$. This is in conflict with the group
property \eqref{m+} and also causes the unmotivated exclusivity of
the unitary brace $\{\Over[1.5]{\sss\statePsi}{\msf n},
\Over[1.5]{\sss\statePhi}{\msf n}\}\,\state\alpha$. There remains the
case $\msf\Delta=\msf n^2+\msf m^2$, and it reduces the scheme to
the form
\begin{equation*}
\mathds1=\pm(1,0)\,,\qquad(\msf N,\msf M)\odot(\msf n,\msf m)=
\pm(\msf{N\,n}- \msf{M\,m}, \msf{N\,m}+\msf{M\,n})
\end{equation*}
with a single symbol $\pm$. It is a simple matter to see that the
choice of sign~$+$ or $-$ leads to the models that are isomorphic in
regard to which of representatives $(+1,0)$ or $(-1,0)$ should be
assigned for the identical replication $\oper[-3]{\mathds I}$. By
virtue of \eqref{unar} it does not matter, and we declare
\begin{equation}\label{riC}
\boxed{\mathds1\DEF(1,0)\,,\qquad(\msf N,\msf M)\odot(\msf n,\msf m)=
(\msf{N\,n}-\msf{M\,m},\msf{N\,m}+\msf{M\,n})}\;.
\end{equation}
This is nothing more nor less than the canonical multiplication of
complex numbers $\msf n+\ri\cdot\msf m=\frak a\in\bbC$, if the
following identifications are performed:
\begin{equation}\label{riC+}
(1,0)\rightleftarrows\mathds1\,,\qquad(0,1)\rightleftarrows\ri\,,\qquad
\{\oplus,\odot\}\rightleftarrows\{+,\cdot\}\,,\qquad(\msf n,\msf m)
\rightleftarrows(\msf n+\ri\cdot\msf m)\;.
\end{equation}

Notice that the known fully matrix (over $\bbR$) equivalent to
\eqref{riC}
\begin{equation*}
(\msf n+\ri\cdot\msf m)\mapsto(\msf n,\msf m)\mapsto
{\mbig[7](}\![6]\array{c}\msf n\\\msf m\endarray\![6]{\mbig[7])}
\mapsto
{\mbig[7](}\![6]\array{r@{}r}\msf n&\;-\msf m\\\msf m&\msf n\endarray
\![6]{\mbig[7])}\,,\qquad
{\mbig[7](}\![6]
\array{r@{}r}\msf n'&\;-\msf m'\\\msf m'&\msf n'\endarray
\![6]{\mbig[7])}=
{\mbig[7](}\![6]\array{r@{}r}\msf N&\;-\msf M\\\msf M&\msf N\endarray
\![6]{\mbig[7])}\circ{\mbig[7](}\![6]
\array{r@{}r}\msf n&\;-\msf m\\\msf m&\msf n\endarray\![6]{\mbig[7])}
\end{equation*}
does directly reflect the above ascertained operator essence
\begin{equation*}
(\widehat{n'\!\!,m'})=(\widehat{N\!\!, M})\circ(\widehat{\,n,m\,})
\end{equation*}
of both the number multiplication $\odot$ and the $\bbC$"=number
itself.

In view of the paramount importance of the $\bbC$"=number field in
\qt\ \cite{benioff, baez, goyal}, let us provide additional
substantiations to the rigidity of the emergence of this specific
number structure, \ie, of the axiom collection \eqref{g+},
\eqref{m+}--\eqref{riC}. Among other things, the transpositions
$\statePsi\rightleftarrows\statePhi$ used above fit more general
reasoning.

\subsection{Involutions and $\tilde\bbC^*$-algebra}\label{C*}

Apart from a freedom in ordering the primitives
$\statePsi\rightleftarrows\statePhi$ in brace
$\{\Over[1.5]{\sss\statePsi}{\msf n},
\Over[1.5]{\sss\statePhi}{\msf m}\}\,\state\alpha$, there is one more
arbitrariness: reappointing them
($\statePsi\rightarrowtail\state\Theta$, \ldots) as elements of the set
$\boT$. However, no physics predetermines any of these degrees of
freedom. For, if other ingoing $\boT$"=elements $\state\Theta$,
$\state\Omega$ were present in \eqref{split'} instead of
$\statePsi$, $\statePhi$, then the theory of semigroup $\frak G$,
strictly, should be declared the segregated theories
$\frak G_{\sss\state{\!\Psi\!}\,\,\state{\!\Phi\!}}$,
$\frak G_{\sss\state{\!\Theta\!}\,\,\state{\!\Omega\!}}$, \etc. It is
clear that the labeling the theories, or a family thereof, is a
manifest absurdity, and they should be thus factorized with respect
to all kinds of ways to label them by $\boT$"=primitives. The
liberation from the $\statePsi,\statePhi$"=icons and reconciliation
of the result with pt.~\hyperlink{R+}{$\red\msf R^\textsf{+}$}
(p.~\pageref{R+}) are then performed by the scheme \lrceil{primitive
has changed} $\rightarrowtail$ \lrceil{a number character is
changing}.

Inasmuch as declaring the $\{\statePsi, \statePhi, \state\Theta,
\ldots\}$ to be ingoing primitives in \eqref{split'} is a replacement
of one to another, any such an appointment boils down to
permutations of no more than \emph{pairs}, with two types
(inner/outer):
\begin{equation}\label{invol}
\oper[-1]{\bo\gimel}_{\sss\Psi\!\Phi}\,\,{:}\quad
(\statePsi,\statePhi)\:\Over[1.8]{\sss\state{\!\!\Psi\!\!}\,
\leftrightarrow\, \state{\!\!\Phi\!\!}}
{\scalebox{2.3}[1]{$\rightleftarrows$}}\:(\statePhi,\statePsi)\,,\qquad\qquad
\Aleph_{\!\sss\Phi\!\Theta}\,\,{:}\quad
(\statePsi,\statePhi)\:\Over[1.8]{\sss\state{\!\!\Phi\!\!}\,
\leftrightarrow\, \state{\!\!\Theta\!\!}}
{\scalebox{2.3}[1]{$\rightleftarrows$}}\:(\statePsi,\state\Theta)\;.
\end{equation}
However, it is immediately obvious that these reappointments change
nothing in the $\cup$"=relationships between \eqref{split'} and are
defined by the structural relations
$\oper[-1]{\bo\gimel}_{\smash{\sss\Psi\!\Phi}}^{\smash2}=
\oper[-3]{\mathds I}$, $\Aleph_{\smash{\!\sss\Phi\!\Theta}}^{\smash2}=
\oper[-3]{\mathds I}$. Then the need to indicate the primitives
themselves, as required, is eliminated, and their symbols may be
thrown away, if semigroup $\frak G$ is properly furnished with the
two abstract involutions $\oper[-1]{\bo\gimel}$ and $\Aleph$. The
$\frak G$ itself, of course, possesses also involution \eqref{unar}
that turns it into the group \itsf{H}, but this involution has
already had a numerical representation \eqref{nmPsi} by signs $\pm$.
To be precise, it suffices to identify here the term ``numerical''
with the group arithmetic of the $\oplus$"=addition \eqref{g+} coming
from the superposition principle realized on pairs
\eqref{oplus}--\eqref{nmPsi}. Therefore the operators' actions
\eqref{invol} should be carried over onto objects defined in
precisely this manner; nothing more needs to be assumed.

Operator $\oper[-1]{\bo\gimel}_{\sss\Psi\!\Phi}$ is immediately
translated into a numerical form independently of the property that
the objects $\{\Over[1.5]{\sss\statePsi}{\msf n},
\Over[1.5]{\sss\statePhi}{\msf m}\}\,\state\alpha$ form a
(semi)group. Indeed, since the swap
$\statePsi\rightleftarrows\statePhi$ in the unordered pair
\begin{equation*}
\oper[-1]{\bo\gimel}_{\sss\Psi\!\Phi}:\qquad
\{\Over[1.5]{\sss\statePsi}{\msf n},
\Over[1.5]{\sss\statePhi}{\msf m}\}\rightarrowtail
\{\Over[1.5]{\sss\statePhi}{\msf n},
\Over[1.5]{\sss\statePsi}{\msf m}\}
=\{\Over[1.5]{\sss\statePsi}{\msf m},
\Over[1.5]{\sss\statePhi}{\msf n}\}\qquad\cdots
\end{equation*}
(the $\state\alpha$"=label is dropped here as superfluous) is
indistinguishable from the permutation of numbers
$\msf n\rightleftarrows\msf m$, the symbols $\statePsi$ and
$\statePhi$ may be thrown away, organizing the numbers themselves
into the ordered pairs
\begin{equation*}
\cdots\quad\hence\quad(\msf n,\msf m)
\mathrel{\Over{\oper{\scs\bo\gimel}}{\rightarrowtail}}
(\msf m,\msf n)\;.
\end{equation*}
When required, the $\state\alpha$"=symbol returns hereinafter.

Let us now proceed to the outer involution
$\statePhi\rightleftarrows\state\Theta$ in \eqref{invol}:
\begin{equation*}
\Aleph_{\sss\Phi\!\Theta}:\qquad\{\Over[1.5]{\sss\statePsi}{\msf n},
\Over[1.5]{\sss\statePhi}{\msf m}\}\rightarrowtail
\{\Over[1.5]{\sss\statePsi}{\msf n},
\Over[1.5]{\sss\state\Theta}{\msf m}\}\;.
\end{equation*}
It is indifferent to the (first) $\statePsi$"=element of the pair,
and, extracting it by the rule
\begin{equation*}
\{\Over[1.8]{\sss\statePsi}{\msf n},
\Over[1.8]{\sss\statePhi}{\msf m}\}=
\{\Over[1.8]{\sss\statePsi}{\msf n},
\Over[1.8]{\sss\statePhi}{0}\} \Tplus
\{\Over[1.8]{\sss\statePsi}{0},
\Over[1.8]{\sss\statePhi}{\msf m}\}\;,
\end{equation*}
the question boils down to finding a representation to the
transformations
\begin{equation*}
\big(\{\Over[1.8]{\sss\statePsi}{\msf n},
\Over[1.8]{\sss\statePhi}{0}\} \Tplus
\{\Over[1.8]{\sss\statePsi}{0},
\Over[1.8]{\sss\statePhi}{\msf m}\}\big)\;\rightarrowtail\;
\big(\{\Over[1.8]{\sss\statePsi}{\msf n}{\cramp'},
\Over[1.8]{\sss\statePhi}{0}\} \Tplus
\{\Over[1.8]{\sss\statePsi}{0},
\Over[1.8]{\sss\state\Theta}{\msf m}{\cramp'}\}\big)\qquad
(\Over[1.5]{?}{\msf n}{\cramp'},
\Over[1.5]{?}{\msf m}{\cramp'})\;.
\end{equation*}

The component $\{\Over[1.8]{\sss\statePsi}{\msf n},
\Over[1.8]{\sss\statePhi}{0}\}$ must go into itself, since the
symbol $\statePsi$ attached to it has not changed. It means that
$\msf n'=\msf n$, and one is left with the task
\begin{equation*}
\{\Over[1.8]{\sss\statePsi}{0},
\Over[1.8]{\sss\statePhi}{\msf m}\}\;
\mathrel{\Over[1.8]{?}{\rightleftarrows}}
\{\Over[1.8]{\sss\statePsi}{0},
\Over[1.8]{\sss\state\Theta}{\msf m}{\cramp'}\}\;.
\end{equation*}
However, operation $\Aleph_{\sss\Phi\!\Theta}$ recognizes only the
primitive's symbols rather than their numbers. That is, replications
$\oper[-1]m\,\{\Over[1.8]{\sss\statePsi}{0},
\pm\Over[1.8]{\sss\statePhi}{1}\}= \{\Over[1.8]{\sss\statePsi}{0},
\pm\Over[1.8]{\sss\statePhi}{\msf m}\}$ do formally commute with
$\Aleph_{\sss\Phi\!\Theta}$. Hence, by omitting the letters
$\{\statePsi,\statePhi,\state\Theta\}$, it will suffice to look for
the representation of $\Aleph$ by numerical pairs $(0,\pm\msf m)$
factorized with respect to replications $\oper[-1]m$, \ie, by the
set $\{(0,1), (0,-1)\}$. It, for its part, remains to be transformed
into itself, and the replication operators $\oper[-1]n$,
$\oper[-1]m$ will recreate the generic case. The identical
transformation $(0,\pm1)\rightarrowtail(0,\pm1)$ is ruled out since
$\Aleph_{\sss\Phi\!\Theta}\ne\oper[-3]{\mathds I}$; therefore,
$(0,\pm1)
\mathrel{\Over[1.1]{\sss\widehat{\bo\aleph}}{\rightarrowtail}}
(0,\mp1)$. Restoring all the symbols that were dropped, the effect
of $\Aleph$ reduces to the sign change for the 2"~nd element of the
coordinate pair:
\begin{equation}\label{aleph}
(\msf n,\msf m)
\mathrel{\Over{\widehat{\bo\aleph}}{\rightarrowtail}}
(\msf n,-\msf m)\;.
\end{equation}

There is no need to change sign for the 1"~st element, as this change
is the operator $-\oper[-3]{\mathds I}\circ\Aleph$. Furthermore, one
observes that the already existing group inversion
$-\oper[-3]{\mathds I}$ coincides with composition
\begin{equation}\label{ri}
(\Aleph\circ\oper[-1]{\bo\gimel})^2=-\oper[-3]{\mathds I}\;,
\end{equation}
and we may even `forget' about (the `old') subtraction, leaving the
equipment
\begin{equation}\label{base}
\{\oplus, \oper[-3]{\mathds I}, \oper[-1]m, \Aleph,
\oper[-1]{\bo\gimel}\}
\end{equation}
of semigroup $\frak G$ as an irreducible set of mathematical
structures over~it.

In this connection, yet another---more formal---""motivation of the
passage \lrceil{semigroup $\rightarrowtail$ group}, and thus of the
superposition principle, does arise. Indeed, the derivation of
$\Aleph$ above engaged the inversion \eqref{unar}, but reappointment
of primitives $\statePhi\rightleftarrows\state\Theta$ in
\eqref{invol} is a fully independent act. Therefore if we forget
about `$(-)$"=copies of the positive pairs' $(0,\msf m)$, then the
involutory nature of automorphism $\Aleph_{\sss\Phi\!\Theta}$ would
still reproduce the semigroup $\frak G$ in numbers by `duplication'
$\msf m\rightarrowtail\pm\msf m$, \ie, create the negative pairs
$(0,-\msf m)$, thus turning $\frak G$ into a group \itsf{H}. An
analogous reasoning on the symbol `$-$' could be cited even earlier,
when the $\bbC$"=field was being derived.

Now, remembering the above"=described move to the binarity of
$\odot$"=multiplication on the $(\msf n,\msf m)$"=pairs, we arrive at
the problem of matching it with structures~\eqref{base}. Clearly,
one needs only to ascertain the functionality of operators
$\oper[-1]{\bo\gimel}$ and $\Aleph$ that were not available yet.

Relation \eqref{ri} immediately gives us the correspondence
$\Aleph\circ\oper[-1]{\bo\gimel}\rightleftarrows\ri$, since
$\ri^2=-1$. Hence, one of these operators, say
$\oper[-1]{\bo\gimel}$, manifests itself in the imaginary unit
$\ri$. The origin of this operator""---""permutation
$\oper[-1]{\bo\gimel}_{\sss\Psi\!\Phi}$ in \eqref{invol}---is the
very same permutation $\statePsi\rightleftarrows\statePhi$ that
generated the $\ri$"=object in algebra \eqref{riC}--\eqref{riC+}. The
second operator, \ie, \eqref{aleph}, as is directly seen, is also
not related to the binary $\oplus$ and $\odot$ but determines the
change $\ri\rightarrowtail -\ri$. This means that the
\qm"=consideration does not just give birth to the field $\bbC$ but
to a division $\tilde \bbC^*$"=algebra, which is equipped with two
\emph{non"=binary} operations
\begin{equation*}
\frak a\mathrel{\Over{\widehat{\bo\aleph}}{\rightarrowtail}}
\frak a^*\,,\qquad
\frak a\mathrel{\Over{\oper{\scs\bo\gimel}}{\rightarrowtail}}
\tilde{\frak a}\;.
\end{equation*}
Informally, it defines all the basic actions on `complex quantities'
and thereby determines a \qm"=extension/"!generalization to the
intuitive and habitual arithmetical manipulations
\eqref{shema'}--\eqref{n-} with real things. Consequently, the four
binary arithmetical operations""---""addition/"!subtraction/"!%
multiplication/"!division""---should be supplemented with the two
unary ones: conjugation $\Aleph$ and swap $\oper[-1]{\bo\gimel}$.

\begin{comment}
A curious observation for the formal complex"=number mathematics is
appropriate here. None of these operators boil down to involution
$-\oper[-3]{\mathds I}$. We mean that each of the pairs
$(\Aleph,-\oper[-3]{\mathds I})$ or
$(\oper[-1]{\bo\gimel},-\oper[-3]{\mathds I})$ is expressible
through $(\oper[-1]{\bo\gimel},\Aleph)$, and not the reverse; see
\eqref{ri}. To put it plainly, the self"=suggested going from the
natural sign"=change (\ie, $-\oper[-2]{1}$ over $\bbR$) to the
inversion of the 2"=dimensional $\oplus$"=addition (\ie,
$-\oper[-3]{\mathds I}$ over $\bbC$) deprives the involution
$-\oper[-3]{\mathds I}$ of its primary character, as it has taken in
the 1"=dimensional domain $\bbR$. Furthermore, the 2"~nd operation
$\oper[-1]{\bo\gimel}$ is, in a sense, more `primitive' even than
the complex conjugation $\Aleph$, as this operation has had to do
with a formal pair $(\msf n,\msf m)$---merely transposes it---and
does not invoke an arithmetic action, as does $\Aleph$ when changing
the sign $\msf m\mapsto-\msf m$ in~\eqref{aleph}.

Relationship between the operators is by the binary multiplication:
$\tilde{\frak a}=\ri\odot\frak a^*$. By virtue of this relation it
makes no odds which one of these unary operators is left for
$\bbC$"=algebra.
\end{comment}

We note---and this is important \cite{br2}---that the observational
statistics $\fr_j$ are unchanged upon both operations
$\oper[-1]{\bo\gimel}$ and $\Aleph$.

\subsection{Naturalness of $\bbC$-numbers}\label{naturalC}

Thus the $\boT$"=set primitives have been entirely banished from the
theory, with the exception of the eigen"=state
$\state\alpha_s$"=markers, which are needed only for
distinguishability (sect.~\ref{S}) in $\scr A$"=observations. These
markers may be interchanged, but permutability
$\state\alpha_j\rightleftarrows\state\alpha_k$ is already reflected
by the superposition's commutativity. Taking now into account the
fact that reassigning the $\state\alpha$"=labels does not touch on
the concept of the number, one infers: the covariance attained above
is exhaustive. As a result, we draw the following conclusion.
\begin{itemize}
\item The coordinate representatives $\{\frak a, \frak b, \frak c,
 \ldots\}$ of states and superpositions thereof \eqref{s1} form the
 complex number field $\tilde\bbC^*$ equipped with the structures of
 conjugation and of swap:
 \begin{equation}\label{inv2}
 (\msf n+\ri\,\msf m)\mathrel{\Over{\!\ds*}{\mapsto}}
 (\msf n-\ri\,\msf m)\,,\qquad (\msf n+\ri\,\msf m)
 \mathrel{\Over[0.3]{\!\ds\mathchar"0365}{\mapsto}}
 (\msf m+\ri\,\msf n)\;.
 \end{equation}
 Statistical weights $\fr_j$ in object \eqref{set} are invariant
 with respect to both the involutions
 $\fr_j(\frak a^*)=\fr_j(\frak a)= \fr_j(\tilde{\frak a})$ for each
 component $\frak a_s$ independently.
\end{itemize}
What is more, the commentary on the primacy of \qm\ over the
abstract arithmetic (see p.~\pageref{comment}) has a logical
continuation.
\begin{itemize}
\item Quantum"=theoretic description invokes no $\bbC$"=numbers; nor does
 it introduce them. It \emph{does create} them together with the
 $\tilde\bbC^*$"=algebra. The $\bbC$"=numbers are in and of themselves
 the \emph{quantum} numbers.
\end{itemize}

This fact is remarkable in its own right because the `2"=dimensional'
numbers arise at the lowest empirical level, not from the need for
solving any mathematical problems. Mathematics is still lacking.
Therefore, pt.~\hyperlink{R+}{$\ds\msf R^\textsf{+}$} (p.~\pageref{R+})
could have even been weakened by replacing \lrceil{homomorphism onto
numbers}, roughly, with the \lrceil{homomorphism onto continuum}.
Our minimal points of departure are replications and the
ingoing/"!outgoing structure of brace \eqref{split'}. The imaginary
part of the complex number""---as a supplement to the real
one---comes, as a rough guide, from the left hand side of the
conception $\statePsi\goto\state\alpha$. The theory does not depend
on the meanings that will be later attached to the physical
concepts""---""observables, measurement, spectra, means, \etc---to
their interpretations or rigorous definitions. At the same time, the
interferential `effects of subtraction and of zeroes' are
intrinsically present within the construct's foundation itself.

Let us add, in conclusion, two more formal vindications of rigidity
of emerging the $\bbC$"=structure. In doing so, one assumes that we
have already had the $\bbR$"=numbers.

Unitary brace contain pairs of the form
$\{\Over[1.8]{\sss\statePsi}{\msf n},
\Over[1.8]{\sss\statePhi}{0}\}$. The binary operations
$\{\oplus,\odot\}$ on their numerical representatives $(\msf n,0)$
are closed and, as easily seen, form a commutative field, which is
isomorphic to $\bbR$. It is a subset of the generic pair set
$(\msf n,\msf m)$. From the operator nature of $\odot$, it follows
that these pairs form a certain distributive ring with general"=group
properties \eqref{m+}--\eqref{m++}. The presence of the field $\bbR$
contained in it tells us that these pairs can be realized by the
elements $\msf n+\msf m\,x$ of, at most, associative algebra $A$ over
$\bbR$. Here, $\msf n,\msf m\in\bbR$, $x$ is a generator of any
ring's element beyond $\bbR$, and the habitual~$+$ replaces the sign
$\oplus$. Multiplication of two such elements
\begin{Align*}
(\msf{n+m}\,x)\odot(\msf{n'+m'}\,x)&=
\msf{n\,n'+(m\,n'+n\,m')}\,x+\msf{m\,m'}\,x^2=\cdots
\intertext{immediately shows that result does not depend on order of
factors, \ie,}
\cdots&=(\msf{n'+m'}\,x)\odot(\msf{n+m}\,x)\;,
\end{Align*}
due to permutability of $\{\msf{n,m,n',m'}\}$ between each other and
of any $x$ with itself. This is a direct consequence of
2"=dimensionality of the algebra $A$; it must be commutative.
Invoking now the well-known Frobenius theorem on associative and
commutative structures containing the field $\bbR$ \cite{vander}, we
arrive once again at a multiplication of the form~\eqref{riC}.
K\"orner puts this point as follows: ``The complex numbers constitute
the largest system of objects that most people are content to call
numbers'' \cite[p.~230]{koerner}.

\subsubsection{Topologies on numbers}
\addcontentsline{toc}{subsection}{\qquad\qquad Topologies on numbers}

Yet another reasoning about exclusivity of $\bbC$"=numbers follows
from matching the topological and algebraic properties of the
general number systems \cite[sect.~27]{pontr}. The case in hand is
the uniqueness and non"=arbitrariness in the emergence of the
topological field $\bbC$; Pontryagin (1932). In our case, we have
two continuums""---""numerical symbols $\msf n$ and $\msf m$, each
of which, by the very method of constructing the
$\obj{\state\Xi}$"=objects \eqref{min}, is equipped only with the
natural ordering $<$. Since we do not have any more math"=structures
yet, the topology, continuity, and limits on each of the continuums
can already be introduced with respect to this relation. For one
example, there is no need to introduce the topology a~priori by
creating the arithmetical operation of multiplication/"!divisibility
of rationals (and a concept of the prime integer), as is conducted
in the $p$-adic approaches to \qm\ \cite{hren2, vladimirov, hren0}.
The `non"=naturalness' of multiplication as compared with addition
was already noted above. Moreover, in the $p$-adic versions for a
numerical domain, the topologically and physically required matching
between the natural ordering, connection, and continuity
\cite[Ch.~4]{pontr} is destroyed, and the approaches themselves
stipulate the existence of the \emph{observations} numbers with a
comprehensive arithmetic. At the same time, questions about the
`structure' of the physical $x$"=space at Planck's scale and about
measurements by rationals (see motivation in \cite{vladimirov,
hren2}) have not yet emerged, because we are not relying on physical
conceptions and are not yet introducing these notions as numerical.
The $x$"=space itself is as of yet absent, and D.~Mermin
\cite{mermin} overtly claims along these lines that ``when I hear
that spacetime becomes a foam at the Planck scale, I don't reach for
my gun''. From the low"=level empiricism standpoint, any objects,
apart from the $\bbR^2$"=continuality and frequencies $\fr$, call for
independent axioms. In turn, the primary nature of the
$\bbR$"=continuality itself follows directly from the boolean
$2^\boT$ (p.~\pageref{bulean}) and $\Sigma$"=postulate of infinity.

\vbox{
\section{State space}\label{statespace}

\flushright\tiny
\textsl{Quantum states \ldots\ cannot be `found out'}\\
\textsc{--- W.~Zurek \cite[p.~428]{saunders}}%
\medskip

\textsl{\ldots\ quantum theory refuses to offer any picture\\
of what is actually going on out there}\\
\textsc{--- D.~Mermin (1988)}}
\smallskip\nopagebreak

\subsection{Linear vector space}\label{lvs}

Once replication $(\widehat{N\!\!,M})$ of brace
$\{\Over[1.5]{\sss\statePsi}{\msf n},
\Over[1.5]{\sss\statePhi}{\msf m}\}\, \state\alpha$ has acquired a
binary character
\begin{equation}\label{pq-}
(\widehat{N\!\!,M})\big(\{\Over[1.5]{\sss\statePsi}{\msf n},
\Over[1.5]{\sss\statePhi}{\msf m}\}\,\state\alpha\,\big)
\quad\hhence\quad\big((\msf N,\msf M)\odot(\msf n,\msf m)\big)
\bet\alpha=\frak a\,\bet\alpha\;,
\end{equation}
the difference between `what is replicated' and `how many times'
disappears. A symbol $\bet\alpha$ of the eigen"=state has been
attached to the abstract $\bbC$"=number $\frak a$. Construing this
point as a quantum analog of re"=choosing (liberation of) the
measurement units (p.~\pageref{QMdim}), we obtain that the two
formal states $\frak a\,\bet\alpha$ and $\frak b\,\bet\alpha$ are
always connected by a certain number operator~$\oper[-3]{\frak p}$:
\begin{equation*}
\frak b\,\bet\alpha=\oper[-3]{\frak p}\,\big(\frak a\,
\bet\alpha\big)\,,\qquad\oper[-3]{\frak p}\rightleftarrows\frak b\odot
\frak a^{\sm1}\;.
\end{equation*}

Manipulating the numbers becomes independent of symbols
$\bet\alpha$. The way to formalize this is to think of generic
states $\frak a\,\bet\Psi\in\itsf{H}$ as the `solid characters'
\begin{equation}\label{finalHomo}
\frak a\,\bet\Psi\rightarrowtail\ket\Xi\in\bbH\;,
\end{equation}
\ie, as the $\ket\Xi$"=elements of a new set $\bbH$, which is
equipped with the $\oper[-3]{\frak p}$"=replication images
represented by the $\frak p$"=family ($\frak p\in\bbC$) of maps
\begin{equation}\label{finalHomo+}
\bbC\Times\bbH \mathrel{\Over[1]{\ds\bcdot}\mapsto}\bbH:\qquad
\frak p\bcdot\ket\Xi=\ketPhi\in\bbH\;,
\end{equation}
and which is obliged to inherit the structure \eqref{pq-}. This
inheritance says that the coordination of $\odot$"=multiplication in
\eqref{pq-} with the replication's $\frak p$"=realization is
performed by a new operation~$\bcdot$ of the unary kind on $\bbH$,
\ie, \eqref{finalHomo+}, which should be subordinated to the rule
\begin{equation}\label{pq}
\frak p\bcdot\big(\frak a\bcdot\ketPsi\big)=
(\frak p\odot\frak a)\bcdot\ketPsi\,\qquad(\frak p,\frak a\in\bbC\,,\quad
\ketPsi\in\bbH)\;.
\end{equation}

Due to this connection between operations $\odot$ and~$\bcdot$, the
latter is usually referred to as `multiplication' as well, however
such an intuition with dropping the word ``unary'' may have an
implication \cite[sect.~6.2]{br3}. An analogous rule had already
occurred in the relationship \eqref{s2} between the $\oplus$"=number
$\bbC$"=structure and the $(\Uplus)$"=group superposition, \ie, when
the multiplicative structures $\{\odot,\bcdot\}$ were not available
yet.

Among replication operators $\oper[-3]{\frak p}$, there exists an
identical transformation
\begin{equation*}
\oper[-3]{\frak p}=\oper[-3]{\mathds I}:\qquad\frak a\,\bet\Psi
\mathrel{\Over{\skew1\widehat{\mathds I}}{\rightarrowtail}}
\frak a\,\bet\Psi\;,
\end{equation*}
to which a symbol of the numerical unity $\frak p=\mathds1$
corresponds. From this, in accordance with
\eqref{finalHomo}--\eqref{finalHomo+}, there follows the rule
\begin{equation*}
\mathds1\bcdot\ket\Xi=\ket\Xi\,,\qquad\forall\,\,\ket\Xi\in\bbH\;.
\end{equation*}

It is clear that the $\smallc({\bcdot}\smallc)$"=multiplication needs
to be agreed with the $\uplus$"=union. Let us make use of the fact
that an object of (quantum) replication may be not only the unitary
brace $\{\Over[1.5]{\sss\statePsi}{\msf n},
\Over[1.5]{\sss\statePhi}{\msf m}\}\,\state\alpha$, which is
equivalent to the eigen"=element $\frak a\,\bet\alpha$, but a
$(\Tplus)$"=sum of the like objects and, in general, any constituents
of quantum ensembles (see p.~\pageref{shema'}). Therefore the
$\oper[-3]{\frak p}$"=replication
\begin{equation}\label{cab}
\oper[-3]{\frak p}\,\big(\frak a\,\bet\alpha\Uplus
\frak b\,\bet\beta\big)= \cdots
\end{equation}
is known to have its twin-sum
\begin{equation}\label{cab'}
\cdots=\frak a'\,\bet\alpha\Uplus\frak b'\,\bet\beta=\cdots
\end{equation}
with certain coefficients $\frak a'$, $\frak b'$.

Let us, for the moment, give back \eqref{cab} to the initial
language of operators/brace according to the scheme
\begin{equation}\label{pab}
(\Under[3]{\ts\frak p}
{\underbrace{\widehat{N,M}}})\,,\qquad
\{\Under[3]{\ts\frak a}
{\underbrace{\Over[1.5]{\sss\statePsi}{\msf n}\cramp',
\Over[1.5]{\sss\statePhi}{\msf m}\cramp'}}\}\,
\state\alpha \Tplus
\{\Under[3]{\ts\frak b}
{\underbrace{\Over[1.5]{\sss\statePsi}{\msf n}\cramp{’’},
\Over[1.5]{\sss\statePhi}{\msf m}\cramp{’’}}}\}\,
\state\beta\;.
\end{equation}
Take also into account a pre"=image of operation~$\Tplus$ on objects
\eqref{cab}, \ie, the~$\Uplus$. Then, \eqref{pab} and content of
sects.~\ref{C}--\ref{C*} certainly shows that the expression
\eqref{cab'} must be of the form
\begin{equation*}
\cdots=(\frak p\odot\frak a)\,\bet\alpha\Uplus
(\frak p\odot\frak b)\,\bet\beta=\cdots\;.
\end{equation*}
Reconverting, by \eqref{pq}, expressions like
$(\frak p\odot\frak a)\,\bet\alpha$ into the operatorial
$\oper[-3]{\frak p}(\frak a\,\bet\alpha)$, we complete the ellipsis
\begin{equation*}
\cdots=\oper[-3]{\frak p}\,\big(\frak a\,\bet\alpha\big)\Uplus
\oper[-3]{\frak p}\,\big(\frak b\,\bet\beta\big)\;.
\end{equation*}
Passing now to the $\frak p$"=number and to the $\ket\Xi$"=objects
\eqref{finalHomo}, \ie, replacing
$\frak a\,\bet\alpha\rightarrowtail\ketPsi$ and
$\frak b\,\bet\beta\rightarrowtail\ketPhi$, one derives an additivity
of operation~$\bcdot$ when acting on a sum:
\begin{equation*}
\frak p\bcdot\big(\ketPsi\+\ketPhi\big)=
\frak p\bcdot\ketPsi\+ \frak p\bcdot\ketPhi\;.
\end{equation*}
Here, the \itsf{H}"=addition $\Uplus$ has been carried over to the
group $\bbH$ as a new symbol~$\+$. This is nothing but a
distributive coordination of the
$\smallc({\bcdot}\smallc)$"=multiplication with the group
addition~$\+$.

In a similar way, through a number operator, one establishes yet
another relation
\begin{equation*}
\frak a\bcdot\ket\Xi\+ \frak b\bcdot\ket\Xi=
(\frak a\oplus\frak b)\bcdot\ket\Xi
\end{equation*}
between $\bcdot$ and~$\+$. Its origin is equivalent to \eqref{s2}.
From the constructs above, it is not difficult to see that we have
examined all the possibilities of $\bbC$"=replicating the
superpositions \eqref{s1} or their constituents, which is why we
have exhausted all the compatibility rules that stem from the two
fundamental operations""---""replication and union~($\Uplus$).

Thus, having considered the passage
\eqref{finalHomo}--\eqref{finalHomo+} as a final homomorphism of the
\itsf{H}"=group elements $\frak a\,\bet\Psi$ onto the objects
$\ket\Xi\in\bbH$, \ie, adjusting the previous concept of a state and
of \Courier{Data\-Source} (p.~\pageref{o+}), we infer the following.
\begin{itemize}
\item The minimal and mathematically invariant bearer of the
 observation's empiricism is an abstract space $\bbH$ of states
 $\ketPsi$ of the system $\cal S$. The structural properties
\begin{Align}
\bbH&\DEF\big\{\ketPsi\,,\ketPhi\,,\ldots\big\}\,\quad&
&\lceil\text{commutative group under operation $\+$}\rceil\;,
\label{axiom1}\\
\bbC&\DEF\{\frak a\,,\frak b\,,\ldots\}\,\quad& &\lceil\text{field of complex
numbers \eqref{g+},
\eqref{nma}--\eqref{riC}}\rceil\;,\notag\\
\oper[-2]{\frak a}\,\ketPsi&\FED\frak a\bcdot\ketPsi\in\bbH\,\quad&&
\begin{array}[t]{@{}l}
\lceil\text{closedness under operation $\bcdot$}\\
\text{\phantom($\;\hhence$ operator automorphism
$\oper[-2]{\frak a}\,\ketPsi$}\rceil\;,
\end{array}
\label{axiom2}
\end{Align}
\begin{gather}
\begin{aligned}
\frak a\bcdot\big(\frak b\bcdot\ketPsi\big)&=
(\frak a\odot\frak b)\bcdot\ketPsi\,,\quad&
\frak a\bcdot\ketPsi\+\frak b\bcdot\ketPsi&=
(\frak a\oplus\frak b)\bcdot\ketPsi\;,\\
\mathds1\bcdot\ketPsi&=\ketPsi\,,&
\frak a\bcdot\big(\ketPsi\+\ketPhi\big)&=\frak a\bcdot\ketPsi\+
\frak a\bcdot\ketPhi
\end{aligned}\label{axiom3}
\end{gather}
of the space coincide with the axioms of a linear vector space
(\lvs) over the field~$\bbC$.
\end{itemize}
Attention is drawn to the fact that this is the first place in our
construct where the word `\emph{linear}' has appeared, and even the
superposition principle, page~\pageref{princ}, was formulated
without using this term. Commutativity is necessary for linearity
but is not confined to this. The `axiom list'
\eqref{axiom1}--\eqref{axiom3} should also be complemented with a
declaration of the global \textsf{D}"=number value \eqref{D} established
above.

In a nutshell, the nature of the quantum state space is twofold:
group superposition \eqref{s1} and operator nature of the
`$\frak a$"=numbering' the elements of the group. It admits the
$\bbC$"=field scalars as operators. Relations \eqref{axiom3} describe
rules of `interplay' between all the objects. It is known that such
formations, while being implemented by a binary algebra of numbers,
turn into the vector spaces and modules \cite[Ch.~5]{kurosh},
\cite[sects.~\textbf{I}(7.1--2), \textbf{II}(13.4)]{vander}. Concerning
the consistency of these rules---say, of numerical distributivity
\eqref{m++}---with relations \eqref{axiom3}, see the
work~\cite{rigby}.

\begin{comment}
A certain oddity is in place. \qm"=empiricism is such that the
standard definition of \lvs\ by the all-too"=familiar axioms
\eqref{axiom1}--\eqref{axiom3} is more `\emph{non}physical' by its
nature than the `generalistic' abstraction of a group with operator
automorphisms of the group \itsf{H}"=structure itself
\cite[sect.~\textbf{I}(4.2)]{burbaki}, \cite{kurosh}. A point like this
might be expected, though. This is because, as noted in
sect.~\ref{super}, meaning all of tokens in \eqref{1} and their
origin are entirely unknown, and linearity of \qm\ is radically
different from other `linearities' in physics.

All told, the appearance sequence of the mathematical structures is
as follows:
\begin{equation*}
\begin{aligned}
&\text{\lrceil{sets, union $\cup$, \ldots}}\;\rightarrowtail\;
\text{\lrceil{semigroup (sect.~\ref{semi})}}\\
&\qquad\rightarrowtail\;
\text{\lrceil{group \itsf{H} (sect.~\ref{ssuper})}}
\;\rightarrowtail\;
\text{\lrceil{numbers \& arithmetics (sect.~\ref{Numbers})}}\\
&\qquad\rightarrowtail\; \text{\lrceil{compatibility of the
group and numbers (sect.~\ref{lvs})}}\\
&\qquad\rightarrowtail\; \text{\lrceil{the abstract \lvs\ and its
bases}}\;.
\end{aligned}
\end{equation*}
This sequence is rigid, such as the box"=diagram in
sect.~\ref{physmath}, so the structure of \lvs\ cannot be weakened
because we have the two fundamental principia
(\hyperlink{II}{\red\textbf{\textsf{II}}} and
\hyperlink{III}{\red\textbf{\textsf{III}}}) in between the semigroup, the
group, and the vector space.
\end{comment}

\subsection{Bases, countability, and infinities}

From a ban on transitions $\state\alpha_s\goto\state\alpha_n$ under
$s\ne n$, it follows that unitary $\state\alpha_s$"=brace \eqref{min}
correspond to vectors $\frak a_s\bcdot\ket{\alpha_s}$ that are
linearly inexpressible through each other. Aside from the general
ensemble brace \eqref{split'}, no other elements exist, and all of
them are in one-to-one correspondence with the vector
representations $\frak a_1\bcdot\ket{\alpha_1}\+\frak a_2\bcdot
\ket{\alpha_2}\+\cdots$. Each such vector has a statistical
pre"=image \eqref{split'}, and vice versa; there are no gaps. This
means that the system of vectors
$\{\ket{\alpha_1},\ket{\alpha_2},\ldots\}$ forms a basis of $\bbH$ as
the basis of \lvs---\emph{basis of an observable $\scr A$ or
$\scr A$"=basis}---and the number of symbols $\ket{\alpha_s}$ is its
dimension: $\dim\bbH=\textsf{D}$. The $\textsf{D}=\infty$ case, just like
anything associated with infinity, cannot be formalized without
topology, and its presence is presumed, but this discussion is
dropped. We just remark that even earlier, when arising the
2"=dimensional continuum, we have silently assumed the
$(\bbR\Times\bbR)$"=product topology on it. This supposition is
natural, inasmuch as it does not involve additional
constructions/"!requirements. Thus if properties
\eqref{axiom1}--\eqref{axiom3} are directly accepted as empirical,
then the mathematical rigors augment them axiomatically on the
outside, because one constructs the mathematical theory.

The micro"=transition $\TRANS[1]{\sss\scr A}$ in sect.~\ref{S} is a
solitary entity. This means that the number of eigen
$\state\alpha_s$"=primitives for an actual instrument may be either
finite or discretely unbounded. We base this on the fact that
continual formations is a product of mathematics rather than
empiricism; see also \cite[p.~35]{ludwig5}. The $\boT$-set, as an
example, is also non"=continual, but that premise may even be given
up, because only a discrete portion of this set is present at
arguments (transitions $\GOTO{\sss\scr A}$). Notice incidentally
that continuum, along with the number, does not feature in the
\zf"=axioms \cite{kurat} but is also created, just as ``an infinity is
actually not given to us at all, but is \ldots\ extrapolated through an
intellectual process.\@'' \cite[p.~55; Hilbert--Bernays]{kleene}; see
also the book \cite{lakoff} for the conceptualization of infinity.
One obtains a countability of the $\scr A$"=basis. Hence it follows a
completeness of $\bbH$ and countability of dimension \eqref{D}, as
of the number \lvs"=invariant:
\begin{equation}\label{D'}
\textsf{D}=2,3,\ldots,\aleph_\textsf{o}\qquad( = \dim\bbH)\;.
\end{equation}

Finally, let us mention the following. The basis is a term that in
no way is present in the abstract axiomatics
\eqref{axiom1}--\eqref{axiom3}, and \lvs\ on its own account does
not contain a motive for introducing that concept. But empirically,
the $\bbH$"=space is arising entirely and ab~initio in all possible
linear combinations over $\ket{\alpha_j}$, \ie, through
$\scr A$"=bases. Because of this, in order for an abstract \lvs\ to
become the quantum state"=space the \lvs\ should be considered as
being accompanied by the concepts bases and changes thereof.
Conforming to such a requirement and the formal existence of a basis
is given by a nontrivial math theorem invoking the axiom of choice
\cite[sect.~\textbf{II}(7.1)]{burbaki}.

\subsection{\itbf{The theorem}}

The states $\ketPsi$ and sums thereof, at the moment, form a formal
family of different elements. Recall that symbols
$\{\approx,\not\approx\}$ in pt.~\hyperlink{R}{\red\textsf{R}}, as from
the end of sect.~\ref{SS}, have been replaced with the standard ones
$\{=,\ne\}$. The physical aspects of \SM\ were being left aside so
far, and, for example, $\ketPsi$ and $\frak c\bcdot\ketPsi$ were the
different vectors of the $\bbH$"=space. However,
\begin{itemize}
\item empiricism (deals with and) yields originally \emph{not states}
 and superpositions thereof \emph{but
 $\ket{\alpha}$"=representations}.
\end{itemize}
It is these representations (alone) that carry information about
statistics $(\fr_1,\fr_2,\ldots)$ through coefficients $\frak a_j$. The
replicative character of $\frak c$"=multipliers and
$\Sigma$"=postulate entail however that the two vectors
$\mathds1\bcdot\ketPsi$ and $\frak c\bcdot\ketPsi$ should correspond
to the one and the same statistics
$\fr_{\!\sss(\scr D)}=(1,0,\ldots)=\skew1\tilde\fr_{\!\sss(\scr D)}$
under an observation $\scr D$ with the eigen collection
$\{\mathds1\,\bet\Psi,\ldots\}$.

Let us write the equalities
\begin{equation}\label{tmpD}
\begin{array}{rcccl}
\fr_{\!\sss(\scr D)}&\twoheadleftarrow\quad
\underbrace{\mathds1\bcdot\ketPsi}&=&
\underbrace{\frak a_1\bcdot\ket{\alpha_1}\+\frak a_2\bcdot
\ket{\alpha_2}\+\cdots}&\twoheadrightarrow\quad\fr_{\!\sss(\scr A)}\;,\\
&\scs\text{observation $\scr D$}&&\scs\text{observation $\scr A$}\\
\skew1\tilde\fr_{\!\sss(\scr D)}&\twoheadleftarrow\quad
\overbrace{\frak c\bcdot\ketPsi}&=&
\overbrace{\frak c\bcdot\big(\frak a_1\bcdot\ket{\alpha_1}
\+\frak a_2\bcdot\ket{\alpha_2}\+\cdots\big)}
&\twoheadrightarrow\quad\skew1\tilde\fr_{\!\sss(\scr A)}
\end{array}
\end{equation}
and look at them in the following order: the first line from right
to left, and the second in the reverse direction. Their right hand
sides are the carriers of some statistics $\fr_{\!\sss(\scr A)}$ and
$\skew1\tilde\fr_{\!\sss(\scr A)}$. The frequencies
$\fr_{\!\sss(\scr A)}=(\fr_1,\fr_2,\ldots)$ come from the number set
$(\frak a_1,\frak a_2,\ldots)$ under the same environments \SM\ that
give rise to the statistics $\fr_{\!\sss(\scr D)}$. But it is also
generated by the representative $\frak c\bcdot\ketPsi$, which is
associated with the same \SM; hence,
$\skew1\tilde\fr_{\!\sss(\scr D)}=\fr_{\!\sss(\scr D)}$. By virtue of
the second equal sign in \eqref{tmpD}, the same \SM\ are associated
with the second $\scr A$"=collection $(\frak c\odot\frak a_1,
\frak c\odot\frak a_2,\ldots)$. Therefore the frequencies
$\skew1\tilde\fr_{\!\sss(\scr A)}$ that emanate from it have to be
identical to those emanating from the first collection
$(\frak a_1,\frak a_2,\ldots)$. That is to say
$\skew1\tilde\fr_{\!\sss(\scr A)}=\fr_{\!\sss(\scr A)}$, and the scale
stretches $\ketPsi\rightarrowtail\frak c\bcdot\ketPsi$ are not
recognized by any $\scr A$"=instruments. A more concise reasoning is
that the quantum replication $\frak c=\msf n+\ri\,\msf m$ may be
viewed as the 1"=dimensional replications
$\oper[-1]n\,(\frak a\bet\alpha)$,
$\oper[-1]{\text\i}\,\,(\frak a\bet\alpha)$ of all the brace
$\frak a_s\bet{\alpha_s}$"=images and of sums such as
$\oper[-1]n\,(\frak a\bet\alpha)\Uplus
(\oper[-1]{\text\i}\circ\oper[-1]m) (\frak a\bet\alpha)$. These
replications do not change the superposition statistics as a whole.

The aforesaid gives birth to a universal""---stronger than $\approx$
and irrespective of instruments""---""observational equivalence
relation
\begin{equation*}
\ketPsi\Approx\const\bcdot\ketPsi
\end{equation*}
on the space $\bbH$, \ie, the `physical' indistinguishability
(sect.~\ref{MM}).

The basis vectors $\ket{\alpha_s}$ and their ($\Approx$)"=equivalents
will be referred to as \emph{eigen vectors/states of instrument}
$\scr A$. Clearly, the concepts of instrument and of
(macro)"=observation (\hyperlink{O}{\red\textsf{O}}) should now be
distinguished. Accordingly, the spectral construction
\eqref{spectr}--\eqref{spectr-} should be corrected. Call the data
set
\begin{equation}\label{spectr+}
\big\{\ket{\alpha_1}_{{\sss\lfloor\!}\alpha_1^{}},\,
\ket{\alpha_2}_{{\sss\lfloor\!}\alpha_2^{}},\, \ldots\big\}\FED
\bo[\scr A\bo]
\end{equation}
the $\bo[\scr A\bo]$-\emph{representative} of instrument $\scr A$ in
$\bbH$. The add-on \eqref{spectr+} does not touch on $\bbH$"=space,
since the spectral labels $\smallc[2]\lfloor\! \alpha_j$ are the
self"=contained objects independently of vectors $\ket{\alpha_k}$.
These labels and their degenerations determine internal properties
of the formalized notion of an instrument \eqref{spectr+}.
Conversely, any state vector $\ketPsi$ or $\frak c\bcdot\ketPsi$ may
be treated as a $\bo[\scr C\bo]$"=representative for an
imagined/actual instrument $\scr C$ (spectrum is arbitrary) and is a
certain $(\+)$-sum of the eigen elements for any other
$\bo[\scr A\bo]$"=representative.

Remembering \eqref{m2}, we arrive at the quantum ``\emph{kinematic
framework}'' \cite{bub}, \ie, at the ultimate conclusion that
determines the pre"=dynamical theory of macroscopic data on
micro"=events.
\begin{itemize}
\item \embf{The core $\bo{(1}$-st\,$\bo)$ theorem of quantum
 empiricism}\eLab{Theorem}
 \begin{enumerate}
 \item[1)] The mathematical representatives of physical observations and
  of preparations are the quantum states $\ketPsi$ and statistical
  mixtures of eigen $\ket{\alpha}$"=states
  \begin{equation}\label{mix'}
  \big\{\ket{\alpha_1}^{{\sss(}\varrho_1^{}\sss)},\;
  \ket{\alpha_2}^{{\sss(}\varrho_2^{}\sss)},\; \ldots\big\}\,,\qquad
  \varrho_1+\varrho_2+\cdots=1\;.
  \end{equation}

 \item[2)] Properties \eqref{axiom1}--\eqref{D'} define objects
  $\ketPsi$ as elements of a (complete separable) linear vector
  space $\bbH$ over the algebra of complex numbers~$\bbC^*$.

 \item[3)] Dimension $\dim\bbH=\textsf{D}\geqslant2$, representing an observable
  quantity ($\textsf{D}<\infty$), is set to the value
  $\msf{max}\,\{|\boT_{\!\sss\scr A}|, |\boT_{\!\sss\scr B}|,\ldots\}=
  \textsf{D}$ as required by an accuracy of the toolkit $\cal O=$
  $\{\scr A$, $\scr B$, \ldots\}. The eigen $\ket{\alpha}$"=vectors for
  each $\bo[\scr A\bo]$"=representative provide a basis of $\bbH$
  ($\textsf{D}<\infty$) independently of spectra~\eqref{spectr+}.

 \item[\hypertarget{4}{4})] The $\scr A$"=bases stand out because the
  \emph{observational} number"=notion has been associated to
  them---""statistics of the micro"=events. The frequencies
  $\fr_k(\frak a)$ are invariant under involutions \eqref{inv2}, and
  states $\ketPsi$ and $\frak c\bcdot\ketPsi$ are statistically
  indistinguishable.

 \item[5)] Rules \eqref{axiom1}--\eqref{D'}, for a fixed
  $\textsf{D}\ne\infty$, are categorical as an axiomatic system; they
  admit no non"=isomorphic models.
\end{enumerate}
\end{itemize}

The words `complete separable' have been supplemented here for
mathematical reasons. This point is partly commented in \cite{br2}
and more fully in \cite{br3}. Indeed, the algebra constructed above
calls for some amendments of a topological nature because all the
construction contains three infinities: continuum $\bbC$, continuum
$\bbH$, and dimension \textsf{D}. In this connection, see the book
\cite{pontr}. The term ``categorical'' may require some explanation
and it is fully given in \cite{br3}. Here, one suffices to mention
the point that one mathematical axiomatical system can in general
have different inequivalent realizations/"!models \cite{kleene,
raseva, stoll}. In turn, the only thing that distinguishes two
vector"=space models between themselves is their dimension~\textsf{D}.

Now, having considered the micro"=destruction arrays with empirical
rather than a formal take on arithmetic, the ideology of creating
the quantitative theory leads to the key feature of quantum
states---""addition thereof---and the quantities under addition `do
amount to' the complex numbers.

\begin{comment}\label{units}
As in sect.~\ref{qm+A}, we draw again attention to a hidden and
(logically) unremovable extension of the physical units' concept.
\par\larger 
\begin{itemize}
\item \smaller``\ldots\ \emph{units}. Despite the rudimentary nature of
 units, they are probably the most inconsistently understood concept
 in all of physics \ldots\ where do units come from?''.
 \\\phantom.\hfill S.~Gryb \& F.~Mercati~\cite[p.~91]{gryb}
\end{itemize}
\smaller 
Surprisingly, the na\"{\i}ve and straightforward conjunction of this
concept with an abstract number seems to contravene the
multiplication arithmetic; but not the addition one. The typical
example illustrates the point:
\begin{equation*}
(2\,\,\textsf{kg})\times(3\,\,\textsf{kg})\ne6\,\,\textsf{kg}\qquad
(3\,\,\textsf{sheep}+5\,\,\psi\ne8\,\textsf{St\"uck})
\end{equation*}
(see also \cite[items (4) and (5) on p.~16]{koerner}). On the other
hand, $2\times(3\,\,\textsf{kg})=6\,\,\textsf{kg}$ and
$(2\,\,\textsf{kg})+(3\,\,\textsf{kg})= 5\,\,\textsf{kg}$, and the \textsf{kg} may be
replaced here with any other entity: the classical m\`etres, the
abstract `Quanten St\"ucke $\psi$', \thelike. They have no any
operational significance, but one cannot get by without them.

The numeral characters acquire their usual abstract"=numerical
meaning""---""mathematization \cite{ludwig5}---only when we throw
(sect.~\ref{qm+A}) the `units' $\lceil\textsf{St\"uck}$, $^{\circ}$C,
\textsf{sheep}, $\psi$, $\ldots\rceil$ out of data like
$\lceil5\,\textsf{St\"uck}$, $5^{\circ}$C, 5\,\textsf{sheep}, $5\,\psi$,
$\ldots\rceil$. The symbol `$5$' in $5^{\circ}$C is the very same~`5'
as in $5\bcdot\ket{\alpha}$. It is pointless without such a
matching/"!abstracting. In the Newtonian spirit (epigraph to
sect.~\ref{Numbers}) the symbol could be defined as follows:
\begin{equation*}
\lceil\text{the abstraction $\msf5$}\rceil\DEF
\frac{\oper[-1]5^{\circ}\text{C}}
{\oper[-1]1^{\circ}\text{C}}= \frac{\oper[-1]5\:{}\ketPsi}
{\oper[-1]1\:\ketPsi}=\cdots
=\frac{\oper[-1]5\:\textsf{unit}}{\oper[-1]1\:\textsf{unit}}\;.
\end{equation*}

It may be even said that creating the number $\msf n$ (from its
operator $\oper[-1]n$) as an abstracted entity reflects a kind of
covariance with respect to attaching the various language tags; no
matter, the real \textsf{St\"uck}, \textsf{sheep}, or the abstract ones
like $\ketPsi$. At the same time, the inversion of this
abstracting""---""attaching the $\lceil\textsf{St\"uck}$, $^{\circ}$C,
\textsf{sheep}, $\psi$, $\ldots\rceil$ to the character~\textsf{5}---is
always \emph{an interpretation of abstraction}: interpretation `the
\textsf{St\"uck}', `$^{\circ}$C"=interpretation the Celsius', \etc. It
is not improbable that this is the only point, when the completed
\qt---\qm/\qft/"!quantum"=gravity yet to be constructed""---will resort
to the word interpretation. See also comments by D.~Darling on
``sheep, fingers, tokens, numbers, things, to ``add'' things,
abstraction --- the process of addition'', \thelike\ on p.~178 in the
book \cite{fuchs4} and in \cite[pp.~263--264]{fuchs3}.
\end{comment}

Incidentally, within this physical and quantum context,
\begin{itemize}
\item the \lvs\ itself should be regarded as no less a primary
 math"=structure than the numbers themselves. Empiricism gives birth
 to both these structures together. Neither of them is more/"!less
 abstract/"!necessary than the other. Behind them is certainly a
 commutative group with operator automorphisms over it, and
 ``numbers'' is just a shortened term for that operators. Therein lies
 their nature (sect.~\ref{arif}).
\end{itemize}
The habitual physics' construct
\lrceil{num\-ber}${}\times{}$\lrceil{physical unit} exemplifies in
effect the simplest (1"=dimensional) \lvs. However, the structure
``the \lvs'', in contrast to the `bare' arithmetic, simply `does not
forget and keeps' an operator nature (unary multiplication~$\bcdot$)
of the structure ``the number'' and its empirical inseparability from
the notion of the unit:
\begin{equation*}
\Under[3]{\ds\text{\smaller[2]vector space}}
{\UnderBrace[2]{\text{$\oper[-2]2\:(\oper[-2]3\:\textsf{unit})\;\hhence\;
2\bcdot(3\bcdot\textsf{unit})$}}}
\quad
\rightarrowtail\cdots\text{\ abstracting\ }\cdots\rightarrowtail\quad
\Under[3]{\ds\text{\smaller[2]arithmetic}}
{\UnderBrace[2]{\text{$(2\times3)\: \textsf{\sout{unit}}$}}}\;.
\end{equation*}
A direct corollary of this point is the fact that
principium~\hyperlink{II}{\red\textbf{\textsf{II}}} can in no way be given
up or disregarded. This would be tantamount to impossibility to
introduce the further empirical (and classical) notion of a physical
unit. The `forgetfulness' of arithmetic about measuring units even
leads to a new way of looking at the classical Pythagoras theorem
\cite[sect.~6]{br3}.

At the moment, it is worthwhile to summarize where we stand. As we
have seen, nothing above and beyond what was used in constructing
the mathematics \eqref{axiom1}--\eqref{D'} is required to explain
the nature and meaning of the quantum state. Besides, we have
obtained not merely a completion of construction~\eqref{Repr}:
\begin{equation*}
{\bo\Oplus}(\frak a_1,\ket{\alpha_1};\frak a_2,\ket{\alpha_2};\ldots)=
\frak a_1\bcdot\ket{\alpha_1}\+\frak a_2\bcdot
\ket{\alpha_2}\+\cdots\;.
\end{equation*}

In the first place, one establishes \emph{a genesis of the quantal
discreteness. Discriminating is an isolated act in the very nature
of the perception process}: `one thing is distinct from another',
`the controlling the minimal begins with a distincting of something
the two', \thelike\ (sect.~\ref{observ}). Accordingly,
``indivisibility, or ``individuality'', characterizing the elementary
processes'' \cite[p.~203; N.~Bohr]{schilpp} must be formalized into
the \emph{`elemental' click}.
\begin{itemize}
\item The classical continuality of the perceptual reality---the
 $(3\,{+}\,1)$"=space, fields $\{u(\bo x,t)$, $\psi(\bo x,t)$, $\ldots\}$,
 and the $\bbR$"=numbers---is a theorization act, whereas the nature
 of the perception fundamentally ``contains an element of
 discontinuity'' \cite[p.~179]{muynck}. The continuality of the
 classical"=physics mathematics we are used to is a `quantum effect'.
\end{itemize}

The theorization does also bear on preparation \SM. For example,
smoothy reducing the interferometer intensity is not an empiricism
but an \emph{imagination} of abstracta the continuity/"!infinity.
Clearly, such an (incorrect) substitution of the perception process
should somewhere be replaced with a `correct
understanding'"!---""introduction of the categories the
\lrceil{isolated micro"=events} \tplus\ \lrceil{(myriads) assemblages
thereof}. Granted, the natural language is able to describe the
discontinuity only in the classical (the energy) terms---Plank's
quantum of action~$\hbar$, although the \emph{quantum discreteness
is not a discretization of} something classical but a discreteness
on its own account.

We also clarify the formalization of measurement/"!preparation and of
genesis of the $\bbC$"=numbers. The well-known
$\smallc({*}\smallc)$"=conjugation operation also finds its origin.
Moreover, it is supplemented with a transposition
$\Re(\frak a)\rightleftarrows \Im(\frak a)$ of the real/"!imaginary
part of the $\bbC$"=number, and this transposition should be regarded
just as natural operation as the conjugation. The emergent concepts
of spectra, of their degenerations and eigen"=states provide a nearly
comprehensive mathematical image of physical observables. The state
becomes devoid of its mysteriousness \cite{leifer, nielsen, ney},
since it is explicitly built in terms of the unique model of the
`statistical' $\ket{\alpha}$"=representatives supplemented with
macroscopic mixtures~\eqref{mix'}.

\vbox{
\section{Numbers, minus, and equality; revisited}\label{minus}

\flushright\tiny

\textsl{\ldots\ quas decet numeris negativis exprimantur,
additio et subtractio\\
consueto more peracta nullis premitur difficultatibus\footnotemark}\\
\textsc{--- L.~Euler} (1735)}%
\footnotetext{\ldots\ if we represent the notions, which are necessary,
by negative numbers, then addition and subtraction \ldots\ are executed
without any difficulty.}
\smallskip\nopagebreak%

\subsection{Separation of the number matters}\label{divis}

The empirical adequacy of \qt\ can be based only on empirical
ensembles (sects.~\ref{A+S},~\ref{whyC}). Creation of their
mathematics tells us, then, that the `quantity of something'
\eqref{shema'}--\eqref{real'} turns into a formal operational
algebra through labeling the operator replications
(sects.~\ref{Dupl}--\ref{arif}) and yields the numbers per~se. At
first, they appear merely as
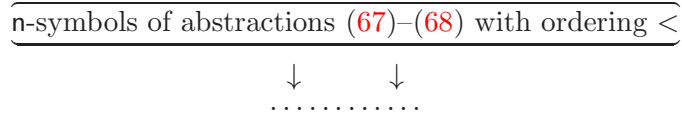
\begin{gather*}
\OverBrace[2]{\UnderBrace[2]{\text{$\msf n$"=symbols of abstractions
\eqref{shema'}--\eqref{real} with ordering $<$}}}\hbox to0ex{\;}\\
\downarrow\qquad\quad\downarrow\\[-2ex]
\ldots\ldots\ldots\ldots
\end{gather*}
and then as internal objects of theory:
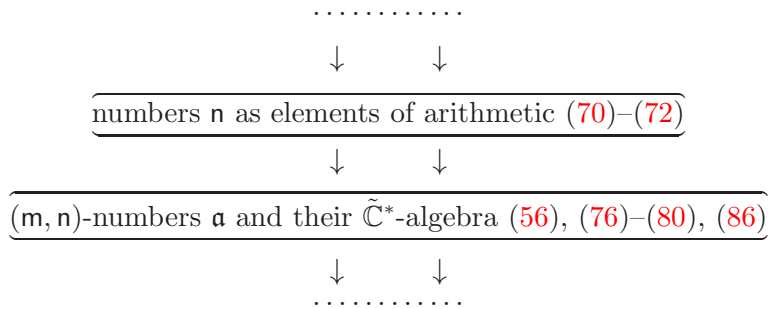
\begin{gather*}
\ldots\ldots\ldots\ldots\notag\\
\downarrow\qquad\quad\downarrow\\
\begin{array}{c}
\OverBrace[2]{\UnderBrace[2]{\text{numbers $\msf n$ as elements of
arithmetic \eqref{distr}--\eqref{n-}}}}\\[1ex]
\downarrow\qquad\quad\downarrow\\[1ex]
\OverBrace[2]{\UnderBrace[2]{\text{$(\msf m,\msf n)$"=numbers
$\frak a$ and their $\tilde\bbC^*$"=algebra \eqref{g+},
\eqref{nma}--\eqref{riC}, \eqref{inv2}}}}
\end{array}\hbox to0ex{.}\\
\downarrow\qquad\quad\downarrow\\[-2ex]
\ldots\ldots\ldots\ldots
\end{gather*}

These steps are necessary and mean that not only are the complex
numbers far from self"=evident, but even the negative ones are; a key
place \eqref{g1}, \eqref{ab} wherein a group arises. All the other
structural points, first and foremost the observational quantities,
may be further produced (even as concepts) only by way of certain
mathematical mappings:
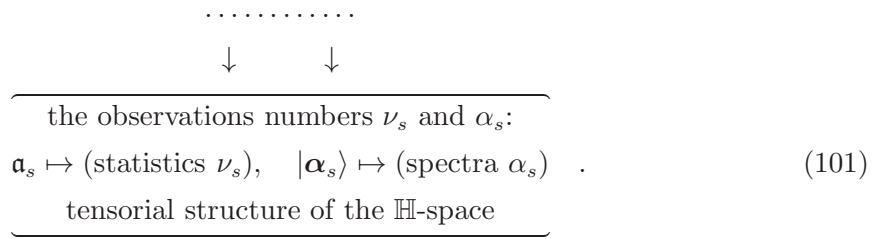
\begin{gather}
\ldots\ldots\ldots\ldots\notag\\
\downarrow\qquad\quad\downarrow\notag\\
\OverBrace[2]{\UnderBrace[2]{
\begin{array}{@{}c@{}}
\text{the observations numbers $\fr_j$ and $\alpha_j$:}\\[1ex]
\frak a\mapsto\lceil\text{statistics }\fr_j\rceil, \quad
\ket{\alpha_j}\mapsto\lceil\text{spectra }\alpha_j\rceil\\[1ex]
\text{tensorial structure of the $\bbH$"=space}
\end{array}}}\hbox to0ex{\quad.}\label{star}
\end{gather}
In other words, if a concept is a numerical one already in
empiricism""---""frequencies, spectra, \etc---then its meaningful
formalization by means of a mathematical definitio can only resort
to mathematics that we have at our disposal: \lvs\ and algebra of
numbers.

Thus numerical quantities in the entire theory are initially divided
up by their emergence mechanism
(\hyperlink{II}{\red\textbf{\textsf{II}}}): the intrinsic abstracta and
reifications \eqref{star}. Without such a division, the
\emph{circular logic is inevitable}, and the above"=mentioned `unit'
treatment of numbers would still be supplemented with the task of
their observational interpretation complicated by
two"=dimensionality. This task would be present in formalism not
merely as a problem but as an inherently intractable challenge.
Actually, \emph{any entity can be identified with numbers}, and this
is why, the quantum empiricism and
principium~\hyperlink{II}{\red\textbf{\textsf{II}}}---""paradigm of the
very number in the physical theory""---insist on the need to pay the
closest possible attention to all these things.

\begin{comment}
In this regard, the situation has a parallel with the familiar
history of electrodynamics of moving bodies; as was pointed out just
before principium~\hyperlink{III}{\red\textbf{\textsf{III}}}. Lorentz's
contraction theory is inconsistent, if the space"=time tags to events
are not linked up to the empirically precise and operationally
defined concepts in different reference frames: clocks,
simultaneity, rigid rods, distances, \thelike.

In quantum case, the chief subject of empirical definition is a
concept of the number and of the `numerical value of \ldots'.
Otherwise, the meaning given to the conception of a quantitative
theory itself has been blurring. The `quantum numbers' $\bbC$ are
built up from the reals, and the latter have an operator nature
(sect.~\ref{arif}). But the complexes $\bbC$, being also operators
and unlike the reals, never act (operationally) on the reified
quantities. They do act on the abstract $\ketPsi$"=elements of the
abstract commutative group~$\bbH$. Recall that this group and
superposition principle were arising before the numbers.
\end{comment}

We have seen now that ``it is quite wrong to try founding a theory on
observable magnitudes[/categories] alone'' \cite[p.~504; Einstein, in
a talk (1926) to Heisenberg]{saunders}, and resorting to the
physical notions---the camouflaged
\hyperlink{M}{\red\textsf{M}}"=observations""---is prohibited; see also
Remark~\ref{two}. The attempts to use statistics at the very
beginning of the theory are known \cite{moyal, hartle, ballentine,
ballentine2, espagnat, ludwig1, ludwig3}, and rightly so; they were
initiated by H.~Margenau (1936) \cite[Ch.~15]{auletta}. However, the
scheme just given is rigid. To obviate the premature appearance of
the very need for an interpretation, the scheme must not be varied.
Being a sequence of steps, it provides in essence an answer to the
principium~\hyperlink{II}{\red\textbf{\textsf{II}}}.

\subsection{Operations on numbers}\label{92}

The last step in this scheme contains, in particular, the map
$\frak a\mapsto\fr$, \ie, measurement \eqref{red}. Its form should
be established in its own right---Born's rule \cite{br2}. To
illustrate, the na\"{\i}ve transformation of negative numbers
$\msf p$ into the actually perceived quantities by a `seemingly
natural' rule such as $|{\pm}\msf p|$ is not correct and does not
follow from anywhere. For the built algebra
\eqref{axiom1}--\eqref{axiom3}, the operation $\pm\msf p\mapsto
|{\pm}\msf p|$ is extrinsic and illegal, it is nowhere to be had.
According to ideology of sect.~\ref{physmath}, not only
objects""---""numbers, vectors, quantities, characteristics,
\etc---but also all the math operations should be created because
one without the other is meaningless. The more so as the numerical
object of the theory""---the complex pair
$(\pm\msf p,\pm\msf q)$---is as yet single, it contains a
principally `non"=materializable' ingredient (sect.~\ref{C}) and
behaves as a whole. With regards to empiricism, the negative and the
$\bbC$"=numbers are equally `nonexistent, fictitious entities', since
the state's mathematics \eqref{axiom1}--\eqref{axiom3} has not been
supplemented with the doctrine of `empirically perceived' quantities
\eqref{star}. As a matter of fact, the step-by-step transformation
of the binary operation $\cup$ to symbols $\uplus$, $\Uplus$,
$\Tplus$ and, finally, to operations $\{\+,\bcdot,\oplus,\odot\}$
does not terminate at states. Algebra \eqref{axiom1}--\eqref{axiom3}
will be further required to create, now, the mathematically correct
calculation rules of the proper observational quantities.

The foregoing is amplified by the fact that
pr.~\hyperlink{II}{\red\textbf{\textsf{II}}} has been involved in the
classical description and in vindication/"!refutation of, say, the
hidden variable theories. Here, numbers are identified with the
reified quantities, and subtraction is taken for granted from the
outset. However, the negative quantities are also being created
here, and they are constructed in the same manner as the `quantum
zero' for the \itsf{H}"=group in sect.~\ref{ssuper}.

Indeed,
\begin{itemize}
\item the instrument indications and physical quantities are not
 numbers, nor the (``pointer'') states;
\end{itemize}
``detector \ldots\ does not measure a field or an $S$"=matrix"=element''
(R.~Haag (2010)). They are no more than notches, and `negative
notches' are introduced prior to mathematics of symbols according to
the following subconsciously intended scheme. The self"=apparent
physical conventionality, which has been calling `an addition' of
two such notches, must produce, in accord with the
supra"=mathematical requirements of physics, what is named `nought,
zero'. Two waves at a point, for instance, compensate each other.
The result is asserted to be identical with a zero the mathematical,
and that is the subtraction.

The classical `explanations' are the ones with use of
compensations/"!subtractions (see Remark~\ref{13}), whereas \emph{the
minus} we have been accustomed \emph{is a fairly abstract
construction in its own right}. J.~Baez and J.~Dolan best reflected
the situation, observing on page~37 of \cite{baez+} that ``half an
apple is easier to understand than a negative apple!''; on the same
page, a good discussion of division is given. In this respect, one
might state that the very classical physics needs an
interpretation""---in terms of strictly positive `the number of
St\"ucke'. Mathematization of empiricism into numbers is not a
distinctive feature of quantum description. However, comprehending
`abstracting the minus sign' is not confined even to this. A word of
explanation is necessary with regard to the situation.

Mathematics formalizes \cite{kurat} the positive/"!negative
$\pm\msf p$ into the pairs' classes $(m,n)$ being equivalent with
respect to an `adding' of the class $(\ell,\ell)$ (the `zero'):
\begin{gather}
(+\msf m)\DEF(m,0) \approx(m+\ell,0+\ell)\,,\qquad
(-\msf n)\DEF(0,n) \approx(0+\ell,n+\ell)\;,\notag\\
\label{pair}
\pm\msf p\quad\hhence\quad(\msf m-\msf n)\DEF(m,n)\approx(m+\ell,n+\ell)\;,
\end{gather}
where $m,n,\ell$ are to be seen as `something strictly positive'.
This `adding' is yet another tacitly assumed and much more abstract
action: addition of objects of some other kind---`positive couples'
$(m,n)$. Technically, at an appropriate place of
sect.~\ref{Numbers}, we had to introduce such classes and to assign
their own algebraic operations for them. The result might be called
the `genuine' arithmetic (of `the positives') and could be enlarged
to the `complete arithmetic' with multiplication and division.

In consequence, the single"=token object $(+\msf m)$ or $(-\msf n)$,
which we perceive as self"=evident
(cf.~pr.~\hyperlink{II}{\red\textbf{\textsf{II}}}), is a highly unobvious
construction""---the generic equivalence-class of two"=token
$(m,n)$"=abstractions \eqref{pair}. Essence of the symbol of a
negative number $(-\msf n)$ is revealed only when contrasting the
two positive ones. Exactly the same situation has occurred when
deriving the superposition principle in \eqref{o+}--\eqref{g1} and
\eqref{ab}.

It is clear that once all the $\pm\msf p$"=numbers, and the `normal'
positive~$+\msf p$'s among, have been formalized into the
equivalences \eqref{pair}, the fact that they possess any `natural
meanings', such as the `operation of the quantity
$\msf p\mapsto|\msf p|$' invented above, becomes more than
unnatural; the abstract class operations that appear out of nowhere.
Similarly with $\bbQ$"=numbers and their $\bbR$"=extension: classes of
equivalent pairs $(\msf n/\msf m)\DEF(n,m)\approx(n\,\ell,m\,\ell)$.

\subsection{Naturalness of abstracta}

We thus infer that rejection or disregard of the similar `naturally
abstract' set"=theoretic models would be tantamount to the rejection
of the minus sign even in the elementary physics. This is an absurd,
but its root is a need for abstracting. On the other hand, the
motivated deduction of these models cannot be replaced with (hidden)
axiomatic assumptions or with ready"=made math"=structures. Such an
ambivalence is, in our view, one reason why the problem with
`decrypting' quantum postulates is so difficult; it touches on the
metamathematical and metaphysical aspects of the very thinking
\cite{lakoff, kleene, chomsky, harin, knott}. The stream of
subconsciously abstractive homomorphisms
\begin{multline}
\lceil\text{pt.~}\hyperlink{S}{\red\textbf{\textsf{S}}}\,,\,\boT\,\rceil
\rightarrowtail\{\state\alpha\,,\statePsi\}
\rightarrowtail\{\statePsi\,,\:\cup\}
\rightarrowtail\cdots\rightarrowtail
\big\{\obj{\Xi}_{\!\sss\scr A}\,,\:
\Over[1.5]{\sss\scr A\!\!}{\uplus}\,;\:\ldots \big\}
\rightarrowtail\cdots\rightarrowtail
\\
\mbig[3]\{\big[\smallmatrix\mu\\\lambda\endsmallmatrix\big]\,
\state\alpha\,,\:
\uplus\mbig[3]\}\rightarrowtail\cdots\rightarrowtail
\{\frak a\,\bet{\Psi}\,,\: \Uplus\} \rightarrowtail
\big\{\{{\Over[1.5]{\sss\statePsi}{\msf m}},
{\Over[1.5]{\sss\statePhi}{\msf n}}\}\,{\state\alpha}\,,\:
\Tplus\big\}\rightarrowtail\{\frak a\bcdot\ket\alpha\,,\:
\+\}\rightarrowtail
\\
\rightarrowtail\big\{\frak a\,,\ketPsi;\:\+,\bcdot\big\}\;\hence\;
\bbH\label{stream}
\end{multline}
is considerable and is always larger than it seems. In
sects.~\ref{Ans}--\ref{minus}, we have described not all of them.
Each such homomorphism is a mapping into a representation by a
model, and for a philosophical discussion of these representations
and the origin of models, see pages 1--230 in \cite{wart}. For
another comment concerning the abstracting/"!realism, we refer to the
first half of a letter from A.~Einstein to H.~Samuel in
\cite[pp.~157--160]{einstein1}; see also \cite{einstein2}.

Thus `difficulties' with complex numbers, stricto sensu, should
already be attributed to the level of the usual negative ones.
Bearing in mind that the minus comes from the equal sign~$=$
\cite{mazur} and the equality comes from the scheme
$\eqref{o+}\rightarrowtail\eqref{g1}$, both the principia
\hyperlink{II}{\red\textbf{\textsf{II}}} and
\hyperlink{III}{\red\textbf{\textsf{III}}} are very important (and
functioning) also in the classical case. In quantum case, they are
just fundamentally unavoidable for the very creation of the theory.
The nature of \qm\ theory, of arithmetic, of complex numbers, and of
their algebras is one and the same.

Transferring the reasoning above to the natural numbers $\bb N$, the
degrees of classical and quantum abstraction become even
indistinguishable. Empirical motivation leads, in one way or
another, to the standard von~Neumann's representation for ordinals
\begin{equation}\label{ordinal}
0\DEF\varnothing\,,\qquad1\DEF\{\varnothing\}\,,\qquad
2\DEF\{\varnothing,\{\varnothing\}\}\,,\qquad
3\DEF\{\varnothing,\{\varnothing \},\{\varnothing,
\{\varnothing\}\}\}\,,\qquad\ldots\;,
\end{equation}
\ie, to using the \zf"=axiom of union: $n+1\DEF \{n\}\cup n$
\cite{kurat}. Therefore, less obvious is the $\bb N$"=numbers
themselves, to be followed by the ordering $<$, topologies,
extensions, generalizations, \etc. The formal characterization of
all of the experimental reduces then to the successive creating from
the set"=theoretic atoms---unions of sets---some direct products
thereof and mappings into other constructions of the same kind.
Hence, both the physical images `being under a ban above' and
auxiliary structures""---""dimensionalities/"!orders, \etc---should
equally become homomorphisms onto certain formal constructions
regardless of the description's classicality/"!quantumness. The
presence of, say, non"=binary operations \eqref{inv2} does not stand
out, because their nature does not differ from the one of habitual
subtraction and of division. All of these are involutory structures
that have been mathematically inherited from the empirical
meta"=requirements: repetitions
(\hyperlink{M}{\red\textsf{M}}"=paradigm), experimental context \SM, and
covariance~\hyperlink{III}{\red\textbf{\textsf{III}}} (sect.~\ref{SS}).

To close the section we add that the distancing the concepts of
state/\Courier{Data\-Source} and of a physical property is the
continuation of a more primary idea---""detaching the proper
macro"=perceptions from what is being \emph{represented}
theoretically \cite{wart, ludwig1, ludwig3} and from conceptualizing
the notions \cite{lakoff}. As B.~Mazur has noted in
\cite[p.~2]{mazur}, ``This issue has been with us \ldots\ forever: the
general question of \emph{abstraction}, as separating what we want
from what we are presented with'', \ie, the separating the `bare'
empiricism from mathematics with $\Sigma$"=limit and the number.

The atomic constituent of perception""---""sensory experience""---is
an elementary quantum event \cite{ludwig3, hart, ulfbeck, ulfbeck2},
and it begins and terminates in
$(\not\approx_{\!\!\!\sss\scr A})$"=distinguishability of
$\state\alpha$"=clicks (sect.~\ref{S}). Any continual is a
`speculative theory' (act of abstracting), not the underlying
empiricism. Therefore all the further matters---""numbers,
arithmetic, cause/"!effect, (non)""inertial reference frames, the
notions of an observer, of a classical event in the Minkowski space,
the spacetime concept itself and coordinates in the relativity
theory (a quantum view of the equivalence principle), device
read-out, tensors, composite systems, symbols $\Otimes$,
\thelike---self"=evident as it may seem, are the math add-ons, which
could originate only in the `$\cup$"=theory' of sect.~\ref{EmpMath}.
Following von~Weizs\"acker, it might be coined the name `Ur"=theory'.
There are no contradictions in observations themselves, whether we
call them macro- or micro"=scopic. Contradictions do arise in the
`mathematicae being constructed'.

\vbox{
\section{About interpretations}\label{interpret}

\flushright\tiny%
\textsl{It is \ldots\ not \ldots\ a question of
a re-interpretation \ldots\ quantum\\
mechanics would have to be objectively false, in order that\\
another description \ldots\ than the statistical one be possible}\\
\textsc{--- J.~von~Neumann} \cite[p.~325]{neumann}%
\medskip

\textsl{\ldots\ one begins to suspect that all the deep questions\\
about the meaning of measurement are really empty}\\
\textsc{--- S.~Weinberg}}
\smallskip\nopagebreak


``At this point in time it appears that a stalemate has been reached
with regard to the interpretation of quantum mechanics'' (E.~Tammaro
\cite[p.~1]{tamm}). A ``stalemate in which each side refuses to cede
territory but is unable to produce a defining argument that would
change the hearts and minds of the opponents'' (M.~Schlosshauer
\cite[p.~227]{schloss3}).

\subsection{Click, again}

The source of the ``foundational skirmishes'' \cite[p.~227]{schloss3}
and the numerous treatments of \qm\ \cite[Ch.~10]{home},
\cite[Ch.~10]{laloe}, \cite{auletta, schloss4, saunders,
zeilinger3}---`the Copenhagen' among---is the fact that the
$\state\alpha$"=event and intuitive sense of the
$\state\Psi$"=primitive (pt.~\hyperlink{S}{\red\textsf{S}}) are a~priori
endowed with physical properties, observational/"!determinative
characteristics of the \Courier{Data\-Source}, and operationality of
the canonical \qm"=concept of the ket"=vector $\ketPsi$. A
representative example in this regard is one of the first sentences
from Everett's PhD: ``The state function $\psi$ is thought of as
objectively characterizing the physical system \ldots\ at all times
\ldots\ independently of our state of knowledge of it''
\cite[p.~3]{everett}, \cite[p.~73]{barrett}; and also, on p.~48,
``The general validity of pure wave mechanics, \textsl{without any
statistical assertions}, is assumed for \textsl{all} physical systems,
including observers and measuring apparata''. And, again the
Everett's: ``The physical `reality' is assumed to be the wave
function of the whole universe itself'' \cite[p.~100]{freire},
\cite[p.~70]{barrett}. However, none of these initially exist. The
primitive $\state\alpha$"=events' abstractions
$\statePsi\GOTO{\sss\scr A}\state\alpha$ are all there is.

An important point is that the eigen $\state\alpha$"=click (of a
photon/"!electron in the \textsc{epr}"=experiment, say) should not be
identified with an $\ket{\alpha}$"=state. The latter is
re"=developable with respect to eigen"=states associated with other
click-sets of any other instruments:
\begin{equation*}
\ket{\alpha_1}=\frak b_1\bcdot\ket{\beta_1}\+
\frak b_2\bcdot\ket{\beta_2}\+\cdots=
\frak c_1\bcdot\ket{\gamma_1}\+
\frak c_2\bcdot\ket{\gamma_2}\+\cdots=\cdots\;,
\end{equation*}
which is why it is logically meaningless to attribute one
$\state\alpha$-click to that which carries the statistics of other
clicks $\state\beta_k$, $\state\gamma_k$, \ldots\ and has nothing in
common with $\state\alpha$. All the more so the click may not be
related with the physical texts---""physically descriptive
collocations such as `the measuring act on Bob's electron reveals
the spin-up state'. As in the `cat case' (sect.~\ref{cat}), the
spin-up here is a click-up rather than a state $\ket{{}{\uparrow}}$.
Similarly, a click (allegedly of a photon) with Alice/Bob has
nothing to do with distance between photons (the locality
`problem'), with speed of light, nor with a kinematic
`understanding' of the photon.

The quantum"=detection micro"=event is not a classical one as we have
been understanding it, say, in the special relativity. The ``click
does not establish the presence of something''
\cite[p.~761]{ulfbeck}, it ``is an elementary act of ``fact
creation.''\,'' \cite[Wheeler]{fuchs3}. The facts and phenomena are
made up of clicks. That is to say, the distinguished
$\state\alpha$"=clicks are not the events spaced at some distant from
each other or at different points in time. These are just clicks
without accounting to them such descriptive notions as distance,
coordinates, point of time $t$, or the picturesque words like
`dead/"!alive/"!\ldots/"!cat'; of course, the click itself has no
size/"!duration. Exempli gratia, the particles at accelerators and
their physical properties are observed not as ``material bodies in
the proper sense of the word'' \cite[p.~62]{heisenberg}---this is
impossible""---but rather through the abstract detector-snaps.
Neither the electron in interferometer nor the Higgs boson at a
collider are observed in a detector as objects that are finite in
extent; they are not observable entities. `A Higgs' is just a
frequency $5\,\sigma$"=histogram at LHC.

Likewise, the math"=properties of the eigen $\ket{\alpha}$ and of the
abstract $\ketPsi$ can in no way be `syncretized' with
$\state\alpha$"=events when they are still being accumulated. The
screen scintillation is not a photon and photon is not a
scintillation . Similarly, ``the arrival of an electron''
\cite[p.~3]{feynman} at the screen does not mean `what is here at
this point in time with a given coordinate is the materialized
particle"=electron'. Accompanying the $\ket\alpha$"=states and
$\state\alpha$"=clicks with the phenomenological and dramatic words
`up/"!down/"!\ldots/"!alive/"!dead' has nothing to do with physics, which is
yet to be created. The click should not be an element of the
language in which $\ketPsi$"=terminology, numbers, and physical
properties have been employed whatsoever.

\subsection{Abstraction the state}\label{SM}

Then something subsequently referred to as a state (the abstract)
and a measurement (the concrete) is created. However, as already
stressed in sect.~\ref{MM}, process of abstracting is a rather
multistage one (sects.~\ref{Ans}--\ref{minus}), and a reduction of
the long sequence \eqref{stream} `for physical reasons' does always
contain phenomenological axioms a~priori. Clearly, in the reverse
direction, we confront hard-to"=disentangle assumptions and the
well-known axiomatic cycle. The physical considerations and
phenomenology should not be present in fundamentals of quantum
mathematics.

To avoid paradoxes with `quantum cats'---``state vector does not
describe \ldots\ a single cat'' \cite[p.~37]{ballentine0}) and ``One
cannot think about it as in a superposition'' \cite[p.~134;
D.~Greenberger]{schloss3}, with `the presence of a particle here and
there', or with `quantum bomb"=testing' (Elitzur--Vaidman)
\cite{auletta, greenstein},
\begin{itemize}
\item it is imperative to keep a severe conceptual differentiation
 \cite[1"~st column]{mermin} between term `the state' and `physically
 sounding' adjectives/"!verbs and the spatiotemporal or cause"=effect
 images.
\end{itemize}
Similarly T.~Maudlin: ``we need to keep the distinction between
mathematical and physical entities sharp. Unfortunately, the usual
terminology makes this difficult'' \cite[p.~129]{maudlin}. Even
indirect usage of the terminology borrowed from the classical
description can be a source of confusing. For example, a so"=called
exchange interaction as a `cause of correlation' between identical
parts of system.

It seems preferable to radicalize the non"=connectivity of these
categories, \ie, to proclaim it a postulate. For instance, boldface
italics in Remark~\ref{Qcat} or selected thesis on
page~\pageref{micro}. At least, the differentiation between them
should not be neglected in reasoning, inasmuch as it seems
unrealistic to change the deeply ingrained \cite[p.~7]{touzalin} and
ill"=defined terminological locutions such as ``photon is in a certain
\emph{state of polarization} \ldots\ \emph{one} photon being in a
particular place'' \cite[p.~5, 9]{dirac}, ``an observable has/"!acquires
a (numerical) value when being measured'' \cite[p.~310;
criticism]{ludwig2}, ``outcome of a measurement'' \cite{auletta, home,
laloe}, ``quantum parallelism'' \cite[p.~282]{home}, or `simultaneous
measurability'; see sect.~\ref{S} and
pr.~\hyperlink{I}{\red\textbf{\textsf{I}}}. (As if we've had some
micro"=physics prior to math; there is nothing a~priori.) With this
mixing, the circular logic pointed out in Remark~\ref{meta} will be
present at all times. See, for instance, pages 29--30 in the work
\cite{foulis0} and notably an emphasized warning by D.~Foulis about
``\!\emph{a mistake, and a serious one\/}!'', including criticism
addressed to von~Neumann on p.~29. This
\begin{itemize}
\item trap of the `\emph{braket}ting the
 ClassPhys'---$\ket{{}\Courier{physical words}}$ or
 $\ket{{}\Courier{in}}/\ket{{}\Courier{out}}$---is the very
 ``somewhere \ldots\ hidden a concept'' that M.~Born spoke of
 (sect.~\ref{physmath}, p.~\pageref{born}); \ie, the mistaken
 `\emph{physicality of $\ket{\psi}$ and of $+$}' in~\eqref{1}.
\end{itemize}

Again (see p.~\pageref{povtor} and sect.~\ref{phys}), even the
indirect attempts to physically characterize the state function or
`reconcile' its non"=classicality with any \emph{observational
prototypes} are hopeless. ``The wave function is in the head and not
in nature''; ascribed to A.~Zeilinger (2014) by A.~Khrennikov. The
function is the very information \Courier{Data\-Source} around which
all sorts of words on physics---""readings, frequencies, objects,
phenomena, particles, events, and other entities""---are only slated
to create.
\begin{itemize}
\item ``We cannot \ldots\ manage to make do with such old, familiar, and
 seemingly indispensible terms'' (Schr\"odinger (1933)) as the
 ``\,``\:physikalische Realit\"at\:'' \ldots. ``\:Realit\"at der
 Aussenwelt\:'', ``\:Real-Zustand eines Systems\:''\,''
 \cite[p.~34]{einstein} in the way we are doing this in classical
 physics, even philosophically. To put it both informally and more
 precisely, the automatic speech---""stereotype""---``the system in a
 state'' (pt.~\hyperlink{S}{\red\textsf{S}}) \cite[criticism]{ludwig3,
 ludwig4} should be dismissed from \qt"=fundamentals, because the
 microscopy of quantum $\state\alpha$"=clicks shows that this
 colloquial habit is an unmeaning collocation.
\end{itemize}
This term may only be a theoretical conventionality in the follow-up
\emph{physical} theory. See also the first sentence in \cite{henry}
and selected theses on page~\pageref{provo} and at the beginning of
sect.~\ref{MM}.

The principled abstractness of the $\ketPsi$"=object
\cite[pp.~27--28]{gottfried} is a core attribute of quantum theory
as contrasted to the classical one. This abstraction cannot be
`struggled', it is not an idealization of something
phenomenological. It is absolute. An interpretative comprehension
like $\ket{{}\Courier{dead}}\+\ket{{}\Courier{alive}}$, even if it
is permissible, may issue only from the $\ket\alpha$"=representations
$\frak a_1\bcdot\ket{\alpha_1}\+
\frak a_2\bcdot\ket{\alpha_2}\+\cdots$, \ie, from a treatment of the
$(\+)$"=addition (of quantum amplitudes) as an accumulation of
$\state\alpha$"=microevents; many `cat boxes'. In other words,
\begin{itemize}
\item \emph{the} interpretation of the quantum state \emph{is its very
 definiendum} \eqref{axiom1}--\eqref{axiom3}. Even with the physical
 terminology created, there may be only one paraphrase for the
 meaning to the state: an abstract element of the
 abstract\footnote{The point~\hyperlink{4}{4}) in \itbf{Theorem}
 determines a supplement""---the number add-on over the utterly
 abstract \lvs.} linear (not Hilbert \cite{br3}) vector space over
 $\bbC^*$.
\end{itemize}
The ``not Hilbert'' here is because the norm and inner"=product are the
extra---""nonessential""---math add-ons over $\bbH$ \cite{br3},
which come from the follow-up introducing the Born statistics
\cite{br2}. In and of itself, the state needs none and knows nothing
of them. These concepts, similarly to the descriptive physical
notions and a measurement, will be required further, not now, for
calculation of observable quantities: math"=calculus of statistics
$\fr_k$ and of means.

We may not blend the fundamentally abstract part of quantum
mathematics""---pre"=physics and the structural properties of
$\bbH$---with those in charge of its observational/"!physical
constituent; \ie, we may not ascribe the ontological status
\cite{leifer, leggett1} to everything. In the strict sense, the
ontology of/and physics, the classical one included, cannot arise
before the statistical processing of quantum micro"=events.
(Parenthetically, the 6"~th Hilbert problem on ``Mathematical
Treatment of the Axioms of Physics'' \cite{accardi+} becomes thus an
ill"=posed problem \cite[sect.~8]{br3}.) The processing itself begins
with the Born rule \cite{br2}.

Continuing scheme \eqref{shema'}, a certain parallel takes place
between the following couples:
\begin{equation*}
\begin{array}{rr@{}c@{}l}
\text{\footnotesize observations' language:}&
\lceil\text{$\bbR$"=numbers}\rceil&\;\tplus\;&
\lceil\text{physical quantities/"!\ldots}\rceil
\\[1ex]&\downarrow\;\uparrow\qquad&&\qquad\downarrow\;\uparrow\\[1ex]
\text{\footnotesize quantum language:}&
\lceil\text{$\bbC$-,~$\ketPsi$"=objects}\rceil&\;\tplus\;&
\lceil\text{physical properties/"!data/"!\ldots} \rceil
\end{array}\;.
\end{equation*}
Just as we are not rasing the question about abstractness/"!treatment
of the \lrceil{$\bbR$"=numbers} in isolation from the
\lrceil{physical quantities/"!\ldots} (Remark~\ref{units}), so also we
should not question a treatment or the physical meaning of the
\lrceil{$\bbC$-,~$\ketPsi$"=objects}. By analogy, being torn away
from the \$"=symbol in $\$5$, the number 5 in and of itself may carry
neither the financial nor any other (`bank/"!(non)commuting/"!\ldots')
treatment; nor does it contain some hidden `microeconomic' content.
The number has no a `retrograde memory'.

The first `summands' in the aforementioned
$\smallc({\tplus}\smallc)$"=conjunctions are the abstracta of
principle. They may exist as the `math-things-in"=themselves' and we
know that they really do just as we are comprehending the existence
of the $\bb N$"=arithmetic that has been constructed in
sect.~\ref{Numbers}. The second `summands' are the interpretative
supplementations in their own rights. If a theory does not spell out
a nature of accounting the second to the first
(pr.~\hyperlink{II}{\red\textbf{\textsf{II}}}), then it is impossible to
find-out/"!guess the `true' interpretation or nonexistent physical
`protosource' of the abstracta $\msf n,\frak a$, and $\ketPsi$
`ab~intra' their algebras \eqref{distr}--\eqref{n-}, \eqref{g+},
\eqref{m+}--\eqref{m+++}, \eqref{inv2},
\eqref{axiom1}--\eqref{axiom3} or from the Hilbert-space
mathematics. See again Remarks~\ref{two}, \ref{units} and warnings
by Ludwig of ``a mistake. \ldots\ false notion that ``mathematical
objects'' must be pictures of physical objects''
\cite[p.~228]{ludwig4} and of a ``reality [of the] word ``state,'' a
reality in which one must not believe!\@'' \cite[p.~78]{ludwig5}.
A.~Peres also makes special note of the analogous: ``\ldots\ physicists
have been tempted to elevate the state vector $\psi$ to the status
of substitute of reality'' \cite[p.~645]{peres0}; and D.~Mermin puts
this as ``a regretable atavistic tendency to reify the quantum state''
\cite[p.~144]{fuchs3}.

\subsection{Measurement `problem'}\label{SM+}

The most representative example is the (in)famous problem of
measurement \cite[sect.~V.4 and Ch.~VI]{neumann}---``tyranny of
thinking of von~Neumann measurements'' \cite[p.~534]{fuchs3} with the
collapse postulate. This is the subject of an ``endless stream of
publications suggesting new theories \ldots\ unending discussions \ldots\
symposia'' \cite[p.~519]{jammer2} and of ``the mountains of
literature'' \cite[p.~118]{ludwig4} containing opposing opinions
\cite[Ch.~11]{jammer2}, \cite{mermin2, kampen, home, laloe, london,
mittel, mittel2}. It is, indeed, the source of questions around
locality in \qm. As we have seen, this problem ``is simply not a
problem at all!\@'' \cite[p.~1013]{fuchs2}. It is a
nonexistent""---``the alleged \ldots\ does not exist as a problem of
quantum theory'' \cite[p.~15]{englert}---as well as a pseudo problem
and a non-issue \cite[p.~79\,(!)]{ludwig3},
\cite[p.~118\,(!)]{ludwig4} \cite{edwards}, because
\begin{itemize}
\item in measurements, nothing either propagates (much less at
 superluminal speeds) nor interacts, nor is anything collapsed
 \cite[sect.~XVII.4.3]{ludwig2}, \cite[p.~328]{laloe},
 \cite{ballentine0, ludwig5}, nullified, localized, there are no
 such things as quantum jumps \cite{zeh}, and no ``pieces'' of the
 wave function are ``cut out'' \cite[p.~57, 158]{ivanov}.
\end{itemize}

It is no exaggeration to say that the need to projective
postulate""---``a fruit of realist thinking''
\cite[p.~172]{muynck}---is much the same as the necessity for the
world ether supporting the electromagnetic waves. All the more so
because such a view of the theme has been present in the literature
for quite a while \cite{everett, hartle, slavnov3, lipkin, klyshko,
gottfried, ludwig3, ludwig4, muynck} even as appeals.
\begin{quote}
``There is nothing \ldots\ problematic about measurement''
\\\phantom.\hfill L.~Ballentine (1996)
\\[1ex]
``\ldots\ there is no collapse of wave packets in reality. Do not
believe in fairy tales!''
\\\phantom.\hfill G.~Ludwig \cite[p.~104]{ludwig5}
\\[1ex]
``A state vector \ldots\ does not evolve continuously between
measurements, nor suddenly ``collapse'' into a new state vector
whenever a measurement is performed''
\\\phantom.\hfill A.~Peres \cite[p.~644]{peres0}
\\[1ex]
``This ``reduction'' \ldots\ is not a new fundamental process, and, \ldots\
has nothing \ldots\ to do with measurement''
\\\phantom.\hfill L.~Ballentine \cite[p.~244]{ballentine2}
\\[1ex]
``The mystifying notions arise from attributing physical reality to
the ``jump'' at a given time $t$''
\\\phantom.\hfill G.~Ludwig \cite[p.~327]{ludwig2}
\\[1ex]
``Really bad books \ldots\ claim that the state of the physical system
\ldots\ collapses into the corresponding $u_n$. This is sheer nonsense.
(Finding appropriate references is left as an exercise for the
reader.)''
\\\phantom.\hfill A.~Peres (2003)
\end{quote}
Englert \cite[p.~8]{englert} does particularly object to the
``folklore that ``a measurement leaves the system in the relevant
eigenstate'' \ldots\ It is puzzling that some textbook authors consider
it good pedagogy to elevate this folklore to an ``axiom'' of quantum
theory''. See also the 2"~nd epigraph to sect.~\ref{S0}.

The point here, put very briefly, is that the measuring `problem' is
one of principle, not of practice. Expressed by Bell's words, ``the
word [measurement] has had such a damaging effect on the discussion,
that \ldots\ it should now be banned altogether in quantum mechanics''
\cite[p.~216]{bell}. J.~Bub and I.~Pitowsky do insist in the book
\cite[p.~453]{saunders} that presumptions ``about the ontological
significance of the quantum state and about the dynamical account of
how measurement outcomes come about, should be rejected as
unwarranted dogmas about quantum mechanics''.

Another example of circular logic is the critiqued \cite{home,
stapp, stapp2} meaning of the phrase ``an ensemble of similarly
prepared systems'' \cite{ballentine, home, saunders}. The revision of
this (by and large correct) idea, as was set forth above, does
actually demonstrate that, like in the ensemble approaches, ``quantum
mechanics is a \emph{statistical} theory'' \cite[p.~2]{muynck},
\cite[p.~123]{ludwig5}, \cite[p.~223]{mises}, \cite{ballentine,
ballentine2, hren1, allah, ludwig3, ludwig5, silverman, moyal} with
a frequency content of the randomness and the classical logic
\cite{ludwig5} but with a different math"=calculus of the statistical
weights. The ``different'' is due to the fact that the theory is not
tied, as in the classical description, to the notion of an
observable quantity, and the $\fr$'s are calculated from the
`other/"!abstract' numbers \cite{br2}. However, for the same reason,
emphasizing a close resemblance with the statistical mechanics
\cite{faddeev, mackey}, \cite[p.~72]{everett} and `explanations'
with playing cards/"!dice, coins/"!balls/"!\ldots/"!urns/"!`socks'
\cite[Ch.~16]{bell} or with the classical phenomena""---unusual as
it may sound---are in error. The case in point is not a drastic
dismissal of the classical ideas, but rather a `quantum audit' of
the classical"=physics language \cite[sect.~8]{br3}. The correct
`audit' of the classical is a \emph{re}creating of this very
classical:
\begin{equation}\label{part}
\begin{aligned}
&\text{\lrceil{\wavy{classical phenomena}}}\rightarrowtail
\text{\lrceil{classical events/objects}}\rightarrowtail
\text{\lrceil{micro world}}\rightarrowtail\\
&\qquad\text{\lrceil{micro-event}}\rightarrowtail \text{\lrceil{quantum
micro-event}} \mathrel{\Over{(!)}\rightarrowtail}
\text{\lrceil{\emph{abstract} click}}\rightarrowtail\\
&\qquad\text{\lrceil{abstraction
$\statePsi\TRANS[1]{\sss\scr A}\state\alpha$}}\rightarrowtail
\text{\lrceil{$\obj{\state\Xi}$"=brace \eqref{obj}, \ldots}}
\rightarrowtail\text{\lrceil{state $\ketPsi$}}\rightarrowtail\\
&\qquad\text{\lrceil{observable concepts}}\rightarrowtail
\text{\lrceil{observable numbers}}
\rightarrowtail\cdots\rightarrowtail\\
&\qquad\text{\lrceil{statistics, the concept of a mean}}\rightarrowtail
\text{\lrceil{state, objects, \ldots, physics}}\rightarrowtail\\
&\text{\lrceil{\wavy{classical phenomena}}}
\end{aligned}
\end{equation}
and, consequently, creation of the classical concept of a measuring
process. Thus, this scheme along with quanta's statistics and
\lvs"=mathematics all add up to a positive answer to Wheeler's
question: ``Is the entirety of existence, rather than being built on
particles or fields of force or multidimensional geometry, built
upon billions upon billions of elementary quantum \ldots, \ldots\ acts of
``observer"=participancy,'' \ldots?'' \cite[p.~286]{schloss3},
\cite{fuchs3}.

\subsection{Interpretations and self-referentiality}

Although we have not yet touched on other significant
attributes""---the means over statistics, operators, and products of
$\bbH$"=spaces will be considered in their own rights---it is clear
that the need to quest for a description in terms of hidden
variables is also eliminated. Even from a formalistic perspective,
the proof of the presence/"!absence \cite{neumann, dmitr, greenstein}
of these `physical' quantities should be attributed to the semantic
conclusions of meta"=theory (\,$=$~physics) \cite{raseva}, \ie, to
theorems \emph{about} formal theory rather than to theorems of its
\emph{inner} calculus. In our case, and more generally, the formal
theory is the syntactical axioms of \qm. The corollaries of such
axioms are inherently unable to lead to statements about
interpretations \cite{raseva}, since theorems \emph{about}
object"=theory itself is not provable by means \emph{of} its
object"=language \cite{shenfield, raseva}. In a word, the nature and
interpretation of axioms are not recovered from the very axioms or
from the replacement thereof by the other ones.

A similar line of reasoning has accompanied \qt\ for quite a while:
``claim that the formalism by itself can generate an interpretation
is unfounded and misleading'' \cite[p.~38]{ballentine0}. It is known
that even the mathematics itself cannot be grounded in a
self"=contained way \cite{gray}\,(!), \cite{frenkel, kleene},
\cite[p.~201]{heisenberg2}. All of this stands in stark contrast
with the known statement of DeWitt to the effect that ``mathematical
formalism of the quantum theory is capable of yielding its own
interpretation'' \cite[pp.~160, 165, 168]{dewitt} or that
``conventional statistical interpretation of quantum mechanics thus
emerges from the formalism itself'' \cite[p.~185]{dewitt}. In
particular, if we take account of the fact that it is not the theory
itself, but only its formal interpretation that determines the very
semantic terms truth/"!falsity/"!provability of sentences (K.~G\"odel).
In turn, ``interpretation \ldots\ allows a certain freedom of choice''
\cite[p.~310]{jammer}. See also \cite{benioff} and specifically
Ch.~III in \cite{kleene}. In other words, the subconscious striving
for `to interpret'"!---and transporting the macro into the
micro---is the very thing that prevents us from truly gaining an
understanding of quantum mathematics.

In any case, the fact that we were initially constructing the
set"=theoretic model (cf.~\cite{benioff}) rather than an
interpretation simply eliminates the problem or, at most, transfers
it into the domain of questions about micro"=transitions
$\GOTO{\sss\scr A}$ and $\boT$"=family as entities being employed
(see Remark~\ref{two}). This is the domain of questions that invoke
the set theory and touch on ontological status of sets at all
\cite[sects.~V.8 and 9\,(!)]{frenkel}. Be this as it may,
logic---""formalized or non---does not allow us to make statements
about statements, much less a statement that refers to itself. The
self"=referentiality (``von~Neumann catastrophe'') is almost the chief
trouble \cite{mittel, muynck} encountered in quantum foundations.

All of this, of course, does not depend on whether interpretation is
built in a strictly formalized form \cite{shenfield} or in a
physically natural one. In effect, the issue of
interpretations""---in the rigorous definition sense
\cite[Ch.~2]{raseva}, \cite{shenfield}, \cite{stoll}---is simply
nonexistent. Accordingly, the demystification of the known and the
quest for ontological interpretations to $\frak a$"=coordinates of
the $\ketPsi$"=vector \cite{auletta, fine, hartle}---the wave
function""---is no longer a problem, and with it disappears the
Feynman question of ``the \emph{only} consistent interpretation of
this quantity'' \cite[p.~22]{feynman}. See also M.~Leifer's review
\cite{leifer} and extensive list of references therein.

\vbox{
\section{Closing remarks}

\flushright\tiny
\textsl{\ldots\ quantum mechanics has been a rich\\
source for the invention of fairy tales}\\
\textsc{--- G.~Ludwig \& G.~Thurler} \cite[p.~122]{ludwig5}%
\medskip

\textsl{I simply do not know how to change quantum mechanics\\
by a small amount without wrecking it altogether \ldots\\
any small change \ldots\ would lead to logical absurdities}\\
\textsc{--- S.~Weinberg (1994)}}
\smallskip\nopagebreak

\subsection{Language and `philosophy of quanta'}\label{phil}

Remembering and continuing sect.~\ref{physmath}, it is generally
tempting to infer that when creating the theory, we may not rest on
any meanings that are tacitly associated with the typical
terminology""---no matter physical or mathematical""---and on the
tacit assumption that customary concepts are substantially correct
\cite{heisenberg2}.

One should also be very cautious about the wording of statements
concerning the phenomena outside the everyday experience. One means
that even the very natural utterances""---`here/there, electron
\emph{with} Alice/Bob' (locality), `big(ger)/small(er)', `let there
be a two"=particle $\cal S$' (quantitative statements), `subsystem
$\cal S_1$ in such-and-such system $\cal S$, consisting of \ldots'
(statements about structure)---are de facto ``(apparently)
``plausible'' conclusions from the observed phenomena''
\cite[p.~334]{ludwig2}. These have comprised an equivalent of a
measurement/"!preparation (\cite[pp.~195--196]{muynck} and
sect.~\ref{mixture}) and of physical (pre)""imagery, and thereby
imitate the way of thinking and schemes of classical mechanics; see
also the 2"~nd epigraph to sect.~\ref{SP}.
\begin{itemize}
\item The ``particle, here/"!there, big/"!small, this/that/another one,
 before/"!after'', \thelike\ are \emph{already} the `illegal'
 observation's numbers of sorts and a premature arithmetization,
 \ie, this is \emph{already} the subconscious quantifying the
 micro"=events or the arrays thereof by a theory, and classical
 \eqref{class}--\eqref{grubo} at that.
\end{itemize}

Reality's attributes are yet only slated to create. Say, when we
decrease the particles in experiments and reach the atomic level, we
still stay in the atomistic paradigm of the particle and of numbers:
the objects having mass, their coordinates, degrees of freedom,
\etc. This is a mistaken intuition. Very informally, we should
`religionize themselves' to the quantum micro"=events, while the
return to the words `particle/"!\ldots/"!macro' must be performed by a new
reasoning mechanism. It comprises, apart from the quantum-\lvs\
apparatus (sect.~\ref{lvs}), the re"=creating the very classical
concept of the particle as schematized in~\eqref{part}.

At the other extreme is an attempt to `hurry up' and bring the
reasoning to a Hilbert"=space theory or to the quantum mixtures
\eqref{mix'}. As in sect.~\ref{2slit}, all this may well be
incorrect \cite{maudlin}. A source of antinomies is in the implicit
implying, \ie, in the eclectic""---this we stress once
again---""confounding the observations, clicks, numbers, physics,
time, math, and imagination, followed by the \emph{uncontrollable}
lexical"=`branching' such as \emph{replacing the symbol $\+$ with a
meaning} taken from reality. For instance, the emerging the word
`simultaneously' in the sequence \lrceil{the $(\+)$"=superposition of
multiple states} $\rightarrowtail$ \lrceil{simultaneously}
$\rightarrowtail$ \lrceil{quantum parallelism} \cite[p.~26]{svozil}.
W.~James has underscored that the ``\!\emph{viciously privative
employment of abstract characters and class names} is, \ldots, one of
the great original sins of the rationalistic mind''
\cite[p.~547]{fuchs3}. This results in the sense messes, well-known
no"~go (meta)""theorems \cite{bell, greenberg}, the locality
`problem' in \qm, and paradoxes such as the \textsc{epr}
\cite[sect.~XVII.4.4]{ludwig2} or the jocular Bell question: ``Was
the world wave function waiting to jump for thousands of millions of
years \ldots\ for some more highly qualified measurer --- with a
Ph.D.?\@'' \cite[p.~117]{bell}, \cite[p.~18]{laloe}, \cite{lipkin,
klyshko}. As to the no-go theorems, Ballentine remarks that ``the
growing number [thereof], combined with some peculiar terminology,
has led to confusion \ldots\ A woefully common feature, \ldots\ each
protagonist had some interpretation of the quantum state in mind,
but never stated clearly what it was'' \cite[pp.~2 and
6]{ballentine4}. Ludwig, echoing Ballentine, asks: ``But what do we
mean by the notion of a state?'' \cite[p.~5]{ludwig1}.

Clearly, the quantum-clicks do not depend on whether personified
homo sapiens interpret the arrays thereof, or a biological observer
such as a ``Heisenberg--Zeilinger dog'' \cite[pp.~171--174]{hren3},
\cite{englert, englert2} does simply perceive. The
observer""---without his ``subjective features'' \cite[pp.~55,
137]{heisenberg2} or ``the anthropomorphic notions ``specifying'' and
``knowing''\,'' \cite[p.~645]{peres0}---is just a formally logical
element \hyperlink{O}{\red\textsf{O}} in theory. Without numbers,
solely a quantitative theory is not possible (sect.~\ref{2slit}),
because the entire terminology becomes indefinite.

Thus, once a mathematics and \emph{unambiguous}
language""---spectra, means, and macroscopic dynamical models---have
been created, not only is there no longer a need to call on the
`otherworldly', eccentric, or anthropic explanation ways, but the
very presence of a certain share of mysticism, of subjectivity, and
of (circum-)""philosophy \cite{ivanov}---``a philosophical
\textsl{\"Uberbau}'' \cite[p.~12]{englert}---in quantum foundations
becomes extremely questionable. Ludwig is much more thoroughgoing in
his assessment of the language games---the ``philosophical
gymnastics'' \cite[p.~79]{ludwig3}.

Eventually, we have no longer any freedom to invent exegeses of the
quantum"=postulate as `a Bible` or ``a sacred text''
\cite[p.~1038]{fuchs3}. Moreover, the liberty to ask questions is no
longer there, since the created object"=language of states, of
spectra, and of frequencies narrows down the entire admissible
lexicon. It is able to generate questions that are not only
ill"=posed but must, as in sect.~\ref{2slit}, be qualified as
``meaningless'' \cite[p.~422]{nahman}. For example, those that are
based on a (human"=beings') \emph{intuitive} taking the term
observation or questions about `the underlying nature of reality'.
As we mentioned earlier, the notion of `a physical level of rigor'
(in reasoning) and the physical justification will not help us with
regard to the grounds of \qt. Another example is the attempts at (or
``to refrain from'') ``tying description to a clear hypothesis about
the real nature of the world'' (Schr\"odinger (1933)) and, in
general, the question of `how it should function' at the
micro"=level. See also \cite[p.~100]{ludwig5} on ``reality''.

In the classical framework, the language sentences are always
interrelated, since \emph{all} of them, one way or another, handle
the observational notions. In the famous Como~address, N.~Bohr had
remarked that ``every word in the language refers to our ordinary
perception''. These notions, in medias res, form our natural speech
when describing experiments but are inadequate in the quantum
\cite{heisenberg2}\,(!). That is, these concepts do not make clear
the fact that behind the \qt\ are some structureless
abstracta""---we believe that these are
$\{\statePsi\GOTO{\sss\scr A}\state\alpha,\; \cup\}$ and procedures
\eqref{stream}---rather than an `improved' physicomathematical
axiomatics or sophisticated math vehicles; \eg, non"=commutative
calculus.

The language intuition usually makes it easy for us to do away with
paradoxes the semantic closedness causes. However, the quantum
situation is just the one of a misuse of the vocabulary, \ie, when
contradictions are inevitable, and this unlimited source of
confusion demands a \emph{controlling the language over itself}. One
does create the other (`relative') languages within itself
\cite{chomsky+}; at first, the language of quantum mathematics and
thereafter the language of math"=physical description and of
classical physics, followed by the language of the semantic
interpretations. This is just what we call the metamathematics and
math"=logic \cite{kleene}, discriminating between metamathematics and
philosophy \cite{raseva}. If this is not the case---the `quantum
conclu\-sions' from thinking (even if partly/"!implicitly) in terms of
physical influences between the classical objects (Deutsch's ``bad
philosophy'')---then we obtain an everlasting source of paradoxes,
since human intuition has roots in the classical world and is a
rather problematic and personal category. A.~Stairs calls upon
``Don't trust intuition'' \cite[p.~256]{stairs} because it is not
meant for \qm.

Inasmuch as the conceptual autonomy in quantum fundamentals is
minimal (sect.~\ref{S0}), the quantum scheme of things must commence
with an extremely `ascetic' language (Remark~\ref{meta}), and it
should be independent of our intuitive knowledge, which ``tend to
declare war on our deductions'' (van~Fraassen). To avoid collisions
between theory and meta"=language, the subconscious striving of the
natural language to include one in the other has to be limited.
Einstein adds also the situations when ``er f\"uhrt dazu, \"uberhaupt
alle sprachlich ausdr\"uckbaren S\"atze als sinnleer zu erkl\"aren''
\cite[p.~33]{einstein}. A.~Leggett's comments on ``pseudoquestions''
and ``gibberish'' at the end of sect.~\ref{super} may then be
strengthened so that the meaninglessness by itself should become a
constitutive element of language, including the language of
`philosophy of quanta'.
\begin{itemize}
\item The\eLab{forbiden} rudimentary quantum (meta)""mathematics creates
 the notion of a \emph{prohibited} statement/"!phrase/"!question, one
 that is devoid of meaning. This are sentences that involve the
 classical analogies in the circumvention of 1) the
 $\ket{\alpha}$"=representatives to the non"=interpretable abstraction
 $\ketPsi$ and of 2) the numerical quantities' nature
 (sect.~\ref{divis}).
\end{itemize}

It is appropriate at this point to quote the 't~Hooft remark ``I go
along with everything [Copenhagen] says, except for one thing, and
the one thing is you're not allowed to ask any questions'' and the
Einstein reasoning on page~669 in the collected articles
\cite{schilpp}: ``One may not merely ask \ldots\ not even ask what this
\ldots\ \emph{means}''. See also Heisenberg's discussion of the problem
\lrceil{language $\rightleftarrows$ concepts} on pages 48--54 in
\cite{heisenberg}, his work \cite{heisenberg+}, the pages 234--235
in \cite{schilpp} with Bohr's appeals regarding the ``necessity of a
radical revision of basic principles for physical explanation \ldots\
revision of the foundation for the unambiguous use of elementary
concepts'', and his comments on words ``phenomena'', ``observations'',
``attributes'', and ``measurements'' on p.~237.

The literature on this subject, even taking only the qualified
sources into account, is vast \cite{aaronson, barrett2, alter,
deutsch, espagnat, fine, fuchs4, greenberg, Hooft, maudlin, mermin,
mermin2, ney, saunders, schloss, schloss3, greenstein, kadom, hren1,
laloe, home, stapp2, auletta} and abounds with
terminology""---``words, \emph{ostensibly} English'' (A.~Leggett
\cite[p.~300; emphasis ours]{laloe})---that defies translation into
the language of events or of concretization: observer's
consciousness, parallel/"!branching universes/worlds, free will,
catalogue of knowledge, world branch, and also such collocations as
rational agent, information (\emph{``Whose''} and ``about
\emph{what}?\@'' \cite{bell}, by ``Bell's sardonic comments''
\cite[p.~262]{home}) has been recorded/"!transmitted/(not)""reached an
observer (Wigner's friend), teleporting a state,
many-minds/"!worlds/"!words, quantum psychology, psycho"=physical
parallelism (in this connection, see \cite[p.~86\,(!)]{chomsky}), and
many other ``bad words'' by Bell. He italicizes them on p.~215 of
\cite{bell}.

Of course, ``without philosophy, science would lose its critical
spirit and would eventually become a technical device''
\cite[p.~800]{auletta} but, on the other hand, ``the concept of the
free will cannot be defined by indications on devices''
\cite[p.~151]{ludwig4}, and ``one must not confuse physics with
philosophy'' \cite[p.~12]{englert}. Furthermore, yet, we should like
to remind a Heisenberg attitude \cite{heisenberg+} on ``a
misconception \ldots\ [and `possibility'] to avoid philosophical
arguments \ldots\ and the way of thinking of \ldots\ physicists who
insisted on not dealing with philosophy''. Namely, ``[w]e can not
avoid using a language bound up with the traditional philosophy''.
One cannot but mention the Rovelli article \cite{rovelli} that is
entirely devoted to this topic. Therefore, ``[i]t must be our task to
adapt our thinking and speaking""---indeed our scientific
philosophy""---to the new situation'' with regard to \emph{abstract}
meaning of the linear quantum addition $\+$ and quantum math
altogether; all of the quotations are from pages 32 and 37--38 of
work \cite{heisenberg+}.

As concerns the attitudes towards \qm---at the suggestion of
M.~Tegmark in the 1990's, there even carried out polls and
statistical analysis of their correlations \cite{zeilinger3}. There
are also known attempts to involve here the biology of
consciousness/"!brain \cite{stapp2, tegmark},
\cite[sect.~6]{schloss6}, \cite[Ch.~9]{schloss}. Regarding them,
however, there have been not merely sceptical but quite the opposite
opinions \cite[sects.~17.5--6]{wiseman},
\cite[sect.~XII.5\,(!)]{ludwig4}. Of special note are Ballentine's
remark ``to stop talking about ``consciousness'' or ``free will''\,'' on
the last page of the preprint \cite{ballentine4} and Popper's
criticism as to ``the alleged \ldots\ \emph{intrusion of the observer,
or the subject, \textup{[or of consciousness]} into quantum theory}
\ldots\ based on bad philosophy and on a few very simple mistakes''
\cite[pp.~11, 17, 42; everything as in the original]{popper} with an
appeal ``to exorcize the ghost called ``consciousness'' or ``the
observer'' from quantum mechanics'' \cite[p.~7]{popper}. ``[Q]uantum
mechanics is a physical theory, not psychology''
\cite[p.~83]{muynck}.

\subsection{Math-`assembler' of quantum theory}

As a result, we gain ``a contribution to philosophy, but not to
physics'' \cite[p.~86]{mackin}. At the same time, the proposed math
`$\cup$"=assembler'
\begin{equation*}
\boxed{\state\alpha_j\not\approx\state\alpha_k\,,\qquad
\statePsi\GOTO[3]{\scr A}\state\alpha\,,\qquad
\text{$\obj{\state\Xi}$"=brace \eqref{split'}\,,\qquad
$(\Vee,\in,\cup)$"=logic \eqref{union}--\eqref{cup}}}
\end{equation*}
is quite sufficient for creating the object"=language. Giving a
natural form to it would be acceptable, however, it is clear that
the set"=theoretic $\cup$"=base of the language cannot be avoided
\cite{benioff, frenkel}. Nevertheless, the syntactically more formal
description of the sequence \lrceil{transitions $\rightarrowtail$
brace $\rightarrowtail$ numbers} is surely of interest until the way
of looking at quanta's mathematics is harmonized with the
math"=logic. This would turn, however, all the above material into a
pure"=logic text, which we eschew in the present work. It is probably
for this reason that the very important and extremely thorough
works\footnote{Pre"=theories, 76 axioms \cite[p.~241]{ludwig3},
ordered sets, morphisms, absence of the word superposition in
monographs \cite{ludwig1, ludwig3}, the (valid) criticism of
``theories of \ldots\ so-called states'' \cite[p.~78]{ludwig5}, \etc.} by
G\"unther Ludwig \cite{ludwig1, ludwig2, ludwig3, ludwig4} and by
his school are often left out of the literature on quantum
foundations. Among other things, in spite of explicit pointing out a
``solution in principle of the measuring problem'' in
\cite[p.~V]{ludwig3} and subtitle ``Derivation of Hilbert Space
Structure'' of \cite{ludwig3}, name of this author has not been
mentioned in the detailed reviews \cite{leifer, schloss4, zurek} and
even in the books \cite{silverman, fuchs4, schloss, mittel,
saunders, ney}.

\subsection{Well, where's probability?}

An answer to this question in \emph{quantum} elements is brief
enough---nowhere. ``There is no probability meter'' \cite[p.~185;
S.~Saunders]{saunders}, and relationship of this concept with
empiricism is unique \cite[p.~46]{ballentine2}---the statistical
proportions $\fr_k$. Cf.~the famous de~Finetti (1970) claim that
``probability does not exist'' \cite{fuchs4} and A.~Khrennikov's
remarks to the effect that ``\!\emph{the only bridge between ``reality''
and our subjective description is given by relative frequencies}''
\cite[p.~139]{fuchs3} and that ``Experimenters are only interested in
\ldots\ frequencies'' \cite[p.~36]{hren9}. Or, more carefully stated by
von~Mises' words,
\begin{quote}
``If we base the concept of probability, \emph{not on} the notion of
relative frequency, \ldots\ at the end of the calculations, the meaning
of the word 'probability' \itbf{is silently changed} from that
adopted at the start to a definition based on the concept of
frequency'' \cite[p.~134; all the emphasis ours]{mises}.
\end{quote}

Indeed, suppose that the word frequencies has been banned
\cite[p.~44]{mittel2} in substantiating the \qt"=elements; so also
have the usage of the words `over/"!repetition/"!\ldots/"!statistics'. Then
the questions do immediately arise: why the Kolmogorovian axiomatic,
and why does it have this very quantification? \ie, why
zero/"!one/"!\ldots/"!positive? why not the $(-1\ldots1)$"=interval? whence the
single"=case probability postulates? \ldots\ subjectivity? Well, what is
the quantification thereof, and what does subjectivity do in the
natural"=scientific theory?
\begin{quote}
``\ldots\ it is very doubtful that quantum probabilities can be
introduced as a measure of our personal belief. Well, it may be
belief, but belief based on frequency information''
\\\phantom.\hfill A.~Khrennikov; V\"axj\"o Conference (2001)
\end{quote}

One way or the other, quantum foundations \emph{would demand an
interpretation} of Kolmogorov's axioms (besides, these are not
categorical in contrast to \lvs), and the latter, in turn, demand
interpreting the concept of the number---an axiomatic add-on
\emph{over} the \zf~theory \cite{kurat}.

Bearing in mind the primary nature of numbers and nontriviality of
their emergence in physical theory (sect.~\ref{arif}), it is not
just impossible to avoid the statistical weights $\fr_k$
\cite[p.~25]{peres}. Logic forbids them from being subsidiary with
reference to probability in any definition: ``probability is the
picture for reproducible frequencies; and it is the [only]
\emph{prescription} for a \emph{correct} experiment''
\cite[p.~144]{ludwig4}. Pauli, among the few, had been ``convinced
that
\begin{itemize}
\item the concept of `probability' should not occur in the fundamental
 laws of a satisfying physical theory''.
 \\\phantom.\hfill (an excerpt from his 1925 letter to Bohr)
\end{itemize}
Ensemble empiricism, for its part, is self"=sufficient, and the only
conventionality within it is an infinite number of repetitions. In
this connection, we cannot agree with a statement of theorem~III in
van~Kampen's work \cite[p.~99]{kampen} and with further comment as
to ``a single system'' and ``calculation of spectra''. At the same time,
for formalizing the infinite, there is an appropriate axiom
\emph{in} the \zf"=theory \cite{kurat, haus}.

To say all this still informally, any non"=statistical/"!non"=ensemble
framework for what we have been calling \qm"=probability does
explicitly or implicitly""---if the expression may be
tolerated""---`parasitize' on statistics by addressing the words
``repetitions, the multiplied, \ldots'' and, at the same time, does
`attract the empirically vague justifications' in terms of
anthropomorphic surrogates: potentiality, tendency, propensity, the
amount of ignorance, subjective uncertainty \cite{saunders} or
likelihood, degree of belief, \thelike\ \cite{fine}. But even from
philosophical point of view ``probability is a deeply troublesome
notion'' \cite[p.~78; L.~Hardy]{schloss3}, that is supported by the
vast literature on this subject \cite{hren0}\,(!), \cite{szabo+}\,(!),
\cite[pp.~41--43]{sudbery}, \cite[Chs.~3--4]{sklar}, \cite{auletta,
accardi, deutsch, hren3}. According to Deutsch, D.~Papineau calls
``this state of affairs \ldots\ a scandal'' \cite[p.~550]{saunders}.

An Einsteinian ``scientific instinct'' \cite[p.~174]{home} against the
probability is very well known \cite{jammer}, and Pauli, again, had
been recollecting his (Einstein's) frequent remarks in this regard:
``One can't make a theory out of a lot of `maybe's' [\,$=$~probably]
\ldots\ deep down it is wrong, even if it is empirically and logically
right''. More to the point, the question of what exactly is meant by
a probability \emph{event}, \ie, ``Probability of \emph{what}
exactly?\@'' \cite[p.~228]{bell} is also a matter of principle. The
answer to it, as seen above, is this: `not of the classical events',
\ie, ``[n]ot of the \ldots\ \emph{being}'' \cite[p.~228]{bell} such as
`\qm"=cats', `particle is here/"!there', `roll of the dice', \thelike.
An excellent text about probability and the aspects of the
probability"=physical constructs is the work \cite{alimov}. Its
`verdict' concerning the treatment of this concept \cite[sects.~4.5
and 8]{alimov} is clearly Misesian \cite{mises}, \ie\ the
``\emph{ensemble} and \emph{frequency}'' \cite[p.~xiii]{hren0}.

Thus to sum up, the philosophy/"!axioms of probability or its `quantum
deformations' should not be present in \emph{quantum} foundations.
There cannot be hidden details underlying the quantum probability
because the ``details'' imply some terminology with a classical
content. Quantum probability is the statistical regularity. It comes
from Kollektivs \cite{mises} of abstracta \eqref{split'} and may
only be a shortened term for the relative ``frequencies in long runs''
(von~Neumann); ``the Einstein hypothesis'' by M.~Jammer
\cite[p.~441]{jammer2}. The realistic/"!physical/"!\ldots/"!pictorial
adjectives and descriptive supplementations to the term ``long runs''
are prohibited. This is why the conventional tractability of the
quantum"=postulates' mathematics, \ie, the calculation of
\emph{probabilities for the classical} events to occur in the
reality""---`alive/"!cat/"!\ldots/"!imploded/"!bomb'---``is not adequate''
neither as a doctrinal point of departure nor as a post-math
interpretation. It presents us with a circulus vitiosus of
re"=exegeses. Fuchs, referring to de~Finetti's words in an interview
with Quanta Magazine (4~June 2015), prognosticates that this
conception ``will go the way of phlogiston''. The ``not adequate'' is a
R.~Haag quotation, and he expresses this ``conviction'', applying it
even to ``the conceptual structure of standard Quantum Theory''
\cite[p.~743]{haag}.

The ultimate conclusion completes Remarks~\ref{two} and \ref{seven}.
If we accept the set"=theoretic eye on things then sect.~\ref{inv},
by all appearances, provides a positive answer to the question about
the rigidity of \qt\ \cite{colbeck}---``change any one aspect, and
the whole structure collapses'' \cite[p.~1]{aaronson}; see also the
second epigraph to this section. At least, it is hard to imagine
\emph{what any other} axiom-free way of turning empiricism into
quantum mathematics would look like, as soon as we abandon the
primitive minimality of the scheme
\begin{equation*}
\text{\lrceil{distinguishable micro $\state\alpha$"=events}} \tplus
\text{\lrceil{ensembles of abstracta
$\statePsi\GOTO{\sss\scr A}\state\alpha$}}\;.
\end{equation*}

\section*{Acknowledgments}

The author wishes to express his gratitude to the QFT"=department
staff of TSU for stimulating conversations. A special word of thanks
is due to Dr.~Ivan Gorbunov and to Professor V.~Bagrov, who
initiated considering the matters on quantum logic
\cite{beltrametti, foulis0, foulis}.

The work was supported by the Tomsk State University Development
Programme (Priority""--2030).

\end{document}